\documentclass[journal]{IEEEtran}
\usepackage{setspace}
\usepackage{cite,graphicx,amsmath,amssymb}
\usepackage{hyperref}
\usepackage{subfigure}
\usepackage{fancyhdr}
\usepackage{mdwmath}
\usepackage{enumitem}
\usepackage{mdwtab}
\usepackage{balance}
\usepackage{xcolor}
\usepackage{bm}
\usepackage{amsthm}
\usepackage{threeparttable}
\usepackage{multirow}
\usepackage{amsmath}
\usepackage{flafter}
\usepackage{makecell}
\usepackage{diagbox}
\usepackage{tcolorbox}
\usepackage{physics}
\usepackage{bbding}
\usepackage{pifont}
\usepackage{algorithm}
\usepackage{algorithmic}
\usepackage{booktabs}
\usepackage{mathtools}
\usepackage{enumitem}

\theoremstyle{definition}
\newtheorem{remark}{Remark}
\newtheorem{theorem}{Theorem}

\newtheorem{lemma}{Lemma}

\newtheorem{corollary}{Corollary}

\makeatletter

\begin{document}
\title{Pinching-Antenna Systems (PASS): A Tutorial}
\author{Yuanwei Liu, \IEEEmembership{Fellow, IEEE}, Hao Jiang, Xiaoxia Xu, Zhaolin Wang, Jia Guo, Chongjun Ouyang, Xidong Mu, Zhiguo Ding, \IEEEmembership{Fellow, IEEE}, Arumugam Nallanathan, \IEEEmembership{Fellow, IEEE}, George K. Karagiannidis, \IEEEmembership{Fellow, IEEE}, \\and Robert Schober, \IEEEmembership{Fellow, IEEE}
\thanks{Yuanwei Liu is with the Department of Electrical and Electronic Engineering, The University of Hong Kong, Hong Kong, and also with the Department of Electronic Engineering, Kyung Hee University, Yongin-si, Gyeonggi-do 17104, Korea (e-mail: yuanwei@hku.hk).}
\thanks{Hao Jiang, Xiaoxia Xu, Jia Guo, and Chongjun Ouyang are with the School of Electronic Engineering and Computer Science, Queen Mary University of London, London, E1 4NS, U.K. (email: \{hao.jiang, x.xiaoxia, jia.guo, c.ouyang\}@qmul.ac.uk).}
\thanks{Zhaolin Wang is with the Department of Electrical and Electronic Engineering, The University of Hong Kong, Hong Kong (email: \{yuanwei, zhaolin.wang\}@hku.hk).}
\thanks{Xidong Mu is with the Centre for Wireless Innovation (CWI), School of Electronics, Electrical Engineering and Computer Science, Queen's University Belfast, Belfast, BT3 9DT, U.K. (x.mu@qub.ac.uk).}
\thanks{Zhiguo Ding is with the School of Electrical and Electronic Engineering (EEE), Nanyang Technological University, Singapore 639798 (zhiguo.ding@ntu.edu.sg).}
\thanks{A. Nallanathan is with the School of Electronic Engineering and Computer Science, Queen Mary University of London, London and also with the Department of Electronic Engineering, Kyung Hee University, Yongin-si, Gyeonggi-do 17104, Korea (mailto:a.nallanathan@qmul.ac.uk)}
\thanks{G. K. Karagiannidis is with the Wireless Communications and
Information Processing (WCIP) Group, Electrical \& Computer Engineering
Dept., Aristotle University of Thessaloniki, 54 124, Thessaloniki, Greece (e-mail: geokarag@auth.gr).}
\thanks{Robert Schober is with the Institute for Digital
Communications, Friedrich-Alexander-University Erlangen-Nurnberg (FAU), Germany (e-mail:
robert.schober@fau.de).}
}
\IEEEspecialpapernotice{(Invited Paper)}
\maketitle
% Authors
\begin{abstract}
Pinching-antenna systems (PASS) present a breakthrough among the flexible-antenna technologies, and distinguish themselves by facilitating large-scale antenna reconfiguration,
line-of-sight creation, scalable implementation, and near-field benefits, thus bringing wireless communications from the ``last mile" to the ``last meter".
To illustrate the benefits of PASS in next-generation wireless networks, a comprehensive tutorial is presented in this paper. 
First, the fundamentals of PASS are discussed, including PASS signal models, hardware models, power radiation models, and pinching antenna (PA) activation methods. 
Building upon this, the information-theoretic capacity limits achieved by PASS are characterized, and several typical performance metrics of PASS-based communications are analyzed to demonstrate its superiority over conventional antenna technologies.
Next, the pinching beamforming design is investigated. 
The corresponding power scaling law is first characterized for the single-waveguide single-user case. For the joint transmit and pinching design in the general multiple-waveguide case, 1) a pair of transmission strategies is proposed for PASS-based single-user communications to validate the superiority of PASS, namely sub-connected and fully connected structures in terms of the connections between the baseband radio frequency chains and waveguides; and 2) three practical protocols are proposed for facilitating PASS-based multi-user communications, namely waveguide switching, waveguide division, and waveguide multiplexing. 
A possible implementation of PASS in wideband communications is further highlighted.
Moreover, the channel state information (CSI) acquisition in PASS is elaborated with a pair of promising solutions, based on pilot-based channel estimation and beam training, respectively.
To overcome the high complexity and suboptimality inherent in conventional convex-optimization-based approaches, machine-learning-based methods for operating PASS are also explored, focusing on selected deep neural network architectures and training algorithms.
Finally, several promising applications of PASS in next-generation wireless networks are highlighted to motivate future works.
\end{abstract}
\begin{IEEEkeywords}
Pinching-antenna systems, beamforming design, channel state information acquisition, flexible-antenna technologies, performance analysis, machine learning.
\end{IEEEkeywords}

\section{Introduction}
With the worldwide deployment of the fifth-generation (5G) communication network \cite{dang2020what}, the upcoming sixth-generation (6G) wireless communication technologies are anticipated to bring a revolutionary leap forward, offering a huge improvement across all dimensions of 5G performance \cite{jiang2021the, you2021towards, wang2023on}. 
Among all the candidate technologies targeting the realization of the 6G vision, multiple-input multiple-output (MIMO) plays an important role, whose very concept was first proposed in the 1990s \cite{winters1984optimum}.
Examining the historical advancement from second-generation (2G) to 5G, MIMO continues to push the boundaries of wireless communication networks by offering considerable multiplexing and diversity gains \cite{yang2015fifty}.

To cater to the key performance indicators (KPIs) in 6G, MIMO has evolved into more advanced forms, from massive MIMO \cite{larsson2014massive}, to gigantic MIMO \cite{bjorson2015enabling}, and finally to continuous aperture arrays (CAPAs) \cite{liu2025capa}.
Although these variants of MIMO show great potential for 6G, their practical implementations are challenging and potentially hindered by three facts: 1)~\emph{high computational complexity}, 2) \emph{heavy channel estimation overheads}, and 3) \emph{challenging in manufacturing}.
To alleviate these problems and expedite real-world deployment, increasing the flexibility of antenna systems has become a technical trend in the communication research community, leading to the emergence of the novel concept of flexible-antenna technologies. 
Generally speaking, flexible-antenna technologies provide the proactive reconfiguration of the wireless channel, thus creating favorable propagation conditions.
Compared to conventional MIMO technologies, the flexible-antenna technologies make channel-related parameters tunable, thereby providing an additional design dimension.
From a historical perspective, there is a long quest towards flexible-antenna designs from the antenna selection in the 2G era \cite{sanayei2004antenna} to meta-surface antennas in the 6G era \cite{faenzi2019metasurface}. 

In recent years, pinching-antenna systems (PASS) have emerged as a promising flexible-antenna technology for 6G, and the first prototype was demonstrated by NTT DOCOMO at the Mobile World Congress (MWC) Barcelona in 2021 \cite{suzuki2022pinching, nttdocomo2022pinching}.
The structure of PASS is simple, consisting of two essential parts: dielectric waveguides and separate dielectric pinching antennas (PAs) (referred to as particles in \cite{nttdocomo2022pinching}).
The waveguides serve as the transmission medium used for carrying intended signals over long distances (typically on the scale of tens of meters) with low attenuation (0.01 dB/m) \cite{ding2025flexible, yeh2000communication}.
By integrating PAs along the waveguides, the signals guided through the waveguides can be radiated to free space at each PA's location. 
Unlike flexible-antenna technologies, such as fluid antenna systems (FAS) \cite{new2024tutorial} and movable antennas (MA) \cite{zhu2024movable}, which offer only wavelength-scale reconfigurability of antenna positions, PASS's waveguides can span tens of meters, and hence, PAs can be activated on numerous candidate positions for transmission/reception of signals. 
This enables PASS to achieve far greater reconfigurability, which facilitates mitigating not only small-scale fading but also large-scale fading \cite{liu2025pinching, yang2025pinching, suzuki2022pinching}.
By way of analogy, PASS can ``pass" intended signals to users by placing PAs close to the users' locations \cite{liu2025pinching}. 
This unique capability has recently garnered considerable attention within the research community \cite{liu2025pinching}.

\subsection{The Road to Flexible-Antenna Systems}
The core idea behind all flexible-antenna technologies is to proactively reshape the transmission conditions in a desired manner.
Having emerged from the path from conventional MIMO to gigantic MIMO, flexible-antenna technologies avoid relying on a large number of expensive and energy-hungry RF chains, thus enabling low-complexity and energy-efficient transmission.
Over the past 20 years, a significant amount of research has been devoted to flexible antenna systems.

The first attempt to make antennas flexible was the antenna selection (AS) proposed by \cite{win1999analysis} in 1999.
The basic idea of AS is to harness diversity gains by switching antenna positions that experience independent fading.
Soon after, the performance limits of AS were rigorously examined in \cite{bachceci2003antenna, molish2004MIMO, sanayei2004antenna, molisch2005capacity}.
From a flexible-antenna standpoint, AS marked the first flexible-antenna prototype, which leverages the customizations of small-scale fadings.
Moreover, AS was formally standardized in the 3rd generation partnership project (3GPP) TS 36.213 (version 8.3.0, Release 8) \cite{3gpp_ts36}.
However, when MIMO systems are deployed in high-frequency bands, such as millimeter-wave (mmWave) and terahertz (THz) \cite{han2022terahertz}, the corresponding channel matrices become sparse and highly correlated. 
This potentially violates AS’s fundamental assumption of independently fading antenna ports, thereby eroding the selection-diversity gain. 

With the advancements in meta-surface technologies \cite{hum2014reconfigurable}, reconfigurable intelligent surfaces (RISs) were introduced to the communication research community \cite{hu2018beyond, wu2019intelligent, huang2019reconfigurable}.
RISs exploit low-cost passive reflecting elements to reconfigure wireless channels in a desired manner \cite{wu2021intelligent, liu2021reconfigurable}.
Compared with AS, RISs provide a virtual line-of-sight (LoS) path for wireless transmission, thereby reconfiguring the wireless channel.
With this unique feature, RISs have evolved to offer more advanced functionalities.
For instance, simultaneously transmitting and reflecting (STAR) RISs \cite{mu2022simultaneously} facilitate signal transmission of both sides of the space surrounding the meta-surface.
Moreover, beyond acting as a reflecting surface, meta-surfaces can also be exploited for transmission purposes, thus giving rise to CAPA \cite{liu2025capa}.
Instead of employing discrete antenna elements, CAPA treats the entire meta-surface as one antenna array with continuous placement of antenna elements.
Thereby, the flexibility of CAPA can be realized by designing a continuously distributed current over the aperture.
With a continuous aperture, CAPA can significantly enhance the throughput and sensing functionality of wireless signals \cite{liu2025capa}. 
Furthermore, to flexibly reconfigure large-scale fading, surface wave communication (SWC) via meta-surfaces was introduced for communication systems for the first time in \cite{wong2021vision}, which utilized surface waves trapped at the interface between two different media to propagate, i.e., signals are sent in a non-radiative manner.

On a parallel track, FAS \cite{new2024tutorial} and MA \cite{zhu2025tutorial} have emerged as two promising technical trends, which realize flexible-antenna design by introducing antenna position flexibility.
Different from meta-surface-based antennas, FAS and MA provide the spatial flexibility by altering the position of antenna ports through activation or mechanically enabled movement \cite{wong2020fluid6g, wong2020fluid, wong2020performance, zhu2024historical}.
Compared to conventional fixed-position MIMO, FAS and MA can reduce the number of antenna elements needed to achieve high throughput, thus lowering the implementation cost.
Recently, MA has evolved into more advanced forms, leveraging three-dimensional positioning and three-dimensional rotation \cite{shao20256d_twc, shao20256d_jsac}, referred to as a six-dimensional movable antenna (6DMA).
To cater for large-scale channel reconfigurability, some research endeavors have been dedicated to allowing for meter-scale movement of antenna elements. 
For instance, in the first attempt in MA, the authors of \cite{zhu2025movable} considered a near-field setup, in which a larger movable region leads to a non-negligible near-field effect and facilitates near-field communications.
Recently, the authors of \cite{fu2025extremely} introduced the concept of extremely large-scale MA, in which the movable area of MA exceeds 100 square meters.
Owing to the extremely large aperture size, i.e., the movable region for MAs, the pathloss can be reduced, thereby enhancing communication performance.
Finally, \cite{lu2025wireless} introduced a novel idea of operating MA from the sky, where UAVs are equipped with MA arrays.
Thanks to the large-scale three-dimensional maneuvers of unmanned aerial vehicles (UAVs), the potential of MA is further unleashed.

Another class of flexible antennas is waveguide-based flexible-antenna technology, in which part of the signal propagates within the waveguide before being radiated into free space \cite{araghi2021holographic}.
Therefore, the signal in this system can be divided into two stages: a guided mode and free-space radiation \cite{rezvani2025energy}.
There are two approaches to realize the transition from the guided mode to free-space radiation: 1) meta-surface-based leaky-wave antennas, and 2) leaky coaxial cable antennas.
In the first category, meta-surface antennas are exploited to radiate wireless signals, where the amplitude of the guided signals can be freely altered for beamforming.
This category includes the reconfigurable holographic surfaces (RHS) \cite{deng2021reconfigurable, yurduseven2017dual} and dynamic meta-surface antennas (DMAs) \cite{nir2021shlezinger}.
In particular, RHS can be regarded as a meta-surface occupied by a large number of radiating elements fed by one or multiple feeds.
DMA, on the other hand, is composed of multiple waveguides arranged in parallel and fed by dedicated radio-frequency (RF) chains. 
Using these antennas, the beam patterns can be manipulated into a desirable fashion through dedicated holographic antenna reconfiguration.
Compared with conventional fully digital and hybrid-beamforming in conventional MIMO, RHS and DMA distinguish themselves by low fabrication cost, high flexibility, and high energy efficiency.
The second approach to waveguide-based flexible antennas uses a leaky waveguide that distributes the radiation along its length, exemplified by leaky coaxial cables (LCX) \cite{jun200theory, delongne1980underground}.
More specifically, LCX is essentially a coaxial transmission line with the outer layer of the conductor being slotted for wave leakage.
Compared to meta-surface-based leaky wave antennas, LCXs can extend to tens of meters and are physically flexible, making them a favorable choice for communication in tunnels, mines, and underground facilities \cite{wu2017an}.
On the other hand, unlike the continuous signal leakage along the waveguide in LCX, PASS controls the signal leakage positions by delicately placing PAs.
Thus, by adjusting the positions of the PAs on the waveguides, which may extend several meters, large-scale fading can be flexibly customized, since the power dissipation within waveguides is negligible.

\subsection{PASS: From ``Last Mile" to ``Last Meter" Communications}
The integration of PASS into telecommunication systems was first investigated in \cite{ding2025flexible}, furnishing theoretical proof of its effectiveness in reconfiguring large-scale fading.
In particular, by adjusting the position of PAs, the electromagnetic (EM) energy or, equivalently, the wireless signals can be emitted at locations close to users.
Thereby, the pathloss experienced by the signals will be reduced by allowing part of the propagation to happen within the low-attenuation waveguides, thereby creating a \emph{near-wire} connection \cite{liu2025pinching, yang2025pinching}.
As such, this approach mitigates the severe pathloss of EM waves, effectively shifting the wireless paradigm from the ``last mile" to the ``last meter."
In the following, we outline the key features of PASS, highlighting its distinct differences from existing flexible-antenna technologies.

\subsubsection{Large-Scale Antenna Reconfiguration}
The most noticeable feature of PASS is its large-scale antenna reconfiguration, implying that the antenna positions can be reconfigured on the scale of meters.
In comparison, other flexible-antenna technologies merely allow for antenna repositioning on a scale of wavelengths, which limits their flexibility.
Thereby, intended signals can be emitted right above the receivers, thus greatly enhancing the received signal power \cite{ouyang2025array} and boosting the energy efficiency of the communication system \cite{wang2025modeling}.
Moreover, if multiple waveguides are deployed, a vast two-dimensional area can be served by PASS with a small number of RF chains.
In light of some pioneering work, a high multiplexing gain can be achieved by carefully selecting the positions of the PAs and designing beamforming vectors \cite{zhao2025waveguide}.

\subsubsection{Line-of-Sight Creation}
With the meter-scale mobility of PAs, PASS can create LoS propagation conditions for transmission, which are a hundred times stronger than non-LoS (NLoS) links \cite{yang2025pinching}
While high-frequency bands offer abundant bandwidth to support massive connectivity, they exhibit little multi-path propagation and vulnerability to blockage.
Overcoming such blockages with conventional flexible-antenna technologies would require physically repositioning over tens of meters, which cannot be realized by existing technologies. 
In particular, FAS and MA can only provide wavelength-scale flexibility, while RISs can only create virtual LoS due to the double fading issue.
As a remedy, PASS dynamically repositions PAs to bypass obstacles, thereby ensuring stable, reliable LoS links and unlocking the full potential of high-frequency bands \cite{yang2025pinching}.

\subsubsection{Scalable Implementation}
In contrast to other flexible antenna technologies, PASS exhibits excellent scalability.
In particular, in conventional flexible antenna systems, adding or removing antennas can be expensive, if not impossible, as the number of antennas is determined once they are manufactured.
On the contrary, scalable implementation is the key feature of PASS, as adding or removing dielectric PAs at selected positions on waveguides is a straightforward step that does not alter the inner structure of the antenna system.
As new users arrive or traffic demands change, additional PAs can be deployed to boost link quality without costly hardware redesign. 
Hence, PASS is a scalable and low-cost solution among the family of flexible antennas.

\subsubsection{Near-Field Advantage}
In addition to the aforementioned features, PASS also introduces near-field advantages.
According to \cite{liu2024near, cui2023near}, the boundary of the near-field region is commonly defined by the Rayleigh distance, which can be mathematically expressed as $2D^2/\lambda$ with $D$ and $\lambda$ being the aperture size of the antenna array and the carrier frequency, respectively.
As waveguides in PASS can extend to tens of meters, the aperture size of a PASS-based antenna array can be extremely large, thus enabling the utilization of near-field channel features even for small numbers of antennas.
This greatly reduces the computational complexity encountered in extremely large antenna arrays (ELAAs), which typically deploy hundreds, even thousands, of antenna elements.

\subsection{Motivations and Contributions}
PASS, as a newly-emerging flexible-antenna technology, can potentially realize the vision of 6G and beyond, as maintained in \cite{liu2025pinching, yang2025pinching}.
However, these existing papers \cite{liu2025pinching, yang2025pinching} provide only a limited perspective regarding the technical details of PASS.
To this end, this tutorial paper is devoted to guiding and inspiring further research endeavors into PASS,  with the following contributions:
\begin{itemize}
    \item We present the fundamentals of PASS, including signal models, hardware, and power radiation models, and PA activation methods. First, we present both LoS and NLoS signal models of PASS, highlighting the distinctive characteristics of in-waveguide propagation compared to conventional free-space propagation. Subsequently, we propose two hardware models for PASS, including a simple directional coupler-based model and a more general multi-port network model, accompanied by a discussion on their respective impacts on PASS signals. 
    Associated with these hardware models, three power radiation models are derived.
    Furthermore, we introduce three PA activation methods: discrete, continuous, and semi-continuous activations, addressing different practical deployment scenarios.

    \item We characterize the information-theoretic capacity limits of PASS for both uplink and downlink transmissions. Furthermore, we investigate several key performance metrics, including ergodic rate, coverage probability, and outage probability, to evaluate the average network performance for PASS. Through theoretical analysis and numerical simulations, we demonstrate that PASS achieves higher spectral efficiency than conventional antenna technologies, with its superiority becoming more pronounced over larger service regions.

    \item We present pinching beamforming optimization technologies for PASS for both the single-waveguide and the multiple-waveguide cases.
    For the single-waveguide single-user case, the corresponding power scaling law is first characterized as the basis for pinching beamforming design. 
    For the general multiple-waveguide case, a pair of transmission structures is first proposed for PASS-based single-user communications, featuring \emph{sub-connected} and \emph{fully-connected} structures in terms of the connections between the baseband RF chains and the waveguides.
    Then, three practical protocols are devised for facilitating PASS-based multi-user communications, namely \emph{waveguide switching}, \emph{waveguide division}, and \emph{waveguide multiplexing}.
    Finally, the potential applications of PASS in wideband systems are further highlighted.

    \item We provide a pair of channel acquisition methods, namely pilot-based channel estimation (CE) and beam training.
    For pilot-based CE, a rank-deficiency issue for CE in PASS is first revealed, which is then solved via sequential activations, compressed sensing, and parameter sensing approaches.
    As an alternative, a beam training protocol tailored for PASS is presented.
    To address this problem, several potential codebook designs and sweeping strategies are proposed.

    \item We present machine learning (ML) assisted approaches for both PASS optimization and channel state information (CSI) acquisition, which help reduce inference computational complexity, mitigate poor local optimal solutions, and improve robustness against environment dynamics. For ML-empowered PASS beamforming design and channel acquisition, we discuss how a suitable deep neural network (DNN) architecture and training algorithm can be selected. 
    In the sequel, ML approaches are exploited to recover high-accuracy multi-path channels for PASS. Furthermore, ML-empowered low-overhead beam training under \emph{pinching alignment}, \emph{pinching tracking}, and \emph{pinching prediction} scenarios are also investigated.
\end{itemize}

\subsection{Notations and Organization}
\emph{Notations}: In this tutorial, for any matrix $\mathbf A$, $[\mathbf A]_{m,n}$, ${\mathbf{A}}^{T}$, ${\mathbf{A}}^{*}$, and ${\mathbf{A}}^{H}$ denote the $(m,n)$-th entry, transpose, conjugate, and conjugate transpose, respectively. The matrix inequality ${\mathbf A}\succeq{\mathbf 0}$ indicates positive semi-definiteness of $\mathbf{A}$. For any vector $\mathbf{a}$, $[\mathbf a]_{i}$ and $\mathrm{diag}\{\mathbf a\}$ denote the $i$-th entry of $\mathbf a$ and the operation to construct a diagonal matrix with $\mathbf{a}$ as the main diagonal entries. 
In addition, $\mathrm{Vect}\{\mathbf{A}\}$, $\mathrm{Blkdiag}\{\mathbf{a}\}$ represent the operation that vectorizes the columns in $\mathbf{A}$, and a block diagonal matrix with the elements of vector $\mathbf{a}$ placed along the main diagonal blocks, and $\mathbf{I}_{M}$ is the identity matrix of size $M$, $\mathbf{0}_M$ is a zero column vector with a dimensional of $M$, $\lVert\cdot\rVert$ denotes the Euclidean norm of a vector, $\lvert\cdot\rvert$ denotes the norm of a scalar, $\mathbb{C}$ stands for the complex plane, $\mathbb{R}$ stands for the real plane, and ${\mathbb{E}}\{\cdot\}$ represents mathematical expectation.
Finally, $\otimes$ and $\odot$ denote the Kronecker product and Hadamard product, respectively, and $\mathrm{j}\triangleq{\sqrt{-1}}$ denotes the imaginary unit. The convex hull of a set is denoted by ${\rm{Conv}}(\cdot)$, and the union operation is denoted by $\bigcup$. The big-O notation is given by $\mathcal{O}(\cdot)$.

The tutorial paper is organized as follows:
Section \ref{sect:fundamentals_of_pass} first presents the fundamentals of PASS.
Section \ref{sect:theoretical_limit} characterizes the information-theoretic capacity limits achieved by
PASS.
In Section \ref{sec:pass-opt}, the pinching beamforming design for PASS is detailed.
Subsequently, CSI acquisition in PASS is elaborated in Section \ref{sect:channel_estimation}.
Then, Section \ref{sect:ml_for_pass} presents ML methods for operating PASS.
Finally, several promising applications for PASS are discussed in Section \ref{sect:applications}, which is followed by the conclusions of this tutorial in Section \ref{sect:conclusions}.

\section{Fundamentals of PASS} \label{sect:fundamentals_of_pass}
In this section, we introduce the fundamentals of PASS, including: 1) basic signal models for PASS, 2) hardware and power radiation models for PASS, and 3) practical PA activation methods for PASS.

\begin{figure}[!t]
\centering
\includegraphics[width=0.35\textwidth]{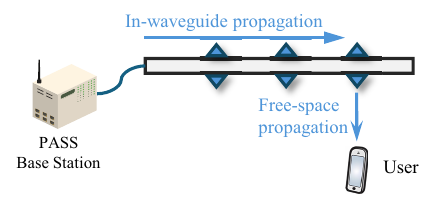}
\caption{Illustration of signal propagation in PASS.}
\label{fig_channel_model}
\end{figure}

\subsection{Signal Model}
For clarity of presentation in this section, we consider a basic point-to-point communication system assisted by a PASS, comprising a single waveguide and $N$ PAs. For simplicity, we assume a narrowband communication scenario at a given carrier wavelength $\lambda$ in free space. Let $s(t)$ denote the complex-valued baseband equivalent of the transmitted signal. As shown in Fig. \ref{fig_channel_model}, this signal is fed into the waveguide, propagates through it, radiates from the PAs positioned along the waveguide, and finally reaches the receiver via free-space propagation. 
Compared to free-space propagation, signal transmission within a waveguide exhibits two notable characteristics:
\begin{itemize}
  \item The signal propagation speed is different from that in free space. Let $c$ denote the speed of light in free space, and $n_{\mathrm{eff}}$ the effective refractive index of the waveguide. Then, the signal speed inside the waveguide is $v_{\mathrm{G}} = c / n_{\mathrm{eff}}$.
  \item {The signal attenuation within the waveguide is negligible over typical PASS deployment distances (on the order of tens of meters). For example, the attenuation factor of a ceramic-ribbon dielectric waveguide can be less than 0.01 dB/m, resulting in only 1 dB of loss over 100 meters~\cite{yeh2000communication, xu2025pinchingantenna}.}
\end{itemize}

\begin{figure*}[!t]
\centering
    \subfigure[]{
        \includegraphics[height=0.25\textwidth]{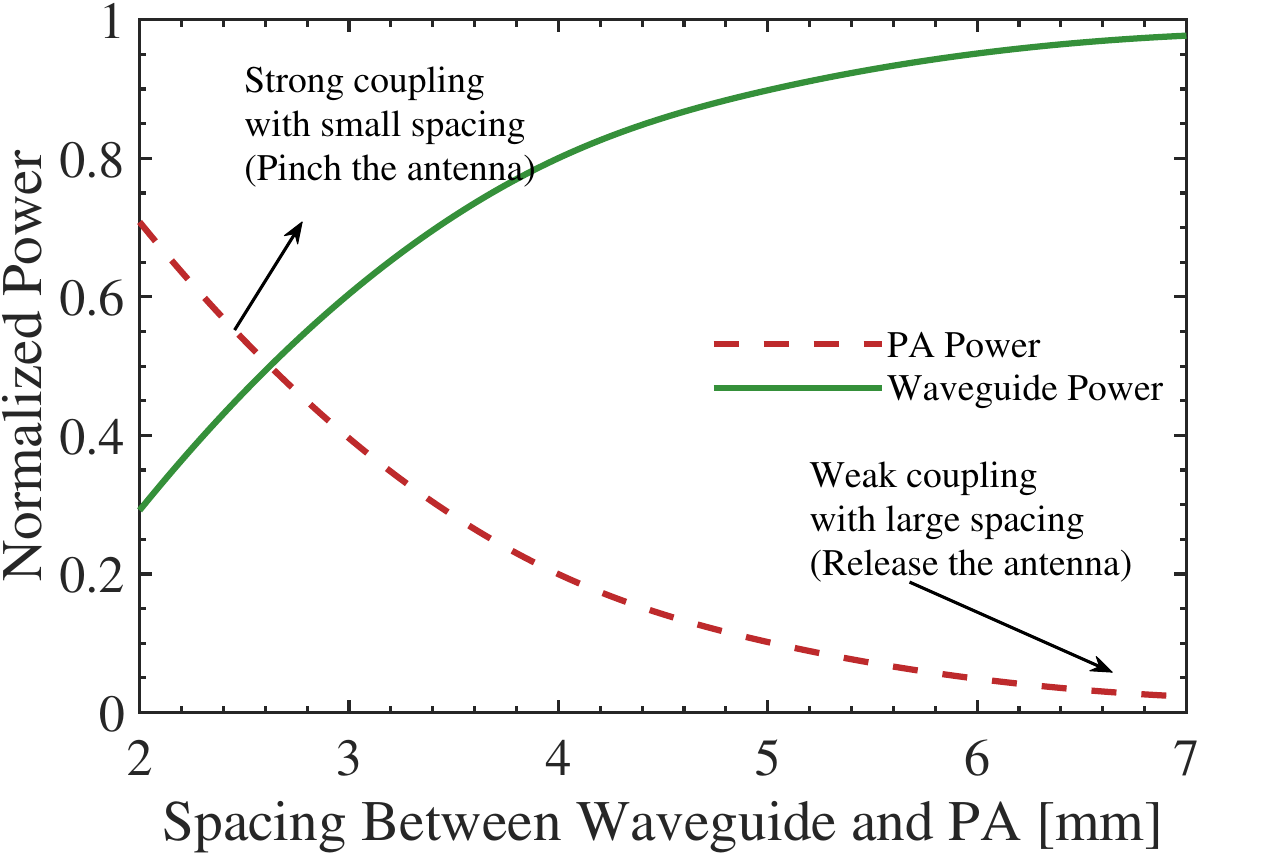}
        \label{coupling_coefficients_2}
    }\qquad \qquad 
    \subfigure[]{
        \includegraphics[height=0.25\textwidth]{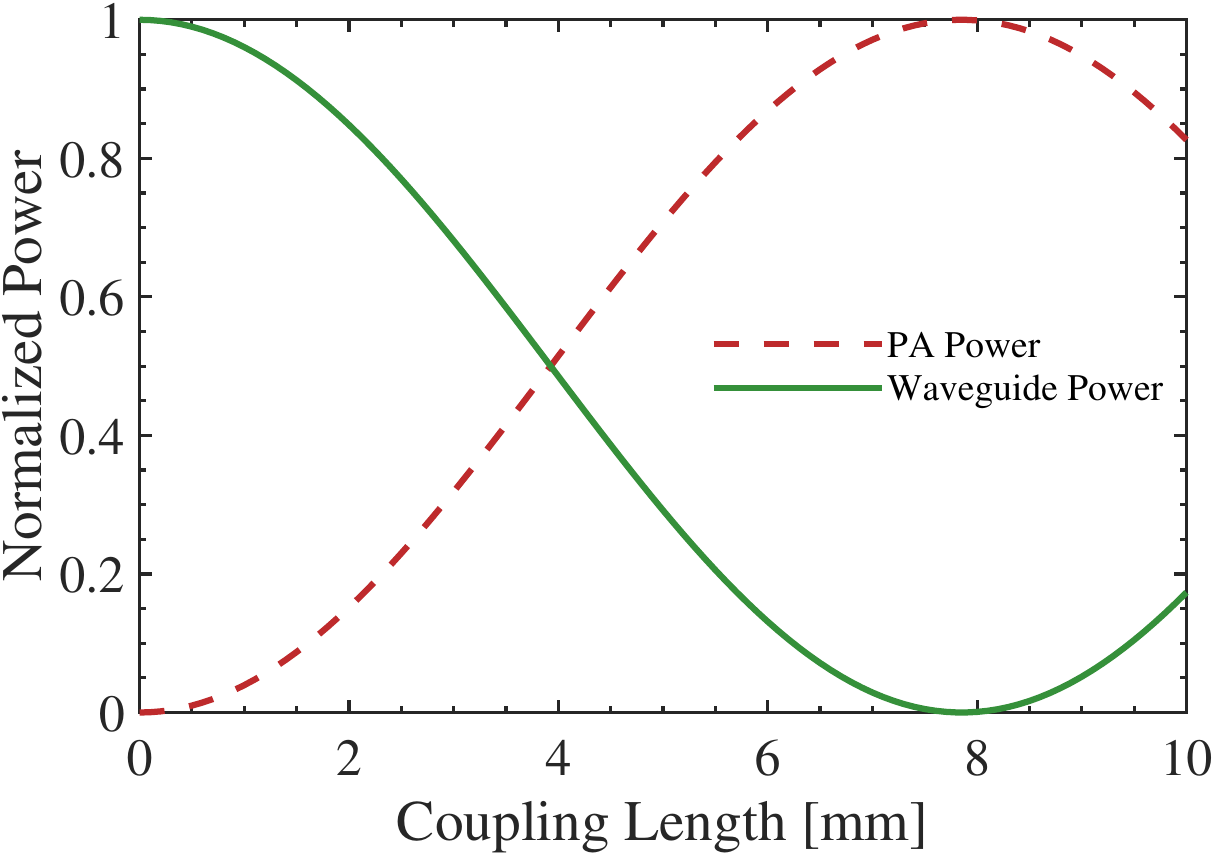}
        \label{coupling_coefficients_1}
    }
\caption{The ratio of power coupled by the PA versus (a) the spacing between waveguide and PA with a fixed coupling length of 5 mm and (b) the coupling lengths with a fixed spacing of 2 mm. The results are generated following \cite{9310539}.}
\end{figure*}

Without loss of generality, we first focus on the signal received via the $m$-th PA, indexed by $m \in \{1, \dots, M\}$. 
Furthermore, for simplicity, the hardware imperfections, such as signal reflections caused by impedance mismatches and the in-waveguide signal attenuation, are initially ignored, and their impacts will be discussed in the subsequent sections. Under these circumstances, the signal received by the user from the $m$-th PA through a LoS channel is given by
\begin{center}
\begin{tcolorbox}[title = Narrowband LoS Signal Model of PASS]
{\setlength\abovedisplayskip{2pt}
\setlength\belowdisplayskip{2pt}
\begin{align} \label{narrowband_LoS_signal_model}
    y_m(t) = \underbrace{\frac{\eta}{r_m} e^{ -\mathrm{j}\frac{2 \pi}{\lambda} r_m}}_{\text{free-space channel}} \times \underbrace{\vphantom{\frac{\eta}{r_m}} \sqrt{P_m} e^{ -\mathrm{j} \frac{2 \pi}{\lambda} n_{\mathrm{eff}} d_m}}_{\text{in-waveguide channel}} s(t),
  \end{align}
}\end{tcolorbox}
\end{center}
where $d_m$ and $r_m$ denote the in-waveguide and free-space propagation distances, respectively, $\eta$ is the free-space channel gain at a reference distance of 1 meter, and $P_m \in [0,1]$ represents the fraction of signal power radiated by the $m$-th PA. Furthermore, both $d_m$ and $r_m$ depend strongly on the position of the $m$-th PA. Consider a 3D Cartesian coordinate system in which the waveguide is parallel to the $x$-axis, and the signal feed point is located at $\mathbf{p}_{\text{feed}} = [0, y_{\mathrm{G}}, z_{\mathrm{G}}]^T$. The position of the $m$-th PA is then given by $\mathbf{p}_m = [x_m, y_{\mathrm{G}}, z_{\mathrm{G}}]^T$, while the position of the receiver is denoted by $\mathbf{r} = [x_{\mathrm{R}}, y_{\mathrm{R}}, z_{\mathrm{R}}]^T$. Based on this setup, the in-waveguide and free-space propagation distances are given by
\begin{align}
  \label{narrowband_LoS_signal_model_dn}
  d_m & = \| \mathbf{p}_m - \mathbf{p}_{\text{feed}} \| = x_m, \\
  \label{narrowband_LoS_signal_model_rn}
  r_m & = \| \mathbf{p}_m - \mathbf{r} \| = \sqrt{(x_m - x_{\mathrm{R}})^2 + \zeta^2},
\end{align}  
where $\zeta^2 = (y_{\mathrm{G}} - y_{\mathrm{R}})^2 + (z_{\mathrm{G}} - z_{\mathrm{R}})^2$.

For a general scenario where both LoS and NLoS paths are present in free space, the received signal from the $m$-th PA can be expressed as
\begin{center}
\begin{tcolorbox}[title = Narrowband Multi-path Signal Model of PASS]
{\setlength\abovedisplayskip{2pt}
\setlength\belowdisplayskip{2pt}

  \begin{align} \label{narrowband_NLoS_signal_model}
    & y_m(t) = h \sqrt{P_m} e^{ -\mathrm{j} \frac{2 \pi}{\lambda} n_{\mathrm{eff}} d_m} s(t), \\
    & h = \underbrace{\vphantom{\sum_{l=1}^L} \frac{\eta e^{-\mathrm{j} \frac{2\pi}{\lambda} r_m}}{r_m} }_{\text{LoS}} \; + \; \underbrace{\sum_{l=1}^L \beta_{m,l} e^{\mathrm{j} \xi_{m,l}}}_{\text{NLoS}}. \label{eq:NLoS_PASS}
  \end{align}

}\end{tcolorbox}
\end{center}
Here, $\beta_{m,l}$ and $\xi_{m,l}$ represent the amplitude and phase shift of the $l$-th NLoS path, respectively.
\begin{remark} \label{remark_1}
  \normalfont
  \emph{(PA Position Reconfiguration and Pinching Beamforming)}
  The above equations reveal that both the signal and channel models depend directly on the positions of the PAs, which can be adjusted dynamically. 
  By optimizing the PA positions, the aggregated wireless channel, i.e., $h$ can be customized to achieve specific objectives, such as signal constructive enhancement and interference suppression.
  Moreover, since the waveguides can span meters in length, PAs can be repositioned over large distances, giving PASS exceptional flexibility to mitigate large‑scale pathloss and small-scale multi-path fading.
  Furthermore, the pinching beamforming design will be elaborated later in Section \ref{sec:pass-opt}.
\end{remark}

\subsection{Hardware and Power Radiation Model}

The advantages of PASS primarily arise from the flexibility of PA positioning, as PASS can be freely attached to and detached from the waveguide. This flexibility relies on the key physical requirement of non-destructive coupling, i.e., the PA must extract energy from the waveguide without requiring a permanent or invasive electrical connection, such as cutting, soldering, or physically altering the waveguide. Moreover, this coupling must remain effective even when the PA is moved along the waveguide. To meet these requirements across typical wireless communication frequencies, such as microwave, mmWave, and THz bands, a promising approach is to design the PA as a contactless directional coupler. Such a design enables signal coupling without hardwired contact, allowing a controllable fraction of the waveguide signal to be extracted. In the following, we first present a physically consistent PA model based on the ideal directional coupler, followed by a more general formulation using a multi-port network model.

\subsubsection{Directional Coupler-based Model} A directional coupler consists of two or more transmission lines, i.e., waveguides, laid so close together that their electromagnetic fields overlap \cite{okamoto2010fundamentals, 9310539}. This overlap lets a fraction of the signal on the main waveguide leak into the neighboring waveguide through near-field evanescent coupling. The coupled signal propagates in the same direction as the signal in the main waveguide. Based on this principle, PAs can be implemented as a short waveguide with one end connected to a radiator or directly opened to the free space. To simplify the model, let us first focus on an ideal directional coupler by making several assumptions: 1) the PA and the main waveguide have identical specifications and are axially uniform; 2) only the fundamental mode exists; 3) the system is lossless, implying that mutual coupling coefficients are complex conjugates of each other by the law of power conservation; 4) butt coupling is neglected; 5) there is perfect impedance matching, resulting in no wave reflections; and 6) the coupling is weak, such that the propagating modes remains unperturbed. Under these assumptions, the coupling between the main waveguide and the $m$-th PA can be described using the following simplified expressions derived from coupled-mode theory (CMT)~\cite{okamoto2010fundamentals, 9310539}:
\begin{align}
  \label{CMT_equations}
  \frac{d A_{\mathrm{G},m}}{d x} = -\mathrm{j} \kappa_m A_{\mathrm{P},m}, \quad   \frac{d A_{\mathrm{P},m}}{d x} = -\mathrm{j} \kappa_m A_{\mathrm{G},m}, 
\end{align}
where $A_{\mathrm{G},m}$ and $A_{\mathrm{P},m}$ represent the complex mode amplitudes in the main waveguide and the $m$-th PA, respectively, and $\kappa_m$ denotes the coupling coefficient. Wave propagation is assumed along the $x$-axis, and without loss of generality, the coordinate $x=0$ marks the starting point of coupling. The coupling length, denoted by $L_m$, is the region where the waveguide and the $m$-th PA are adjacent. Given that only the main waveguide is excited, i.e., $A_{\mathrm{G},m} =1$ and $A_{\mathrm{P},m} = 0$ at $x= 0$, the power relationship between the waveguide and the $m$-th PA can be obtained from \eqref{CMT_equations} as follows: 
\begin{align}
  \label{CMT_exchange_equations}
  P_{\mathrm{G},m} & = |A_{\mathrm{G},m}|^2 =  \cos^2 (\kappa_m L_m), \\
  P_{\mathrm{P},m} & = |A_{\mathrm{P},m}|^2 =  \sin^2 (\kappa_m L_m).
\end{align}   
Thus, if the input power to the $m$-th PA coupling region is normalized to unity, the fraction of power radiated by the $m$-th PA is $\sin^2(\kappa_m L_m)$, and the remaining power continuing along the waveguide is $\cos^2(\kappa_m L_m)$. It is important to note that, as PAs are sequentially positioned along the waveguide, the power radiated by the $m$-th PA is influenced by the power exchange coefficients of all preceding PAs. Consequently, the cascaded power radiation relationship can be expressed as follows:
\begin{center}
\begin{tcolorbox}[title = General Power Radiation Model of PASS]
{\setlength\abovedisplayskip{2pt}
\setlength\belowdisplayskip{2pt}

  \begin{align} \label{general_power_radiation_rule}
    \mathcal{P} = \left\{ P_m \; \left| \; \begin{matrix*}[l]
      P_m = \delta_m^2 \prod_{i=1}^{m-1} \left( 1 - \delta_i^2 \right)\\
      \delta_m = \sin(\kappa_m L_m), \forall m
    \end{matrix*} \right. \right\}
  \end{align}

}\end{tcolorbox}
\end{center}

\begin{remark}
    \normalfont
    \emph{(Controllable Power Radiation)}
    From \eqref{CMT_exchange_equations}, it can be seen that the power coupled into the PA can be adjusted by tuning the coupling coefficient $\kappa_m$ and the coupling length $L_m$. The coupling coefficient $\kappa_m=\Omega_{0}  e^{-\sqrt{\gamma_{0}^{2}-\frac{4\pi^2}{\lambda^{2}}n_{\mathrm{clad}}^2}S_{m}}$ is influenced by several factors, including the waveguide and PA modes captured by coefficient $\Omega_{0}$, wavelength $\lambda$, and the spacing $S_{m}$ between the waveguide and the $m$-th PA \cite{xu2025pinchingopt,little1997microring}. 
    As depicted in Fig.~\ref{coupling_coefficients_2}, a smaller spacing (e.g., 2 mm) results in stronger coupling, enabling a substantial portion of the power to be transferred to the PA, while a slightly larger spacing (e.g., 7 mm) significantly reduces coupling. This emphasizes the necessity of firmly positioning the PA against the waveguide for efficient operation. Compared to the coupling coefficient, the influence of the coupling length is more direct. According to \eqref{CMT_exchange_equations}, the signal power oscillates periodically between the waveguide and the PA with increasing coupling length $L_m$. Complete power coupling ($P_{\mathrm{G},m} = 1$) occurs at $L_m = \pi / (2\kappa_m)$. 
\end{remark}

Based on the above analysis, the power coupled into the PA can be regulated by controlling the spacing and coupling length. 
The coupling length can be simply controlled by fabricating PAs with specific physical lengths, but this method has limited applicability for different PASS deployments (e.g., different numbers of activated PAs).
To enable flexible power radiation control, a waveguide-PA spacing configuration scheme, which controls the power radiation $P_{m}$ by adjusting the spacing $S_{m}$ in an element-wise manner, was introduced in \cite{xu2025pinchingopt}.  
However, this dynamic control relies on precise mechanical adjustments, leading to increased hardware complexity and cost. 
A practical approach is to pre-configure the PAs to achieve a fixed, desired power radiation. In the following, we introduce two practical power radiation models. 
The first one is the \emph{equal power radiation} model \cite{wang2025modeling}, shown as follows:
\begin{center}
\begin{tcolorbox}[title = Equal Power Radiation Model of PASS]
{\setlength\abovedisplayskip{2pt}
\setlength\belowdisplayskip{4pt}
  \begin{align}
    \label{equal_power_radiation_rule}
    & \mathcal{P} = \left\{ P_m \; \left| \; \begin{matrix*}[l]
      P_1 = P_2 = \cdots = P_M \triangleq P_{\text{eq}}\\
      \delta_m = \sqrt{\frac{P_{\text{eq}}}{1 - (m-1) P_{\text{eq}}}}, \forall m
    \end{matrix*} \right. \right\}. 
  \end{align}
}\end{tcolorbox}
\end{center}
Specifically, the equal power radiation model pre-configures each PA according to the rule in \eqref{equal_power_radiation_rule}, ensuring that all PAs radiate with the same efficiency. The second model is the \emph{proportional power radiation model} \cite{wang2025modeling}, shown as follows:
\begin{center}
\begin{tcolorbox}[title = Proportional Power Radiation Model of PASS]
{\setlength\abovedisplayskip{2pt}
\setlength\belowdisplayskip{4pt}
  \begin{align}
    \label{proportion_power_radiation_rule}
    & \mathcal{P} = \left\{ P_m \; \left| \; \begin{matrix*}[l]
      P_m = \delta_{\text{eq}}^2 \left( 1 - \delta_{\text{eq}}^2 \right)^{m-1}\\
      \delta_1 = \delta_2 = \cdots = \delta_M \triangleq \delta_{\text{eq}}
    \end{matrix*} \right. \right\}.
  \end{align}
}\end{tcolorbox}
\end{center}
In contrast to the equal power radiation, the proportional power radiation model can be easily implemented by applying identical configurations to all PAs, as specified in \eqref{proportion_power_radiation_rule}, thereby reducing manufacturing complexity.

\begin{table*}[!t]
\centering
\caption{Hardware Model Comparison of PASS}
\resizebox{\linewidth}{!}{
\begin{tabular}{l|c|c|c|c}
\hline
\textbf{Hardware Model} 
& \textbf{Signal Complexity} 
& \textbf{Reconfigurability} 
& \textbf{Generality} 
& \textbf{Typical Use Cases}   \\
\hline

Directional Coupler 
& Low
& Limited 
& Narrow 
& Theoretical analysis; low-complexity system design  \\
\hline

Multiport Network
& High 
& Full 
& Board 
& Experimental evaluation; advanced hardware designs  \\
\hline

\end{tabular}
}
\label{tab:hardware_comparison}
\end{table*}

\begin{figure}[!t]
\centering
\includegraphics[width=0.48\textwidth]{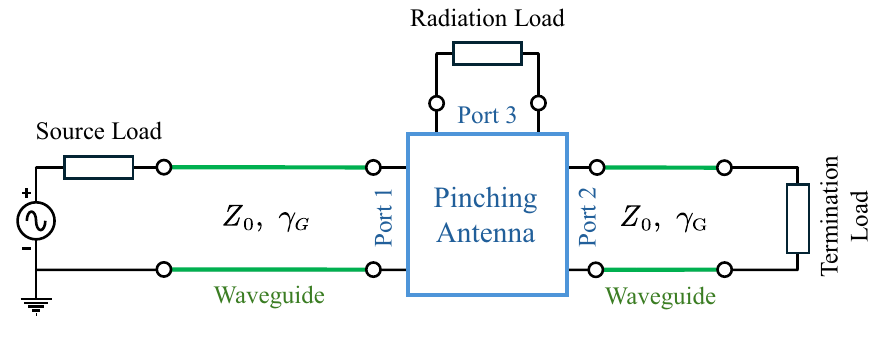}
\caption{Multiport network-based model for the PA.}
\label{fig_multiport_PA}
\end{figure}

\subsubsection{Multiport Network-based Model}
In the previous sections, the PA was modeled as a directional coupler under several idealized assumptions. However, in practice, these assumptions may not hold perfectly due to factors such as manufacturing tolerances and material imperfections, leading to issues like signal reflection caused by impedance mismatch. Moreover, this model restricts the implementation of the PA to a single, specific design. To address these limitations, a higher-level model is needed, which can account for hardware impairments and accommodate a broader range of practical implementations. 
Multiport network theory (MNT) offers a powerful analytical framework for this purpose \cite{pozar2021microwave, 6880934}. 
MNT models each network element as a multi-port circuit with appropriately loaded ports, enabling systematic analysis and optimization of complex electromagnetic systems based on circuit-theoretic principles.

For the purpose of exposition, we first focus on a simple case where only a single PA is applied to the waveguide. Given its function, the PA can be modeled as a three-port network, with two ports interfacing with the waveguide and one port responsible for radiating signals into free space, as illustrated in Fig.~\ref{fig_multiport_PA}. According to circuit theory, the waveguide, which is essentially a transmission line, can be characterized by its per-unit-length parameters, including inductance $L$, capacitance $C$, resistance $R$, and conductance $G$. The propagation coefficient and characteristics impedance of the waveguide are given by $\gamma_{\mathrm{G}} = \sqrt{({\mathrm{j}} \omega L + R) ({\mathrm{j}} \omega C + G)}$ and $Z_0 = \sqrt{({\mathrm{j}} \omega L + R)/ ({\mathrm{j}} \omega C + G)}$, respectively, where $\omega = 2 \pi f$ is the angular frequency and $f$ is the signal frequency. If an input signal $s$ is applied at one end of the waveguide, the output at the other end is a attenuated and phase-shifted version of $s$, as $s \cdot e^{-\gamma_G L_{\mathrm{G}}}$ is the output at the other end, where $L_{\mathrm{G}}$ denotes the waveguide length. Furthermore, the behavior of the PA itself can be characterized using a scattering matrix $\mathbf{S} \in \mathbb{C}^{3\times 3}$, given by
\begin{equation}
  \mathbf{S} = \begin{bmatrix}
    S_{11} & S_{12} & S_{13} \\
    S_{21} & S_{22} & S_{23} \\
    S_{31} & S_{32} & S_{33} \\
  \end{bmatrix}.
\end{equation} 
Here, $S_{ij}$ represents the complex voltage transmission coefficient from port $j$ to port $i$. In accordance with the law of energy conservation, the scattering matrix must satisfy the inequality
\begin{equation} \label{S_matrix_constraint}
  \mathbf{S}^H \mathbf{S} \preceq \mathbf{I}_3.
\end{equation}
Impedance mismatches at each PA port ($S_{ii} \neq 0$ for $i \in \{1,2,3\}$) and the load ports introduce undesired signal reflections. Using the multiport network model and circuit theory, the relationship between the radiated PA signal $y_{\text{rad}}(t)$ and the narrowband source signal $s(t)$, accounting for imperfect impedance matching, can be expressed as follows:
\begin{figure}[!t]
\centering
\includegraphics[width=0.43\textwidth]{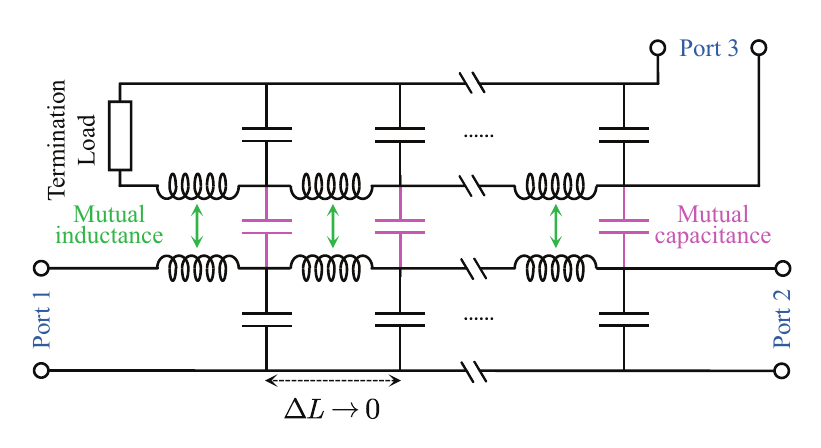}
\caption{Lumped equivalent circuit of the three-port PA based on directional coupler \cite{mongia1999rf}.}
\label{fig_multiport_PA_circuit}
\end{figure}
\begin{center}
\begin{tcolorbox}[title = Power Radiation Model Under Imperfect Impedance Matching]
{\setlength\abovedisplayskip{2pt}
\setlength\belowdisplayskip{2pt}
  \begin{align}
  \label{multiport_voltage_relation}
    y_{\text{rad}}(t) = \frac{e^{\gamma_{\mathrm{G}} L_1} \mathbf{e}_3^H \left( \mathbf{I}_3 + \mathbf{S} \right)\left( \mathbf{I}_3 - \mathbf{\Gamma} \mathbf{S} \right)^{-1} \mathbf{e}_1}{\mathbf{e}_1^H \left(\mathbf{I}_3 + e^{-2\gamma_{\mathrm{G}}L_1} \mathbf{S}\right) \left( \mathbf{I}_3 - \mathbf{\Gamma} \mathbf{S} \right)^{-1} \mathbf{e}_1} s(t).
  \end{align}

}\end{tcolorbox}
\end{center}
Here, $L_1$ and $L_2$ denote the lengths of the waveguide sections connected to port 1 and port 2, respectively. The matrix $\mathbf{\Gamma} = \mathrm{diag}{ \Gamma_{\mathrm{S}} e^{-2\gamma_{\mathrm{G}} L_1}, \Gamma_{\mathrm{L}} e^{-2\gamma_{\mathrm{G}}L_2}, \Gamma_{\mathrm{R}} }$ represents the reflection coefficients, where $\Gamma_{\mathrm{S}}$, $\Gamma_{\mathrm{L}}$, and $\Gamma_{\mathrm{R}}$ are the reflection coefficients at the source, termination, and radiation loads, respectively. The vectors $\mathbf{e}_1 = [1, 0, 0]^T$ and $\mathbf{e}_3 = [0, 0, 1]^T$ are used as entry selection vectors.
A detailed derivation of this relationship, along with the definitions of the involved matrices and vectors, is provided in Appendix~\ref{Appendix_1}. It is important to note that the complexity of the multiport network-based model can grow exponentially when multiple PAs are attached to the waveguide, due to the complex cascaded reflection and interaction effects among the PAs. More specifically, the power radiated by each PA is determined not only by the original power in the waveguide, but also by the power reflected from the other PAs.
If all ports of the PA are perfectly matched ($S_{11} = S_{22} = S_{33} = 0$) and the impedance matching at all loads is ideal ($\mathbf{\Gamma} = \mathbf{0}$), equation \eqref{multiport_voltage_relation} simplifies to
\begin{equation} \label{multiport_voltage_relation_simplified}
  y_{\text{rad}}(t) = e^{-\gamma_{\mathrm{G}}L_1} S_{31} s(t).
\end{equation}
In this case, no reflections occur. Consequently, when multiple PAs are deployed, the corresponding signal models remain effectively identical to the simplified models in \eqref{narrowband_LoS_signal_model} and \eqref{narrowband_NLoS_signal_model}.
In Table \ref{tab:hardware_comparison}, we compare the different hardware models.

\begin{remark}
  \normalfont
  \emph{(Scattering Matrix Characterization)}
  The key component of the multiport network-based model is the scattering matrix $\mathbf{S}$, which can be characterized using several approaches. For ideal or simplified structures, the scattering matrix can be derived directly from physical principles. For instance, in the case of the ideal directional coupler model discussed earlier, the simplified expression in \eqref{multiport_voltage_relation_simplified} applies, with $S_{31} = -\mathrm{j} \sin(\kappa L_c)$, where $L_c$ denotes the coupling length. At the circuit level, one can also represent the system using a lumped equivalent circuit, as shown in Fig. \ref{fig_multiport_PA_circuit}, enabling an analytical yet physically consistent characterization of the scattering matrix. These analytical methods are particularly useful for theoretical analysis. To obtain accurate numerical values of the scattering matrix for real-world systems, full-wave electromagnetic simulators, such as COMSOL and High Frequency Structure Simulator (HFSS), are typically employed. These tools allow for the precise definition of material properties, geometric structures, boundary conditions, and frequency ranges, and solve Maxwell's equations numerically to extract the scattering parameters. In experimental settings, the scattering matrix can be directly measured using a vector network analyzer (VNA) by connecting it to the ports of the device under test. Furthermore, from the perspective of multiport network theory, the reconfigurability of the PA can be captured by modeling $\mathbf{S}$ as a reconfigurable scattering matrix, implemented through an equivalent tunable impedance network \cite{9514409}. Under the assumption of an ideally reconfigurable impedance network, the scattering matrix $\mathbf{S}$ can be optimized subject to the constraint in \eqref{S_matrix_constraint} to maximize overall system performance. However, such reconfigurability can also result in increased hardware complexity and cost, as well as challenges for system optimization due to the complex signal model given in \eqref{S_matrix_constraint}.
\end{remark}

\begin{table*}[!t]
\centering
\caption{Comparison of Activation Methods for PASS}
\resizebox{\linewidth}{!}{
\begin{tabular}{l|c|c|c|c}
\hline
\textbf{Method} 
% & \textbf{Control} 
& \textbf{Flexibility} 
& \textbf{Response Time} 
& \textbf{Hardware Implementation} 
& \textbf{Characteristics} \\
\hline

Discrete 
% & Centralized
& Fixed positions 
& Fast ($\mu$s–ms) 
& Electromagnets/RF switches 
& Low cost, fast switching, scalable \\
\hline

Continuous 
% & Centralized 
& Fully continuous 
& Slow (ms–s) 
& Servo motors, sensors, controllers 
& Highest spatial resolution, differentiable control \\
\hline

Semi-continuous 
% & Centralized
& Local refinement 
& Medium (ms) 
& Electromagnets, micro-actuators 
& Balanced resolution and complexity \\
\hline
\end{tabular}
}
\label{tab:activation_method_comparison}
\end{table*}

\subsubsection{Discussion on Uplink Signal Model}
{We note that the downlink signal model of PASS, such as \eqref{narrowband_LoS_signal_model}, can be clearly characterized using either the coupler-based or multiport network-based formulation, where signals injected at the waveguide feed point radiate passively through the activated PAs. In contrast, formulating a tractable and physically consistent uplink signal model is far more challenging. This difficulty arises because PAs can passively receive EM signals from free space into the waveguide. When multiple PAs are deployed along the same waveguide, the signals captured by one PA may re-radiate through other PAs as they propagate toward the feed point. This inter-antenna radiation (IAR) effect complicates the uplink analysis and renders the signal model mathematically intractable. Consequently, most existing studies on uplink PASS either focus solely on single-PA deployments \cite{zeng2025energy}, thereby avoiding IAR, or neglect the IAR effect altogether in multi-PA scenarios \cite{tegos2024minimum}, leading to oversimplified and physically inconsistent models. To date, no tractable and physically consistent uplink signal model for multi-PA PASS has been established, leaving progress in this direction limited.

Recently, the Segmented Waveguide-enabled Pinching-Antenna System (SWAN) was proposed to address this issue \cite{ouyang2025uplink}. The SWAN architecture employs multiple short dielectric waveguide segments arranged end-to-end. Unlike a single long waveguide, these segments are not physically connected. Instead, each segment has its own feed point through which signals are injected into or extracted from the waveguide and relayed to the base station (BS) via wired connections, such as optical fibers or high-quality coaxial cables. By activating only a single PA within each segment, SWAN effectively eliminates the IAR effect, thereby achieving uplink-downlink channel reciprocity and enabling a tractable, IAR-free uplink signal model. This architecture thus allows the system to harness multi-PA array gain while preserving analytical tractability.}
\subsection{PA Activation Method}
\begin{figure}[!t]
    \centering
    \includegraphics[width=0.6\linewidth]{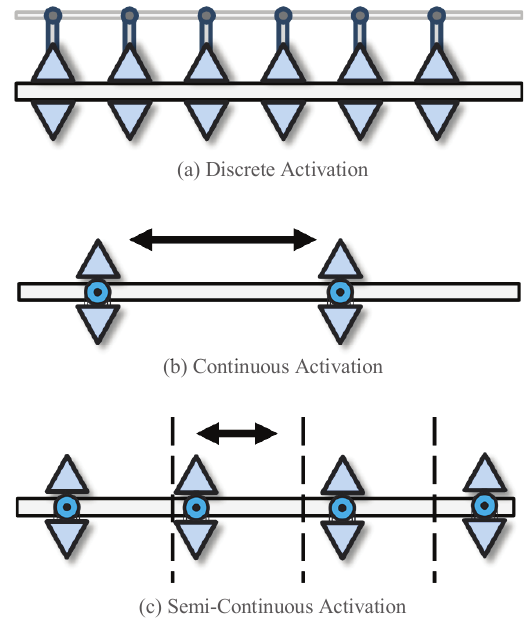}
    \caption{Three practical activation methods of PASS.}
    \label{fig:activation_methods}
\end{figure}

PASS introduces a new class of reconfigurable transceivers that interface electromagnetic waves between guided media and free space through spatially distributed coupling units. A fundamental design aspect of PASS lies in the activation method used to control the PAs. The activation method determines not only where and how PAs are physically positioned along the waveguides but also how they participate in the wireless link. 
We categorize the current activation methods into three distinct modes, i.e., \textit{discrete}, \textit{continuous}, and \textit{semi-continuous} activation, as illustrated in Fig. \ref{fig:activation_methods}. We compare these activation methods in terms of physical feasibility, control complexity, spatial resolution, and suitability for mobile wireless scenarios, as is summarized in Table \ref{tab:activation_method_comparison}.

Furthermore, when multiple PAs are attached to the waveguide, the power radiated by each PA is determined not only by the original waveguide power but also by the power reflected from the other PAs. This is a result of complex cascaded reflection and interaction effects among the antennas. To address the resulting complexity in performance evaluation, a cascaded three-port analysis framework was proposed in \cite{wang2025multiport}. This framework not only simplifies the analysis but also demonstrates that optimal performance is achieved when reflection at each PA is eliminated.

\subsubsection{Discrete Activation} 
For discrete activation, a number of PAs, shown in Fig. \ref{fig:activation_methods}{(a)}, are uniformly-spaced, pre-installed along the waveguide, and a small subset of the PAs is activated when needed. The  discrete set of PAs' positions is given below: 
\begin{center}
\begin{tcolorbox}[title = Discrete Activation Constraint of PASS]
{\setlength\abovedisplayskip{2pt}
\setlength\belowdisplayskip{2pt}
  \begin{align} \label{discrete_activation_constraint}
    \mathcal{X} = \left\{ \left. x_m = \frac{m x_{\max}}{M-1} \; \right| \; 
    \begin{matrix*}[l]
      x_{m} - x_{m-1} \ge \Delta_{\min},\\
      m \in \{ 0, 1, \dots, M-1 \}
    \end{matrix*}
    \right\}.
  \end{align}

}\end{tcolorbox}
\end{center}
The discrete activation can be implemented using programmable electrical control (e.g., using low-cost electromagnets), which adjust the spacing between the PAs and the waveguide, thus changing the coupling strength to activate/deactivate PAs. 
The discrete activation enables fast reconfiguration and low hardware complexity without relying on mechanical modules, which is scalable for large-scale PASS design. However, its spatial resolution is limited by the discrete spacing of the PAs, making it less effective for precise beamforming or near-field sensing applications.

\subsubsection{Continuous Activation} For continuous activation, each PA is mounted on a slide track that is installed along with each waveguide. Hence, the location of each PA $m$ along the waveguide parallel to the $x$-axis satisfies the following conditions:
\begin{center}
\begin{tcolorbox}[title = Continuous Activation Constraint of PASS]
{\setlength\abovedisplayskip{2pt}
\setlength\belowdisplayskip{2pt}

  \begin{align}
    \label{continuous_activation_constraint}
    & \mathcal{X} = \left\{ x_m \; \left| \; \begin{matrix*}[l]
      0 \le x_{m} \le x_{\max},\\
      x_{m} - x_{m-1} \ge \Delta_{\min}, \\
      m \in \left\{0,1,\dots,M-1\right\}
    \end{matrix*} \right. \right\}, 
  \end{align}

}\end{tcolorbox}
\end{center}
where $x_{\max}$ and $\Delta_{\min}$ denote the maximum deployment range and the minimum spacing of PAs, respectively. 
From a hardware perspective, each PA requires a linear actuator (e.g., servo or stepper motor), position sensor, and real-time control logic. 
Compared to discrete activation, the continuous activation method provides the highest spatial resolution and enables fully differentiable optimization, 
but the required closed-loop actuators significantly increase activation latency and control complexity. 
In mobile scenarios, mechanical/electronic delay to adjust PAs' locations may cause outdated configurations under fast channel variations. 
Hence, continuous activation is best suited for slowly varying environments,  such as indoor communications, automated manufacturing, and industrial Internet of Things (IIoT) systems, where beam tracking precision outweighs speed.

\subsubsection{Semi-Continuous Activation} 
Semi-continuous activation seeks to balance the high resolution of continuous activation and the low complexity of discrete activation. The waveguide is divided into $M-1$ segments with a uniform interval. Each PA is pre-located at one segment and supports the local micro-adjustment with an offset $u_{m}\in\left[0, u_{\max}\right]$. This leads to the following feasible set of PA positions: 
\begin{center}
\begin{tcolorbox}[title = Semi-Continuous Activation Constraint of PASS]
{\setlength\abovedisplayskip{2pt}
\setlength\belowdisplayskip{2pt}
  \begin{align} \label{semicontinuous_activation_constraint}
    & \mathcal{X} \!=\! \left\{ x_m \!=\! \frac{mx_{\max}}{M-1} \!+\! u_{m} \; \left| \; \begin{matrix*}[l]
      0 \le u_{m} \le u_{\max},\\
      x_{m} - x_{m-1} \!\ge\! \Delta_{\min},\\
      m \!\in\! \left\{0,1,\dots,M-1\right\}
    \end{matrix*}
    \right.\right\} 
  \end{align}

}\end{tcolorbox}
\end{center}
This hybrid scheme enables fast segment-level control with local refinement. Hence, it can retain pinching beamforming precision, while reducing the activation latency and hardware cost compared to the fully continuous structure.
For instance, a dual scale antenna deployment is proposed by the authors in \cite{gan2025dual}. However, its coordination complexity and physical integration (e.g., actuation precision) require careful system-level co-design.

\section{Information-Theoretic Limits and Performance Analysis of PASS} \label{sect:theoretical_limit}
Having introduced the fundamental principles of PASS, we now characterize its information-theoretic limits along with several key performance metrics to evaluate its performance superiority over conventional antenna technology. For analytical tractability, we focus on the case of equal power allocation as defined in \eqref{equal_power_radiation_rule}, and assume continuous activation of the PAs, as specified in \eqref{continuous_activation_constraint}.
\subsection{Information-Theoretic Limits}
In this section, we characterize the information-theoretic limits of PASS-enabled multiuser channels by analyzing their capacity region, along with the achievable rate regions under practical orthogonal multiple access (OMA) schemes, including time division multiple access (TDMA) and frequency division multiple access (FDMA). Unlike conventional systems using fixed-location antennas, the capacity of PASS is determined by the activated positions of the PAs.  

For conciseness, in this section, we focus on a communication scenario where a single pinched waveguide serves two single-antenna users. The waveguide is aligned along the $x$-axis, and a total of $M$ PAs are activated along its aperture. 

\subsubsection{Uplink PASS}\label{Section: PASS-Enabled Uplink Channels}
Let $s_k\sim{\mathcal{CN}}(0,1)$ denote the desired information symbol for user $k\in\{1,2\}\triangleq{\mathcal{K}}$. The received signal at the BS is given by
\begin{align}\label{Signal_Model_Uplink}
{{y}}=\sum_{m=1}^{M}{\rm{e}}^{-{\rm{j}}\frac{2\pi}{\lambda}n_{\rm{eff}}x_m}\left(\sum_{k=1}^{2}h({\mathbf{r}}_{k},{\mathbf{p}}_m)\sqrt{{\mathsf{P}}_k}{{s}}_k+z_m^{\rm{U}}\right),
\end{align}
where $h({\mathbf{r}}_{k},{\mathbf{p}}_m)\triangleq\frac{\eta}{\lVert{\mathbf{r}}_{k}-{\mathbf{p}}_m\rVert}{\rm{e}}^{-{\rm{j}}\frac{2\pi}{\lambda}\lVert{\mathbf{r}}_{k}-{\mathbf{p}}_m\rVert}$ denotes the spatial response between user $k$ and the $m$th PA, $z_m^{\rm{U}}\sim{\mathcal{CN}}(0,\sigma^2)$ is the additive white Gaussian noise at the $m$th PA, and $\sigma^2$ is the noise power. The transmit power of user $k$ is denoted by ${\mathsf{P}}_k$. The location of user $k$ is $\mathbf{r}_k = [x_{\mathrm{R},k}, y_{\mathrm{R},k}, z_{\mathrm{R},k}]^T$, and the position of the $m$th PA is ${\mathbf{p}}_m=[x_m,y_{\rm{G}},z_{\rm{G}}]^{T}$ for $m\in{\mathcal{M}}\triangleq\{1,\ldots,M\}$. For each user $k$, let $R_k^{\rm{U}}$ denote its achievable uplink rate.

For a given pinching beamformer ${\mathbf{x}}\triangleq[x_1,\ldots,x_M]^{T}$, the capacity region of PASS is known to form a pentagon; see \cite[{\figurename} 4.7]{el2011network}. It includes all rate pairs $(R_1^{\rm{U}},R_2^{\rm{U}})$ satisfying the following constraints \cite{el2011network}:
\begin{subequations}\label{PASS_Uplink_Capacity_Region_Constraint}
\begin{align}
&R_k^{\rm{U}}\leq \log_2(1+{\mathsf{P}}_k\lvert h_k({\mathbf{x}})\rvert^2/\sigma^2),k\in{\mathcal{K}}, \label{PASS_Uplink_Capacity_Region_Constraint1}\\
&R_1^{\rm{U}}+R_2^{\rm{U}}\leq \log_2(1+({\mathsf{P}}_1\lvert h_1({\mathbf{x}})\rvert^2
+{\mathsf{P}}_2\lvert h_2({\mathbf{x}})\rvert^2)/\sigma^2),
\label{PASS_Uplink_Capacity_Region_Constraint2}
\end{align}
\end{subequations}
where $h_k({\mathbf{x}})\triangleq\frac{1}{\sqrt{M}}\sum_{m=1}^{M}{\rm{e}}^{-{\rm{j}}\frac{2\pi}{\lambda}n_{\rm{eff}}x_m}h({\mathbf{r}}_{k},{\mathbf{p}}_m)$ denotes the effective channel gain of user $k\in\{1,2\}$. The corresponding capacity region is characterized as follows:
\begin{align}\label{PASS_Uplink_Capacity_Region_Specific}
{\mathcal{C}}^{\rm{U}}({\mathbf{x}})\triangleq\{(R_1^{\rm{U}},R_2^{\rm{U}})\left\lvert R_1^{\rm{U}}\geq0,R_2^{\rm{U}}\geq0,\eqref{PASS_Uplink_Capacity_Region_Constraint1},\eqref{PASS_Uplink_Capacity_Region_Constraint2}\right.\}.
\end{align}

The capacity region ${\mathcal{C}}^{\rm{U}}({\mathbf{x}})$ can be achieved by carrying out \emph{successive interference cancellation (SIC)} or \emph{joint} decoding at the BS \cite{el2011network}. Under SIC, the BS first decodes one user's message while treating the other user's signal as interference, and then subtracts it before decoding the second message. Let $\bm\pi$ denote the decoding order, with ${\bm\pi}=[2,1]\triangleq{\bm\pi}^{\rm{\uppercase\expandafter{\romannumeral1}}}$ indicating that user $1$ is decoded after user $2$, and ${\bm\pi}=[1,2]\triangleq{\bm\pi}^{\rm{\uppercase\expandafter{\romannumeral2}}}$ otherwise. By incorporating \emph{time sharing} among different ${\mathbf{x}}$'s, the overall capacity region of the PASS-enabled uplink channel is given by the convex hull over all feasible configurations \cite{el2011network}:
\begin{center}
    \begin{tcolorbox}[title = Uplink Channel Capacity of PASS]
    {\setlength\abovedisplayskip{2pt}
    \setlength\belowdisplayskip{2pt}
    \begin{align}\label{PASS_Uplink_Capacity_Region_General_Exhausitive}
{\mathcal{C}}^{\rm{U}}\triangleq{\rm{Conv}}\left(\bigcup\nolimits_{{\mathbf{x}}\in{\mathcal{X}}}{\mathcal{C}}^{\rm{U}}({\mathbf{x}})\right),
\end{align}
    }\end{tcolorbox}
\end{center}
where the feasible set ${\mathcal{X}}$ is given in \eqref{continuous_activation_constraint}. 
\subsubsection*{Single-Pinch Case}
For the single-pinch case with $M=1$, the capacity region can be directly obtained by varying the pinched position $x_1$ over the interval $[0,x_{\max}]$. As proven in \cite{ouyang2025capacity}, the capacity-achieving position must lie within the interval $[x_{\mathrm{R},1},x_{\mathrm{R},2}]$. Therefore, the capacity region is given by
\begin{align}
{\mathcal{C}}^{\rm{U}}={\rm{Conv}}\left(\bigcup\nolimits_{x_1\in[x_{\mathrm{R},1},x_{\mathrm{R},2}]}{\mathcal{C}}^{\rm{U}}(x_1)\right)\triangleq
{\mathcal{C}}_{\rm{S}}^{\rm{U}}.
\end{align}

\begin{figure}[!t]
\centering
\includegraphics[width=0.35\textwidth]{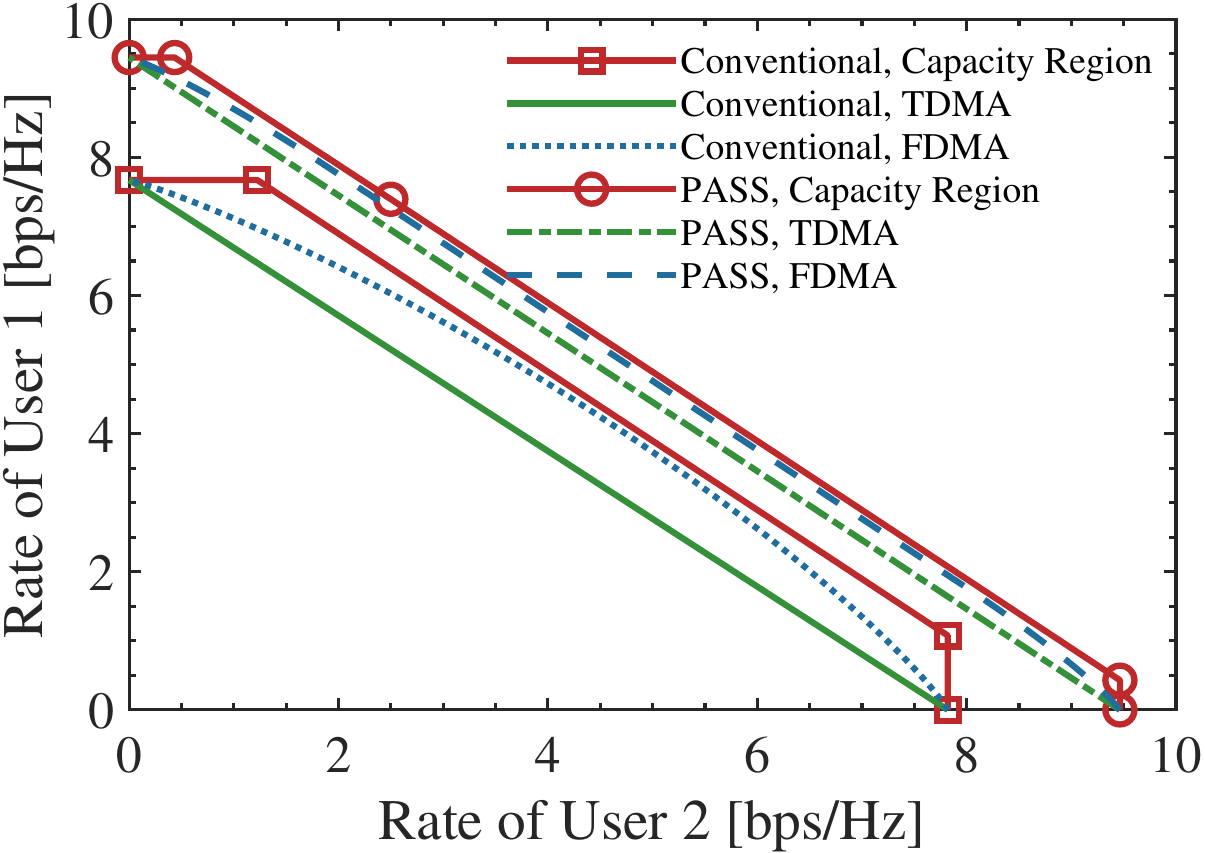}
\caption{Capacity/rate region comparison for PASS-enabled uplink channels in the single-pinch case. The simulation parameters used can be found in \cite{ouyang2025capacity}.}
\label{Figure_SP_MAC_Capacity_Rate_Region}
\end{figure}

The TDMA and FDMA rate regions can be similarly derived, where the users transmit in orthogonal time slots or frequency bands, thereby eliminating inter-user interference. Detailed derivations of these rate regions are provided in \cite{ouyang2025capacity}. For convenience, let ${\mathcal{R}}_{{\rm{S}},{\rm{T}}}^{\rm{U}}$ and ${\mathcal{R}}_{{\rm{S}},{\rm{F}}}^{\rm{U}}$ denote the TDMA and FDMA rate regions, respectively, in the single-pinch case. {\figurename} {\ref{Figure_SP_MAC_Capacity_Rate_Region}} compares the capacity and TDMA/FDMA achievable rate regions of the single-pinch PASS with those of a conventional fixed-antenna system. It is observed that ${\mathcal{R}}_{{\rm{S}},{\rm{T}}}^{\rm{U}}\subseteq{\mathcal{R}}_{{\rm{S}},{\rm{F}}}^{\rm{U}}\subseteq{\mathcal{C}}_{\rm{S}}^{\rm{U}}$, which has also been analytically proven in \cite{ouyang2025capacity}. The figure also demonstrates that PASS outperforms the fixed-antenna system, offering both larger achievable rate regions and an expanded capacity region.
\subsubsection*{Multiple-Pinch Case}
When $M>1$, an exhaustive search over the feasible set ${\mathcal{X}}$ becomes computationally infeasible, especially for large $M$. For each ${\mathbf{x}}$, the Pareto boundary of the capacity region ${\mathcal{C}}^{\rm{U}}({\mathbf{x}})$---excluding \emph{time-sharing} points---can be achieved via SIC \cite{el2011network}. Motivated by this, we first focus on characterizing the union of all SIC-achievable Pareto-optimal rate pairs over ${\mathbf{x}}\in{\mathcal{X}}$. \emph{Time sharing} is then applied across these rate pairs to form the complete capacity region. 

{To characterize the complete Pareto boundary, we adopt the \emph{rate-profile} method introduced in \cite{mohseni2006optimized,zhang2010cooperative}. Specifically, any rate tuple on the Pareto boundary of the achievable rate region can be obtained by solving the following sum-rate maximization problem for a fixed decoding order $\bm\pi$ and rate-profile factor $\alpha\in[0,1]$ \cite{mohseni2006optimized,zhang2010cooperative,zhang2021intelligent}:
\begin{subequations}\label{Rate_Profile_Uplink_General}
\begin{align}
\max_{{\mathbf{x}},R}~&R\\
{\rm{s.t.}}~&{\mathbf{x}}\in{\mathcal{X}},{\mathcal{R}}_{1}^{\rm{U}}({\bm\pi},{\mathbf{x}})\geq \alpha R,{\mathcal{R}}_{2}^{\rm{U}}({\bm\pi},{\mathbf{x}})\geq (1-\alpha) R,
\label{Rate_Profile_Uplink_General_Constraint2}
\end{align}
\end{subequations}
where ${\mathcal{R}}_{1}^{\rm{U}}({\bm\pi},{\mathbf{x}})\triangleq\log_2\left(1+\frac{{\mathsf{P}}_{[{\bm\pi}]_1}\lvert h_{[{\bm\pi}]_1}({\mathbf{x}})\rvert^2}{{\mathsf{P}}_{[{\bm\pi}]_2}\lvert h_{[{\bm\pi}]_2}({\mathbf{x}})\rvert^2+\sigma^2}\right)$ and ${\mathcal{R}}_{2}^{\rm{U}}({\bm\pi},{\mathbf{x}})\triangleq\log_2\left(1+\frac{{\mathsf{P}}_{[{\bm\pi}]_2}\lvert h_{[{\bm\pi}]_2}({\mathbf{x}})\rvert^2}{\sigma^2}\right)$ denote the achievable rates of the first and second decoded users, respectively. The rate-profile factor $\alpha$ specifies the rate ratio between the first decoded user and the overall sum-rate. For a given $\alpha$, let the optimal solution to problem \eqref{Rate_Profile_Uplink_General} be denoted as $R_{\rm{sum}}$. Then, the corresponding Pareto-optimal rate tuple is $(\alpha R_{\rm{sum}},(1-\alpha)R_{\rm{sum}})$. Geometrically, this point represents the intersection between a ray with direction vector $[\alpha,1-\alpha]^{T}$ and the Pareto boundary of the rate region (see \cite[Fig. 1]{zhang2010cooperative}). By varying $\alpha$ over the interval $[0, 1]$, the entire Pareto boundary of the achievable rate region can thus be obtained.}

\begin{figure}[!t]
\centering
\includegraphics[width=0.35\textwidth]{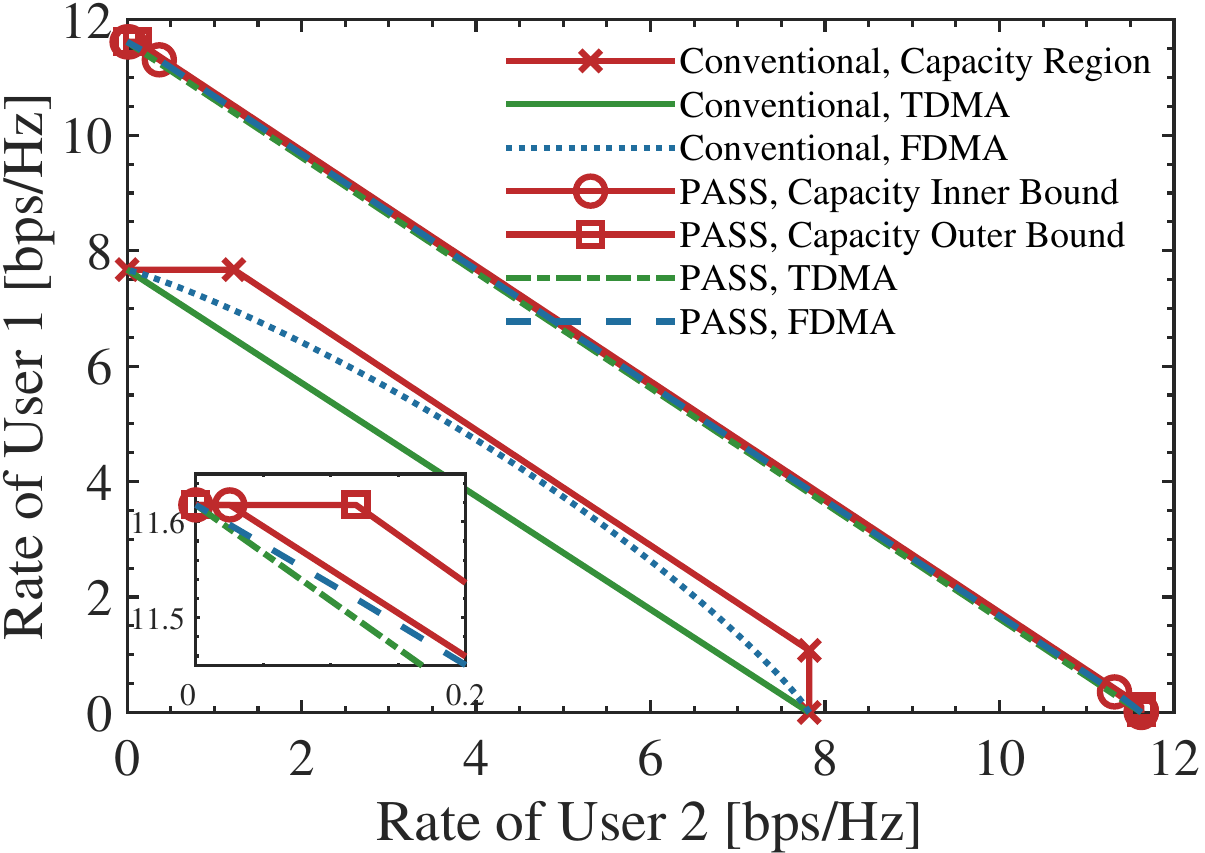}
\caption{Capacity/rate region comparison for PASS-enabled uplink channels in the multiple-pinch case. The simulation parameters used can be found in \cite{ouyang2025capacity}.}
\label{Figure_MP_MAC_Capacity_Rate_Region}
\end{figure}

Let ${\mathbf{x}}_{{\bm\pi}}^{\alpha}$ denote the solution to the above problem, and let ${\mathcal{R}}_{\pi_k}^{\rm{U}}({\mathbf{x}}_{{\bm\pi}}^{\alpha})$ represent the corresponding rate of the $k$th decoded user. Then, the full capacity region is given by
\begin{align}\label{Uplink_Channel_Capacity_Region_Rate_Profile_Basic}
{\mathcal{C}}^{\rm{U}}={\rm{Conv}}\left(\bigcup\nolimits_{\alpha\in[0,1],{\bm\pi}\in\{{\bm\pi}^{\rm{\uppercase\expandafter{\romannumeral1}}},
{\bm\pi}^{\rm{\uppercase\expandafter{\romannumeral2}}}\}}{\mathcal{C}}_{{\bm\pi}}^{\rm{U}}({\mathbf{x}}_{{\bm\pi}}^{\alpha})\right)\triangleq
{\mathcal{C}}_{\rm{M}}^{\rm{U}},
\end{align}
where ${\mathcal{C}}_{{\bm\pi}}^{\rm{U}}({\mathbf{x}}_{{\bm\pi}}^{\alpha})\triangleq\{(R_1^{\rm{U}},R_2^{\rm{U}})\left\lvert R_k^{\rm{U}}\in[0,{\mathcal{R}}_{k}^{\rm{U}}({\mathbf{x}}_{{\bm\pi}}^{\alpha})],k\in{\mathcal{K}}\right.\}$. Since problem \eqref{Rate_Profile_Uplink_General} is non-convex and NP-hard, the element-wise alternating optimization method from \cite{ouyang2025capacity} can be employed to obtain a high-quality solution, yielding an \emph{achievable inner bound} ${\mathcal{C}}_{\rm{M}}^{\rm{U}}$. Additionally, an \emph{outer bound} ${\mathcal{C}}_{{\rm{M}},{\rm{OB}}}^{\rm{U}}$ can also be derived using the Cauchy-Schwarz inequality \cite{ouyang2025capacity}, where these bounds satisfy ${\mathcal{C}}_{{\rm{M}},{\rm{IB}}}^{\rm{U}}\subseteq{\mathcal{C}}_{\rm{M}}^{\rm{U}}\subseteq{\mathcal{C}}_{{\rm{M}},{\rm{OB}}}^{\rm{U}}$.

We next consider the achievable rate regions of TDMA and FDMA in the multiple-pinch case, which are denoted by ${\mathcal{R}}_{{\rm{M}},{\rm{T}}}^{\rm{U}}$ and ${\mathcal{R}}_{{\rm{M}},{\rm{F}}}^{\rm{U}}$, respectively. The TDMA rate region is derived by employing the antenna position refinement method proposed in \cite{xu2025rate} to maximize the per-user data rate. For the FDMA case, the rate region is characterized using the rate-profile approach, where an element-wise alternating optimization algorithm, as developed in \cite{ouyang2025capacity}, is used to compute an inner bound ${\mathcal{R}}_{{\rm{M}},{\rm{F}},{\rm{IB}}}^{\rm{U}}$ on ${\mathcal{R}}_{{\rm{M}},{\rm{F}}}^{\rm{U}}$. More detailed derivations can be found in \cite{ouyang2025capacity}.

{\figurename} {\ref{Figure_MP_MAC_Capacity_Rate_Region}} illustrates the capacity region and achievable rate regions for the multiple-pinch PASS. It can be observed that the derived inner bound closely aligns with the outer bound, indicating the tightness of both bounds. A comparison between {\figurename} {\ref{Figure_MP_MAC_Capacity_Rate_Region}} and {\figurename} {\ref{Figure_SP_MAC_Capacity_Rate_Region}} further reveals that increasing the number of PAs significantly enlarges the capacity and rate regions. {Notably, {\figurename} {\ref{Figure_MP_MAC_Capacity_Rate_Region}} shows that the capacity region of the multiple-pinch PASS is approximately triangular. This shape arises because, at the two SIC corner points, the pinching beamformer is optimized for the second decoded user, which results in a high rate for that user and a seriously degraded rate for the first decoded user. Consequently, the overall region closely resembles the triangular TDMA region. This explains why the TDMA and FDMA achievable regions nearly coincide with the capacity region. We thus conclude that in multiple-pinch PASS, both TDMA and FDMA are nearly capacity-achieving, demonstrating their near-optimality in practical implementations.}

\subsubsection{Downlink PASS}
We now extend the capacity and achievable rate region analysis of the uplink PASS to the downlink PASS by leveraging the \emph{uplink-downlink duality} framework \cite{jindal2004duality}. For ease of exposition, we consider a dual-channel setup in which all downlink channels are assumed to be identical to their uplink counterparts. Under this assumption, the received signal at user $k$ is given by
\begin{align}\label{Signal_Model_Downlink}
y_k=h_k({\mathbf{x}}){{s}}+z_{k},\quad k\in{\mathcal{K}},
\end{align}
where $s=\sqrt{{\mathsf{P}}_1}s_1+\sqrt{{\mathsf{P}}_2}s_2$ is the signal transmitted by the BS, with ${\mathsf{P}}_k$ and $s_k$ denoting the transmit power and information symbol for user $k$, respectively; $z_k\sim{\mathcal{CN}}(0,\sigma^2)$ denotes the receiver noise at user $k$. The total transmit power is constrained by ${\mathsf{P}}_1+{\mathsf{P}}_2\leq {\mathsf{P}}$. Let $R_k^{\rm{D}}$ denote the achievable downlink rate of user $k$. The capacity region of the two-user downlink channel is achieved using \emph{dirty paper coding (DPC)}, which allows the BS to pre-cancel interference \cite{el2011network}. In our considered case, the capacity region can also be achieved by employing superposition coding combined with SIC decoding at the users, i.e., by utilizing power-domain nonorthogonal multiple access (NOMA) \cite{el2011network}.

Based on the \emph{uplink-downlink duality}, the capacity region of the downlink channel is equivalent to the union of the capacity regions of its dual uplink channels. This equivalence holds under identical effective channels $(h_1({\mathbf{x}}),h_2({\mathbf{x}}))$, the same noise power $\sigma^2$, and all power allocations $({\mathsf{P}}_1,{\mathsf{P}}_2)$ that satisfy ${\mathsf{P}}_1+{\mathsf{P}}_2={\mathsf{P}}$. Formally, for any given pinching beamformer ${\mathbf{x}}\in{\mathcal{X}}$, the downlink capacity region is given by \cite{jindal2004duality}
\begin{align}
{\mathcal{C}}^{\rm{D}}({\mathbf{x}})\triangleq\bigcup\nolimits_{{\mathsf{P}}_1+{\mathsf{P}}_2={\mathsf{P}}}{\mathcal{C}}^{\rm{U}}({\mathbf{x}}),
\end{align}
as illustrated in \cite[{\figurename} 2]{jindal2004duality}, where ${\mathcal{C}}^{\rm{U}}({\mathbf{x}})$ (see \eqref{PASS_Uplink_Capacity_Region_Specific}) denotes the capacity region of the corresponding uplink channel for power allocation $({\mathsf{P}}_1,{\mathsf{P}}_2)$ and pinching configuration ${\mathbf{x}}\in{\mathcal{X}}$. 

\begin{figure}[!t]
\centering
    \subfigure[Single-pinch case.]
    {
        \includegraphics[width=0.35\textwidth]{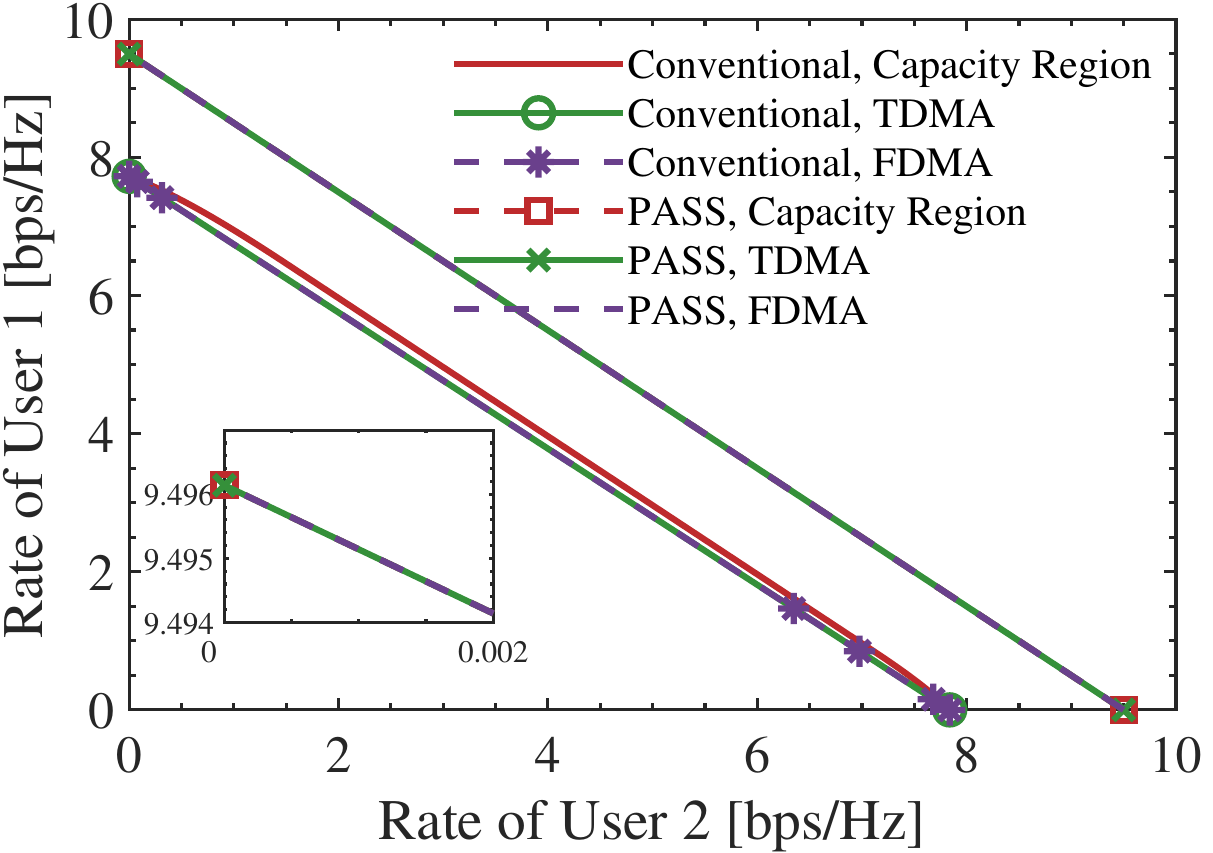}
	   \label{Figure_SP_BC_Capacity_Rate_Region}
    }
    \subfigure[Multiple-pinch case.]
    {
        \includegraphics[width=0.35\textwidth]{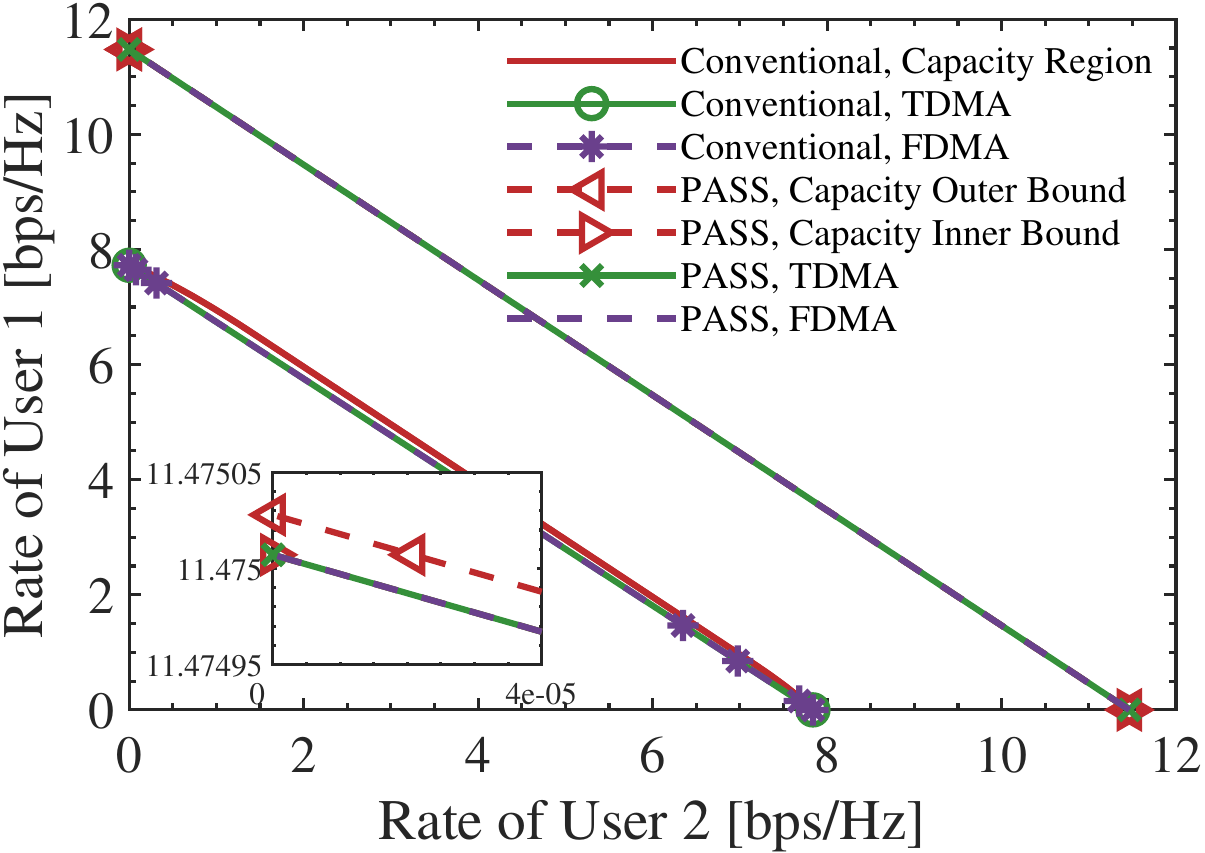}
	   \label{Figure_MP_BC_Capacity_Rate_Region}
    }
\caption{Capacity/rate region comparison for PASS-enabled downlink channels. The simulation parameters used can be found in \cite{ouyang2025capacity}.}
\label{Figure: BC_Capacity_rate_region_comparison_Multiple_Pinch_PASS}
\end{figure}

By considering \emph{time sharing} among different ${\mathbf{x}}\in{\mathcal{X}}$, the overall downlink capacity region of the PASS is given by \cite{ouyang2025capacity}
\begin{center}
    \begin{tcolorbox}[title = Downlink Channel Capacity of PASS]
    {\setlength\abovedisplayskip{2pt}
    \setlength\belowdisplayskip{2pt}
    \begin{subequations}\label{Downlink_Capacity_Region_General}
\begin{align}
{\mathcal{C}}^{\rm{D}}&\triangleq{\rm{Conv}}\left(\bigcup\nolimits_{{\mathbf{x}}\in{\mathcal{X}}}{\mathcal{C}}^{\rm{D}}({\mathbf{x}})\right)\\
&={\rm{Conv}}\left(\bigcup\nolimits_{{\mathsf{P}}_1+{\mathsf{P}}_2={\mathsf{P}}}{\mathcal{C}}^{\rm{U}}\right).
\end{align}
\end{subequations}
    }\end{tcolorbox}
\end{center}
Using this duality-based approach, we can directly derive the capacity and TDMA/FDMA achievable rate regions of the downlink single-pinch PASS from their uplink duals. Similarly, the capacity and achievable rate regions, along with their respective upper and lower bounds, for the downlink multiple-pinch PASS can also be obtained. For brevity, the detailed derivations are omitted here and can be found in \cite{ouyang2025capacity}.

{\figurename} {\ref{Figure_SP_BC_Capacity_Rate_Region}} and {\figurename} {\ref{Figure_MP_BC_Capacity_Rate_Region}} illustrate the downlink capacity and rate regions of PASS for the single-pinch and multiple-pinch scenarios, respectively. In both cases, PASS significantly outperforms conventional fixed-antenna systems in terms of the achievable rate regions. Moreover, the considered OMA schemes (TDMA and FDMA) are observed to be nearly capacity-achieving. Finally, the relationships among the capacity and the rate regions for the downlink PASS are consistent with those for the uplink PASS. 
\subsection{Performance Analysis}
Having characterized the information-theoretic limits, we now analyze several typical performance metrics to evaluate the practical performance of PASS. Given that the earlier capacity analysis has demonstrated that OMA schemes (e.g., TDMA/FDMA) have the capabilities to approach the capacity limits, we focus on a single-antenna typical user served within a single time-frequency resource block without inter-user interference. This scenario effectively captures the standalone performance of PASS in practical deployments.

The performance advantage of PASS over conventional fixed-antenna systems primarily stems from the extended coverage enabled by waveguides and the flexibility in the placement of the PAs. This advantage becomes particularly significant when accounting for the randomness of the users' location. Consequently, an important research direction for PASS lies in the application of \emph{stochastic geometry} to evaluate average network performance under random user distributions.
\subsubsection{Ergodic Rate}
In this section, we analyze the \emph{ergodic rate} achieved by PASS, assuming a user terminal is \emph{uniformly distributed} within a given service region. The resulting average network performance is then compared with that of conventional fixed-antenna systems to quantitatively demonstrate the gains enabled by PASS. 

Consider a typical user which is uniformly distributed within a square region of size $D\times D$, where the square region is centered at the origin with its sides aligned with the $x$- and $y$-axes. A single pinched waveguide is deployed oriented along the $y$-axis, and spans the entire square region. For simplicity, only one PA is activated. To maximize the received SNR, the antenna is positioned at ${\mathbf{p}}=[x_{\mathrm{R}},0,z_{\rm{G}}]^T$, which leads to SNR $\gamma=\frac{P}{\sigma^2}\frac{\eta}{\lVert{\mathbf{p}}-{\mathbf{r}}\rVert^2}=\frac{P}{\sigma^2}\frac{\eta}{y_{\mathrm{R}}^2+z_{\rm{G}}^2}$. Taking into account the user’s random position, the ergodic rate is defined as follows:
\begin{center}
    \begin{tcolorbox}[title = Ergodic Rate of PASS]
    {\setlength\abovedisplayskip{2pt}
    \setlength\belowdisplayskip{2pt}
\begin{align}
{\mathcal{R}}_{\rm{PASS}}\triangleq{\mathbb{E}}_{{\mathbf{r}}}\left\{\log_2\left(1+\frac{P}{\sigma^2}\frac{\eta}{\lVert{\mathbf{p}}-{\mathbf{r}}\rVert^2}\right)\right\}.
\end{align}
    }\end{tcolorbox}
\end{center}
This expectation can be explicitly calculated as follows:
\begin{subequations}
\begin{align}
{\mathcal{R}}_{\rm{PASS}}=\frac{1}{D}\int_{-\frac{D}{2}}^{\frac{D}{2}}
\log_2\left(1+\frac{P}{\sigma^2}\frac{\eta}{y^2+z_{\rm{G}}^2}\right){\rm{d}}y.
\end{align}
\end{subequations}
A closed-form expression for ${\mathcal{R}}_{\rm{PASS}}$ and its high-SNR approximation are available in \cite{ding2025flexible}. 

\begin{figure}[!t]
\centering
\includegraphics[width=0.35\textwidth]{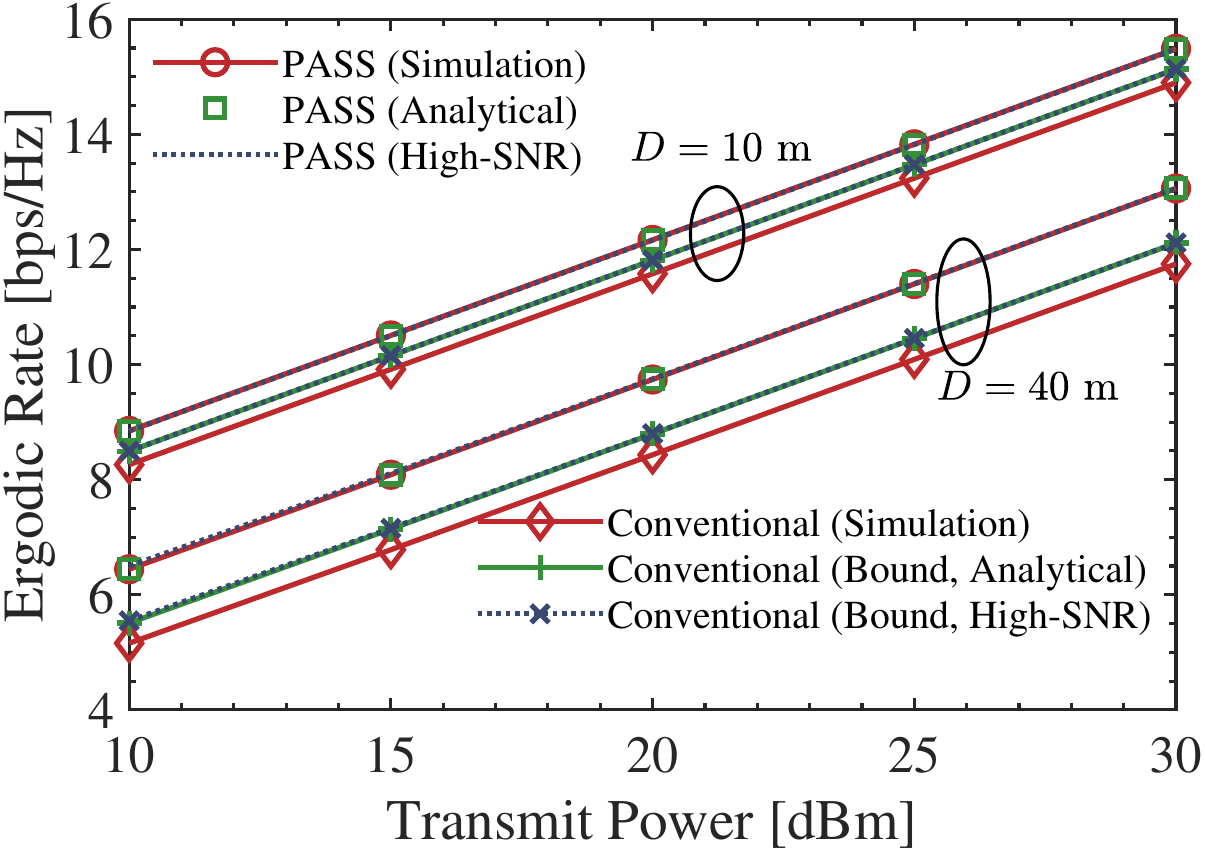}
\caption{Ergodic rates achieved by PASS, with a single PA and a single waveguide. The simulation parameters used can be found in \cite{ding2025flexible}.}
\label{Figure_Ergodic_Rate_CaseI}
\end{figure}

For a conventional fixed-antenna system with the antenna positioned at ${\mathbf{p}}_{\rm{con}}=[0,0,z_{\rm{G}}]^{T}$, the ergodic rate is given by
\begin{subequations}
\begin{align}
{\mathcal{R}}_{\rm{CON}}&\triangleq{\mathbb{E}}_{{\mathbf{r}}}\left\{\log_2\left(1+\frac{P}{\sigma^2}\frac{\eta}{\lVert{\mathbf{p}}_{\rm{con}}-{\mathbf{r}}\rVert^2}\right)\right\}\\
&=\frac{1}{D^2}\int_{-\frac{D}{2}}^{\frac{D}{2}}\int_{-\frac{D}{2}}^{\frac{D}{2}}
\log_2\left(1+\frac{P}{\sigma^2}\frac{\eta}{x^2+y^2+z_{\rm{G}}^2}\right){\rm{d}}x{\rm{d}}y.
\end{align}
\end{subequations}
Since a closed-form expression for ${\mathcal{R}}_{\rm{CON}}$ is generally unavailable, the authors of \cite{ding2025flexible} derived a tractable \emph{upper bound} and its high-SNR approximation for analytical insights. Based on this result, it is shown in \cite{ding2025flexible} that the high-SNR performance gap between PASS and the conventional system satisfies: 
\begin{equation}\label{Comparision_PASS_CONV_Rate}
\begin{split}
\lim_{\frac{P}{\sigma^2}\rightarrow\infty}({\mathcal{R}}_{\rm{PASS}}-{\mathcal{R}}_{\rm{CON}})&\geq\frac{1}{\ln{2}}-\frac{4z_{\rm{G}}}{D\ln{2}}\tan^{-1}\left(\frac{D}{2z_{\rm{G}}}\right)\\
&+\frac{4z_{\rm{G}}^2}{D^2}\log_2\left(1+\frac{D^2}{4z_{\rm{G}}^2}\right).
\end{split}
\end{equation}
This lower bound is proven to be strictly positive and monotonically increasing in $\frac{D}{z_{\rm{G}}}$ \cite{ding2025flexible}. This implies that PASS always outperforms conventional fixed-antenna systems in terms of the ergodic rate at high SNR, and the achieved performance gain grows with the service area size. This advantage is attributed to PASS's ability to create strong LoS links and reduce large-scale path loss by flexibly positioning the PAs.

{\figurename} {\ref{Figure_Ergodic_Rate_CaseI}} depicts the ergodic rate in terms of the transmit power. The results show that PASS significantly outperforms the conventional fixed-antenna system in terms of the ergodic rate. This gain is primarily due to PASS's ability to reduce the user's path loss by dynamically adjusting the antenna position. Moreover, as $D$ increases, the performance gap widens, which highlights PASS's superior scalability in serving larger areas through stronger LoS connections and effective mitigation of the large-scale path loss. 
\subsubsection{Coverage Probability}
Another important performance metric for evaluating the average network performance is the \emph{coverage probability}, which is defined as the likelihood that a typical user achieves a target SNR of at least $\gamma_0$ \cite{andrews2011tractable}, i.e., $\Pr(\gamma>\gamma_0)$. Thus, the coverage probability can also be interpreted as the success probability of the typical transmission/link averaged over all spatial links \cite{haenggi2009interference}. For the considered PASS, the coverage probability is given by
\begin{center}
    \begin{tcolorbox}[title = Coverage Probability of PASS]
    {\setlength\abovedisplayskip{2pt}
    \setlength\belowdisplayskip{2pt}
\begin{align}
{\mathcal{P}}_{\rm{PASS}}^{\rm{c}}\triangleq{\Pr}\left(\frac{P}{\sigma^2}\frac{\eta}{\lVert{\mathbf{p}}-{\mathbf{r}}\rVert^2}>\gamma_0\right).
\end{align}
    }\end{tcolorbox}
\end{center}
Following the derivations in \cite[Proposition 2]{tyrovolas2025performance}, a closed-form expression for ${\mathcal{P}}_{\rm{PASS}}^{\rm{c}}$ can be obtained. On the other hand, the coverage probability for a conventional fixed-antenna architecture is given by ${\mathcal{P}}_{\rm{CON}}^{\rm{c}}\triangleq{\Pr}\left(\frac{P}{\sigma^2}\frac{\eta}{\lVert{\mathbf{p}}_{\rm{con}}-{\mathbf{r}}\rVert^2}>\gamma_0\right)$. As deriving an explicit closed-form expression for ${\mathcal{P}}_{\rm{CON}}^{\rm{c}}$ is challenging, the authors of \cite{tyrovolas2025performance} provided extensive simulation results demonstrating that PASS outperforms conventional fixed-antenna systems in terms of network coverage. This performance advantage becomes even more pronounced as the service region enlarges, which is consistent with the results shown in {\figurename} {\ref{Figure_Ergodic_Rate_CaseI}}. However, a rigorous and insightful theoretical comparison between ${\mathcal{P}}_{\rm{PASS}}^{\rm{c}}$ and ${\mathcal{P}}_{\rm{CON}}^{\rm{c}}$ remains an open problem and presents a promising direction for future research.
\subsubsection{LoS Blockage and Outage Probability}
The results presented earlier are based on the assumption that PASS can establish stable LoS links between PAs and users. However, in highly scattered environments, such as indoor areas with obstacles, such as walls or large furniture, or outdoor scenarios with trees, vehicles, or buildings, LoS links may be likely to be obstructed. A widely used approach to model this effect is to introduce the concept of \emph{LoS blockage probability}, which quantifies the likelihood that a LoS link between transceivers is blocked. Specifically, the probability of maintaining an LoS link is modeled as a function of the transceiver distance, given by $\Pr({\rm{LoS}})={\rm{e}}^{-\beta r} $, where $r$ denotes the distance and $\beta$ is the LoS blockage parameter that reflects the obstacle density in the environment \cite{thornburg2016performance}. Since PASS enables a reduction in transceiver distance, it inherently lowers the LoS blockage probability, thereby enhancing the reliability of wireless links. 

Consider a single-pinch PASS that employs one PA to serve a single-antenna user. Assume that the user is uniformly distributed within a rectangular region with side lengths $D_x$ and $D_y$ along the $x$-axis and $y$-axis, respectively, where the waveguide is aligned with the longer side $D_y$. The LoS blockage-aware received SNR can be modeled as follows:
\begin{align}
\gamma=\frac{P}{\sigma^2}\frac{\varepsilon\eta}{\lVert{\mathbf{p}}-{\mathbf{r}}\rVert^2},
\end{align}
where $\varepsilon\in\{0,1\}$ is a binary indicator representing the presence of a LoS link. Specifically, $\varepsilon=1$ if a LoS link exists between the PA and the user; otherwise, $\varepsilon=0$. According to the previously adopted LoS blockage model \cite{thornburg2016performance}, we have $\Pr(\varepsilon=1)={\rm{e}}^{-\beta \lVert{\mathbf{p}}-{\mathbf{r}}\rVert}$.

\begin{figure}[!t]
\centering
\includegraphics[width=0.35\textwidth]{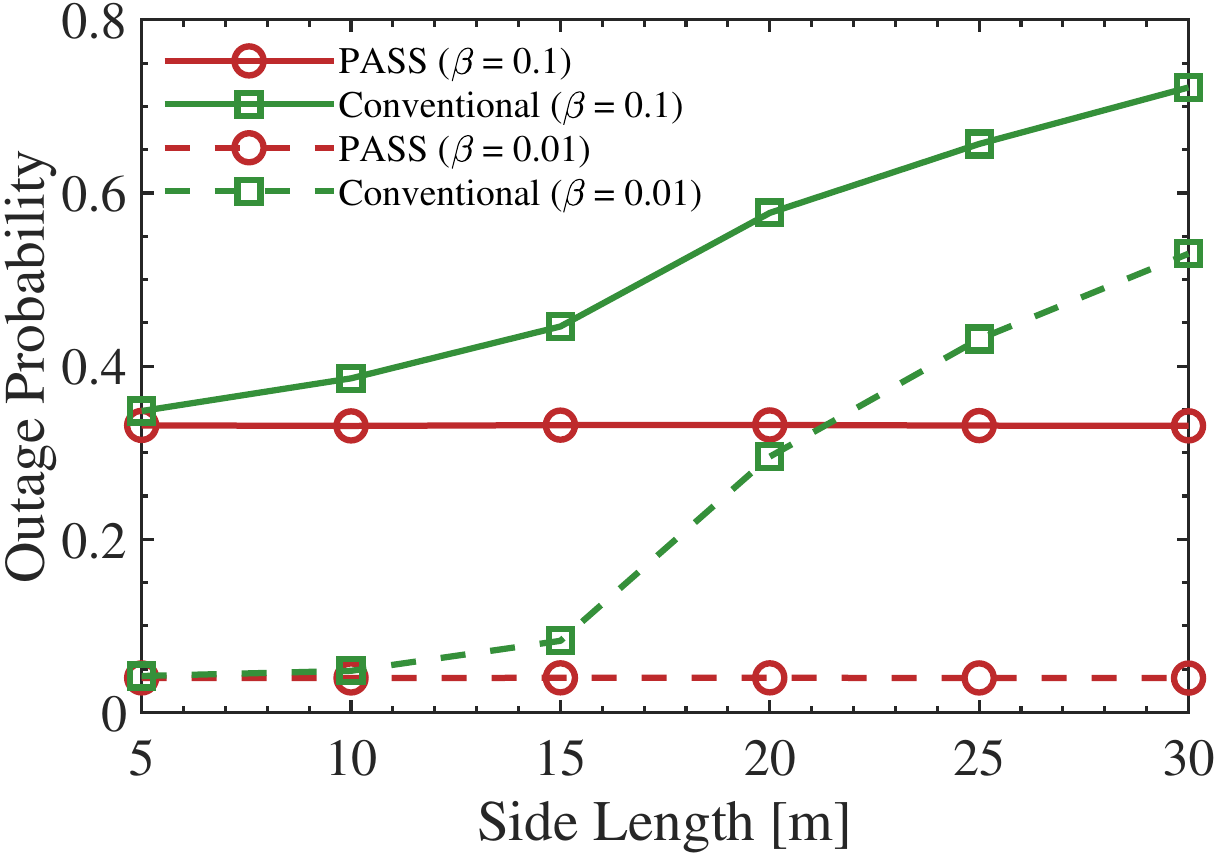}
\caption{Outage probabilities achieved by PASS, with a single PA and a single waveguide for $D_{y}=10$ m. The other simulation parameters used can be found in \cite{ding2025blockage}.}
\label{Figure_OP}
\end{figure}

Given this model, the optimal pinched location to maximize the received SNR remains at the user's projection along the waveguide, i.e., ${\mathbf{p}}=[x_{\rm{R}},0,z_{\rm{G}}]^T$, which yields $\lVert{\mathbf{p}}-{\mathbf{r}}\rVert=\sqrt{y_{\rm{R}}^2+z_{\rm{G}}^2}$. However, due to potential LoS blockage, communication may suffer outages when the instantaneous capacity falls below a target rate $R_{\rm{target}}$. In this context, the system performance is better quantified by the \emph{outage probability}, which is defined as follows:
\begin{align}\label{OP_Calculation_Expression_Basic}
{\mathcal{P}}_{\rm{PASS}}^{\rm{o}}\triangleq\Pr(\log_2(1+\gamma)<R_{\rm{target}}).
\end{align}
This probability can be explicitly calculated as follows:
\begin{center}
    \begin{tcolorbox}[title = Outage Probability of PASS]
    {\setlength\abovedisplayskip{2pt}
    \setlength\belowdisplayskip{2pt}
\begin{equation}\label{OP_Calculation_Expression}
\begin{split}
{\mathcal{P}}_{\rm{PASS}}^{\rm{o}}&=\int_{-\frac{D_y}{2}}^{\frac{D_y}{2}}\int_{-\frac{D_x}{2}}^{\frac{D_x}{2}}\frac{\left(1-{\rm{e}}^{-\beta \sqrt{y^2+z_{\rm{G}}^2}}\right)}{D_xD_y}{\rm{d}}x{\rm{d}}y\\
&+\iint_{{\mathcal{D}}}\frac{{\rm{e}}^{-\beta \sqrt{y^2+z_{\rm{G}}^2}}}{D_xD_y}{\rm{d}}x{\rm{d}}y,
\end{split}
\end{equation}
    }\end{tcolorbox}
\end{center}
where ${\mathcal{D}}\triangleq\{(x,y)|y^2+z_{\rm{G}}^2>\tau_1^2,y\in[-\frac{D_y}{2},\frac{D_y}{2}],x\in[-\frac{D_x}{2},\frac{D_x}{2}]\}$ and $\tau_1=\sqrt{\frac{\eta P}{\sigma^2(2^{R_{\rm{target}}}-1)}}$. A detailed derivation of this result is presented in Appendix \ref{Appendix_OP_Calculation_Expression}. Moreover, it has been shown in \cite{ding2025blockage} that in the high-SNR regime, the outage probability can be approximated as follows:
\begin{align}
{\mathcal{P}}_{\rm{PASS}}^{\rm{o}}\approx 1-f_{\rm{b}}(-D_y/2,D_y/2),
\end{align}
where $f_{\rm{b}}(a,b)\triangleq\frac{1}{D_y}\int_{a}^{b}{\rm{e}}^{-{\beta}\sqrt{y^2+z_{\rm{G}}^2}}{\rm{d}}y$.

The outage probability achieved by a conventional fixed-antenna system can be calculated as follows:
\begin{align}
{\mathcal{P}}_{\rm{CON}}^{\rm{o}}=\Pr(\log_2\left(1+\frac{P}{\sigma^2}\frac{\eta\varepsilon}{\lVert{\mathbf{p}}_{\rm{con}}-{\mathbf{r}}\rVert^2}\right)<R_{\rm{target}}),
\end{align}
where ${\mathbf{p}}_{\rm{con}}=[0,0,z_{\rm{G}}]^{T}$ denotes the fixed antenna location. As shown in \cite{ding2025blockage}, a high-SNR approximation of this outage probability is given by
\begin{align}
{\mathcal{P}}_{\rm{CON}}^{\rm{o}}\approx \int_{-\frac{D_y}{2}}^{\frac{D_y}{2}}\int_{-\frac{D_x}{2}}^{\frac{D_x}{2}}\frac{\left(1-{\rm{e}}^{-\beta \sqrt{x^2+y^2+z_{\rm{G}}^2}}\right)}{D_xD_y}{\rm{d}}x{\rm{d}}y.
\end{align}
For comparison, we define the high-SNR performance gap as $\Delta_{\rm{b}}\triangleq\lim_{\frac{P}{\sigma^2}\rightarrow\infty}({\mathcal{P}}_{\rm{CON}}^{\rm{o}}-{\mathcal{P}}_{\rm{PASS}}^{\rm{o}})$. It is rigorously proven in \cite{ding2025blockage} that $\Delta_{\rm{b}}>0$. This implies that, in the high-SNR regime, the outage probability achieved by PASS is strictly lower than that of the conventional fixed-antenna system. Moreover, it has been shown that $\Delta_{\rm{b}}$ increases monotonically with $D_x$ \cite{ding2025blockage}. This observation is consistent with the earlier findings for the ergodic rate, further highlighting the performance advantage of PAs in environments with broader service areas.

{\figurename} {\ref{Figure_OP}} compares the outage probabilities achieved by PASS and the conventional fixed-antenna system. As shown, PASS outperforms the conventional fixed-antenna system in terms of outage probability. This gain is primarily due to PASS's ability to reduce \emph{both large-scale path loss and LoS blockage} by dynamically adjusting the antenna position. As the rectangle's length $D_x$ increases, the performance advantage of PASS over the fixed-antenna system becomes even more pronounced. This is because increasing the area's length enhances the average user distance from the fixed antenna, thereby increasing path loss and blockage likelihood. In contrast, PASS maintains consistent user proximity by dynamically placing the antenna near the user, ensuring constant path loss and link blockage levels as long as the width is fixed.
\subsection{Discussion and Outlook}
We have characterized the fundamental capacity limits and analyzed several key performance metrics of PASS to highlight its advantages over conventional fixed-antenna systems. Our analysis demonstrates that PASS can effectively reduce transceiver distances, thereby lowering \emph{large-scale path loss} and decreasing the likelihood of \emph{LoS blockage}. The established analytical framework is expected to provide meaningful insights into the efficient design of PASS-enabled networks. However, several open research challenges remain, which we summarize below.
\begin{itemize}
  \item \emph{General Information-Theoretic Limits:} The preceding analysis characterized the capacity limits of uplink and downlink PASS-enabled channels in a two-user scenario with a single pinched waveguide. This framework can be extended to more general settings involving an arbitrary number of users by employing the rate-profile approach and accounting for all possible decoding orders. Another promising direction is the investigation of the capacity regions in multiple-waveguide scenarios, which would necessitate the joint design of baseband and pinching beamforming. Additionally, exploring the information-theoretic limits of other multiuser PASS-enabled networks---such as interference channels and wiretap channels---remains an important avenue for future research.
  \item \emph{Multiuser PASS Performance Analysis:} While we have developed a comprehensive performance evaluation for the single-user, single-waveguide case, extending this analysis to multiuser and multiple-waveguide environments is essential. In such setups, analyzing ergodic rates, coverage probabilities, and outage probabilities under different linear beamforming strategies, including zero-forcing (ZF), maximal ratio transmission (MRT), and minimum mean-squared error (MMSE), is an important next step. However, this extension poses significant analytical challenges due to inter-user interference. Tractable modeling may require simplified channel assumptions, and preliminary efforts in this direction have been initiated in \cite{ding2025flexible,ding2025blockage}.
  \item \emph{EM Coupling-Aware Performance Limits:} In multiple-antenna PASS implementations, a minimum inter-antenna spacing is typically used to mitigate EM coupling. However, recent research shows that EM coupling can be exploited to form super-directive and super-wideband beams, enhancing array performance \cite{marzetta2019super} and widening operational bandwidth \cite{akrout2023super}. To realize these benefits in PASS, it is necessary to analyze its EM-coupling-aware performance. This research challenge lies at the intersection of information theory and EM theory, where multi-port circuit theory has recently emerged as a promising analytical tool to bridge the gap between these domains \cite{ivrlavc2010toward}. Nonetheless, research on this topic is still in its infancy and warrants substantial further investigation.
\end{itemize}
\section{Pinching Beamforming Optimization}\label{sec:pass-opt}

In this section, we discuss the beamforming optimization for PASS. Specifically, we first introduce the fundamental principles of pinching beamforming, providing insights into the associated power scaling law. We then examine the integration of pinching beamforming with conventional transmit beamforming, followed by extensions to multi-user scenarios and wideband OFDM systems.

\subsection{Pinching Beamforming Basis and Power Scaling Law}
\label{Optimization_SISO}
As discussed in \textbf{Remark \ref{remark_1}}, the position of the PAs can alter wireless channels.
In what follows, we will present how to leverage this characteristic to enhance the throughput, referred to as pinching beamforming.
To elaborate on the concept of pinching beamforming, let us first consider a simple case with a single waveguide and a single downlink communication user. The waveguide is fed by a single RF chain. According to \eqref{narrowband_LoS_signal_model}-\eqref{narrowband_LoS_signal_model_rn}, the overall signal received at the user under narrowband and LoS conditions is given by 
\begin{align}
  y(t) = & \sum_{m=1}^M y_m(t) + z(t) \nonumber \\
  = & \mathbf{h}^H(\mathbf{x}) \mathbf{g}(\mathbf{x}, \boldsymbol{\rho}) s(t) + z(t),
\end{align}
where $z(t) \sim \mathcal{CN}(0, \sigma^2)$ is the additive white Gaussian noise, and $\mathbf{h}(\mathbf{x}) \in \mathbb{C}^{M \times 1}$ and $\mathbf{g}(\mathbf{x}, \boldsymbol{\rho}) \in \mathbb{C}^{M \times 1}$ denote the free-space and in-waveguide channel vectors, respectively, given by 
\begin{align}
  & \mathbf{h}(\mathbf{x}) = \left[ \frac{\eta e^{-\mathrm{j} \frac{2 \pi}{\lambda} r_1}}{r_1},\dots,\frac{\eta e^{-\mathrm{j} \frac{2 \pi}{\lambda} r_M}}{r_M} \right]^H, \\
  & \mathbf{g}(\mathbf{x}, \boldsymbol{\rho}) = \left[ \sqrt{P_1} e^{-\mathrm{j} \frac{2\pi}{\lambda} n_{\mathrm{eff}} x_1},\dots,\sqrt{P_M} e^{-\mathrm{j} \frac{2\pi}{\lambda} n_{\mathrm{eff}} x_M} \right]^T,
\end{align}  
with $\mathbf{x} = [x_1,\dots,x_M]^T$ and $\boldsymbol{\rho} = [P_1,\dots,P_M]^T$ denoting the PA position vector and the power radiation vector, respectively. Note that the distance $r_n$ in $\mathbf{h}(\mathbf{x})$ is a function of $x_n$, as specified in  \eqref{narrowband_LoS_signal_model_rn}. The overall received power at the communication user is given by
\begin{align}
  P_r = & \left|\mathbf{h}^H(\mathbf{x}) \mathbf{g}(\mathbf{x}, \boldsymbol{\rho})\right|^2 P_t \nonumber \\
  = & \left| \sum_{m=1}^M \frac{\eta \sqrt{P_m} e^{-\mathrm{j} \frac{2\pi}{\lambda} \left(\sqrt{(x_m - x_{\mathrm{R}})^2 + \zeta^2} + n_{\mathrm{eff}} x_m\right) }}{\sqrt{(x_m - x_{\mathrm{R}})^2 + \zeta^2}}  \right|^2 P_t,
\end{align}
where $P_t = \mathbb{E}[ |s(t)|^2 ]$ denote the overall transmit power. 
Thus, the maximum achievable rate in bits per second per Hertz (bps/Hz) of the considered single-user single-waveguide system is given by $\gamma = \log_2(1 + P_r/\sigma^2)$. To maximize the communication performance, we aim to maximize the array gain by optimizing the pinching beamforming, which can be achieved by selecting the appropriate $\mathbf{x}$, i.e., the position of the PAs, and, when possible, adjusting the power allocation vector $\boldsymbol{\rho}$. The corresponding pinching beamforming optimization problem can be formulated as  
\begin{center}
    \begin{tcolorbox}[title = Pinching Beamforming Optimization]
    {\setlength\abovedisplayskip{2pt}
    \setlength\belowdisplayskip{2pt}
    \begin{subequations} \label{BF_problem_1}
      \begin{align}
        \max_{\mathbf{x}, \boldsymbol{\rho}} \quad & \left|\mathbf{h}^H(\mathbf{x}) \mathbf{g}(\mathbf{x}, \boldsymbol{\rho})\right|^2 \\
        \mathrm{s.t.} \quad & \boldsymbol{\rho} \in \mathcal{P}, \; \mathbf{x} \in \mathcal{X}.
      \end{align}
    \end{subequations}
    }\end{tcolorbox}
\end{center}
In this problem, $\mathcal{P}$ denotes the feasible set of power allocations, defined by \eqref{general_power_radiation_rule}, \eqref{equal_power_radiation_rule}, or \eqref{proportion_power_radiation_rule}, and $\mathcal{X}$ represents the feasible set of PA positions, characterized by \eqref{continuous_activation_constraint}, \eqref{discrete_activation_constraint}, or \eqref{semicontinuous_activation_constraint} for different activation methods. In the sequel, we focus on the pinching beamforming design via the optimization of the PA positions $\mathbf{x}$ and consider the fixed power radiation model to simplify the analysis, but keep in mind that optimizing the power radiation model can further enhance the communication performance.

To gain analytical insights into the optimization problem, we consider the case of equal power radiation \eqref{equal_power_radiation_rule} and continuous activation \eqref{continuous_activation_constraint} of the PAs. The power radiated by each PA is given by $P_m = 1/M$. Under these conditions, {problem \eqref{BF_problem_1} can be solved using the antenna position refinement algorithm proposed in \cite{xu2025rate}. For simplicity of notation, we assume $M$ is an even integer and present the following theorem, which characterizes the optimal solution of problem~\eqref{BF_problem_1}.

\begin{figure}[!t]
  \centering
  \includegraphics[width=0.4\textwidth]{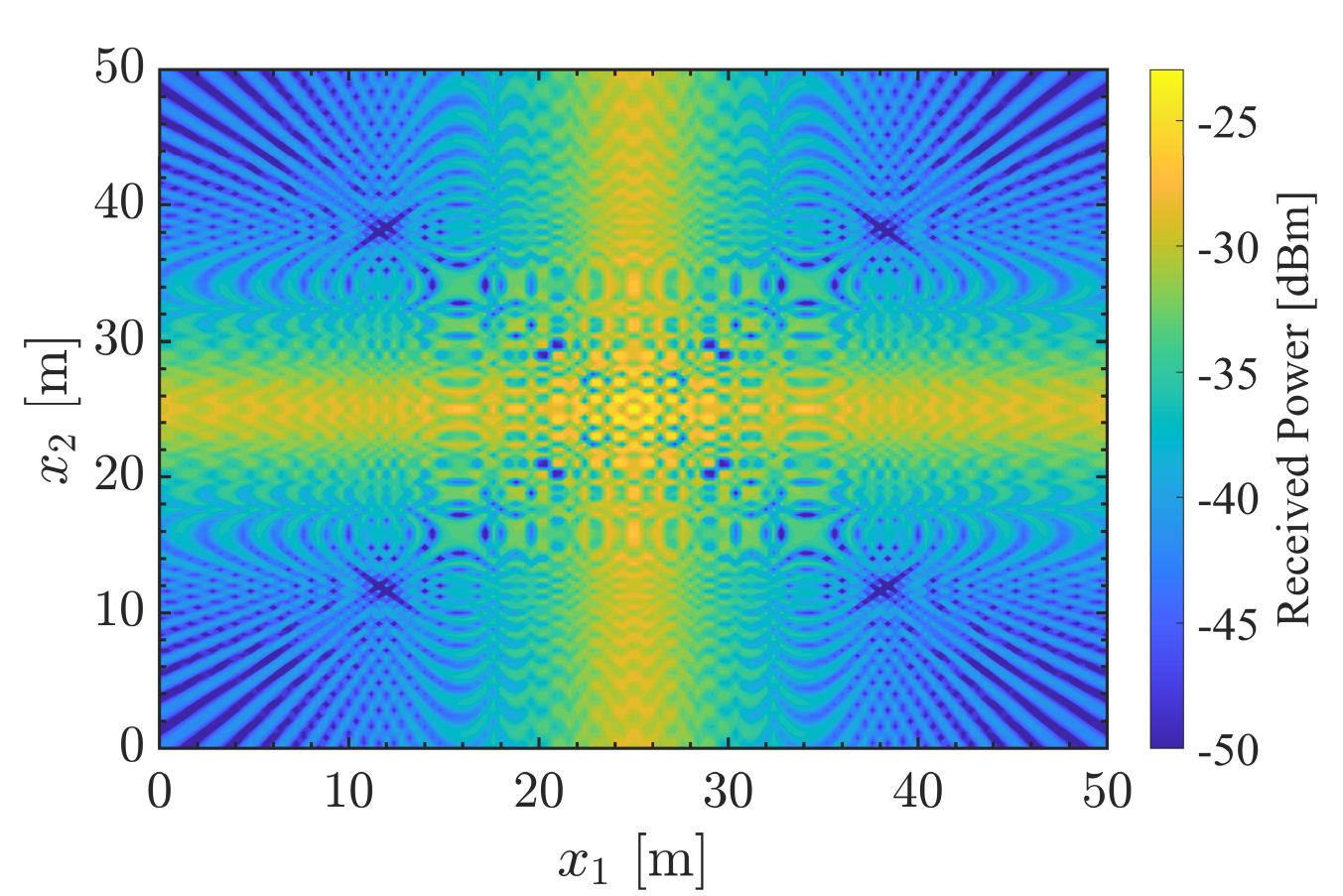}
  \caption{Illustration of the received power with respect to the $M=2$ PA positions on the waveguide, when $P_t = 30 $ dBm. The remaining system parameters follow the setup in \cite{wang2025modeling}. It can be observed that the objective function has a large number of local optima, many of which exhibit significant gaps to the global optimum.}
  \label{Figure_SISO_Obj_function}
\end{figure}

\begin{theorem} \label{theorem_scaling_law}
 \emph{(Maximum Receive Power \cite{ouyang2025array})} With equal power radiation \eqref{equal_power_radiation_rule} and continuous activation \eqref{continuous_activation_constraint}, the maximum value of the receive power $P_r$ is tightly approximated by 
  \begin{equation} \label{maximum_receive_power}
    P_{r,\max} \approx \frac{2 \eta^2 P_t}{\zeta \Delta_{\min}} f_{\mathrm{ub}} \left( \frac{M \Delta_{\min}}{2 \zeta} \right),
  \end{equation}
  where $f_{\rm{ub}}(x)\triangleq\frac{\ln^2(\sqrt{1+x^2}+x)}{x}$. 
\end{theorem}

\begin{IEEEproof}
  Please refer to Appendix \ref{theorem_scaling_law_proof}.
\end{IEEEproof}

Based on the optimal value given in \textbf{Theorem \ref{theorem_scaling_law}} as well as the discussion in Appendix \ref{theorem_scaling_law_proof}, the following \emph{power scaling law} of PASS as $M\rightarrow\infty$ can be obtained: 
\begin{center}
  \begin{tcolorbox}[title = Power Scaling Law of PASS]
  {\setlength\abovedisplayskip{2pt}
  \setlength\belowdisplayskip{2pt}
    \begin{align}
    \lim_{M \rightarrow \infty} P_{r,\max} \simeq {\mathcal{O}}\left(\frac{\ln^2{M}}{M}P_t \right).
    \end{align}
  }\end{tcolorbox}
\end{center}

\begin{figure}[!t]
\centering
\includegraphics[width=0.35\textwidth]{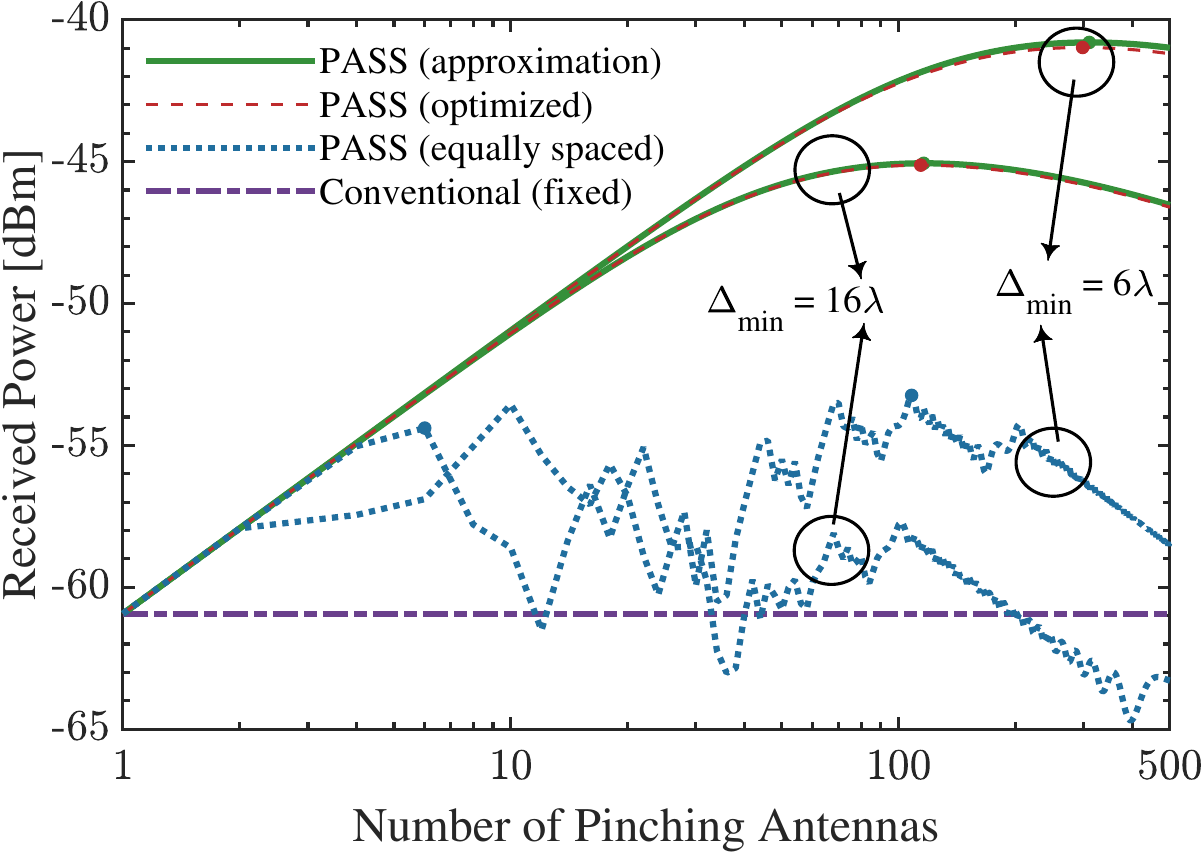}
\caption{The maximum received power achieved by the PAs with $P_{t}=10$ dBm. The other simulation parameters used can be found in \cite{ouyang2025array}.}
\label{Figure_Array_Gain_Number}
\end{figure}

\begin{remark}\label{remark_number}
  From the above power scaling law, it is easy to see that $\lim_{M \rightarrow \infty} P_r = 0$. This result indicates that simply increasing the number of PAs does not guarantee continuous improvements in array gain. As $M$ increases, the power per antenna ${P_t}/{M}$ decreases, and most PAs become too distant from the user to contribute effectively. Therefore, there exists an optimal number of PAs that maximizes the received power in PASS.
  Since \eqref{maximum_receive_power} offers a tight approximation for the array gain, we analyze it to derive an approximate expression for the optimal number of PAs. By evaluating the derivative of $f_{\rm{ub}}(x)$ and numerically solving $\frac{d}{d x}f_{\rm{ub}}(x)=0$ using the bisection method, we find that $f_{\rm{ub}}(x)$ is maximized at $x\approx3.32$. Therefore, the optimal number of PAs satisfies 
  \begin{equation}
    M^{\star} \approx \frac{6.64 \zeta}{\Delta_{\min}}
  \end{equation}
\end{remark}

In the previous discussion, we derived a tight approximation of the maximum received power for a specific scenario. However, to approach this maximum value and effectively address the pinching beamforming problem under general conditions, we need to solve non-convex optimization problem \eqref{BF_problem_1}. For arbitrary power radiation models, this optimization is particularly challenging, as the PA positions $\mathbf{x}$ are intricately coupled in both the objective function and the constraints. Additionally, as shown in Fig. \ref{Figure_SISO_Obj_function}, the objective function is highly multimodal, exhibiting numerous local optima with a large gap to the global optimum. Consequently, gradient-based methods are ineffective for this problem. An efficient alternative is the element-wise optimization approach, where each PA position is optimized sequentially while fixing the positions of the other PAs. In this approach, each individual PA position can be optimized through simple one-dimensional search methods \cite{wang2025modeling} or more sophisticated search methods tailored for PASS \cite{xu2025rate} to avoid convergence to poor local optima.

{\figurename} {\ref{Figure_Array_Gain_Number}} illustrates the receive power as a function of the number of antennas for different values of $\Delta_{\min}$. For comparison, results are shown for equally spaced PAs, an optimized configuration (from \cite{xu2025rate,ouyang2025array}), and the approximation of the maximum receive power. As can be seen from this graph, the maximum receive power in both cases is non-monotonic with respect to the number of antennas, and there exists an optimal number of antennas. Notably, the derived approximation aligns closely with the optimized receive power.

\begin{figure}[!t]
\centering
\includegraphics[height=0.20\textwidth]{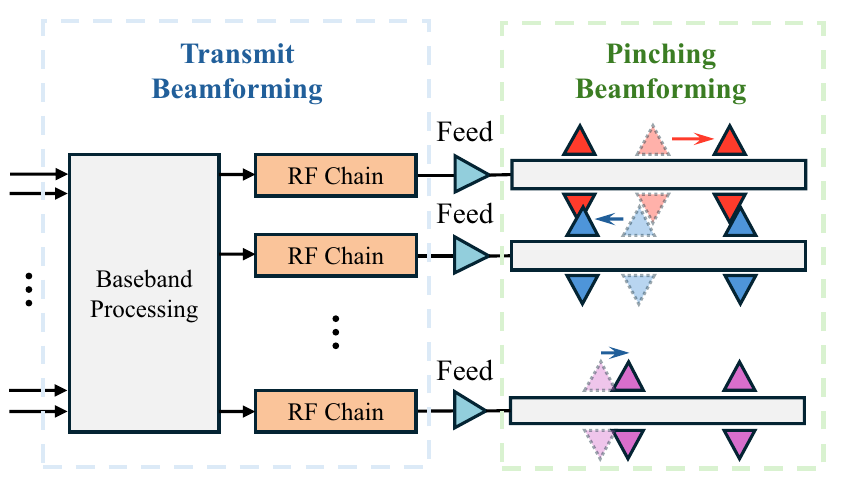}
\caption{Sub-connected architecture for joint transmit and pinching beamforming}
\label{Figure_subconnected}
\end{figure}

\subsection{Joint Beamforming Design for the Single-user Case} \label{sec:SU-MISO}

We now consider a more complex single-user multi-waveguide PASS-based communication system comprising $N$ waveguides, each equipped with $M$ PAs, serving a single-antenna downlink communication user. All waveguides are assumed parallel to the $x$-axis, with the feed point of the $n$-th waveguide located at $\mathbf{p}_{\text{feed},n} = [0,y_{\mathrm{G},n}, z_{\mathrm{G},n}]^T$. Thus, the position of the $m$-th PA on the $n$-th waveguide is given by $\mathbf{p}_{nm} = [x_{nm}, y_{\mathrm{G},n}, z_{\mathrm{G},n}]^T$. Consequently, the distances from the $m$-th PA on the $n$-th waveguide to its feed point and the user are respectively:
\begin{align}
  d_{nm} & = \|\mathbf{p}_{nm} - \mathbf{p}_{\text{feed},n}\| = x_{nm}, \\
  r_{nm} & = \|\mathbf{p}_{nm} - \mathbf{r}\| = \sqrt{(x_{nm} - x_{\mathrm{R}})^2 + \zeta^2_{n}},
\end{align}  
where $\zeta^2_{n} = (y_{\mathrm{G},n} - y_{\mathrm{R}})^2 + (z_{\mathrm{G},n} - z_{\mathrm{R}})^2$. For this configuration, the free-space and in-waveguide propagation vectors from the $n$-th waveguide to the receiver are given by
\begin{align}
  & \widetilde{\mathbf{h}}(\mathbf{x}_n) = \left[ \frac{\eta e^{-\mathrm{j} \frac{2 \pi}{\lambda} r_{n1}}}{r_{n1}},\dots,\frac{\eta e^{-\mathrm{j} \frac{2 \pi}{\lambda} r_{nM}}}{r_{nM}} \right]^H, \\
  & \mathbf{g}(\mathbf{x}_n, \boldsymbol{\rho}_n) = \Big[ \sqrt{P_{n1}} e^{-\mathrm{j} \frac{2\pi}{\lambda} n_{\mathrm{eff}} x_{n1}},\nonumber \\
  &\hspace{3cm}\dots,\sqrt{P_{nM}} e^{-\mathrm{j} \frac{2\pi}{\lambda} n_{\mathrm{eff}} x_{nM}} \Big]^T,
\end{align}
where $\mathbf{x}_n = [x_{n1},\dots,x_{nM}]^T$ and $\boldsymbol{\rho}_n = [P_{n1},\dots,P_{nM}]^T$ represent the positions and radiation power of the PAs, respectively. The system employs a total of $N_{\mathrm{RF}}$ RF chains for feeding the signals. Depending on how these RF chains are connected to the waveguides, we propose two transmission structures for the joint beamforming design in PASS as follows.

\subsubsection{Sub-connected Architecture}
In the sub-connected architecture shown in Fig. \ref{Figure_subconnected}, each RF chain feeds only a single waveguide, resulting in $N_{\mathrm{RF}} = N$. Let $\mathbf{w} = [w_1,\dots,w_N]^T$ denote the transmit beamforming vector, where $w_n$ is associated with the $n$-th RF chain. Hence, under narrowband and LoS conditions, the received signal at the communication user is expressed as:
\begin{align} \label{single_user_MISO_signal}
  y(t) & = \sum_{n=1}^N \widetilde{\mathbf{h}}^H(\mathbf{x}_n) \mathbf{g}(\mathbf{x}_n, \boldsymbol{\rho}_n) w_n s(t) + z(t)\nonumber \\
  & = \mathbf{h}^H(\mathbf{X}) \mathbf{G}(\mathbf{X}, \mathbf{P}) \mathbf{w} s(t) + z(t),
\end{align}    
where $\mathbf{h}(\mathbf{X}) = [\widetilde{\mathbf{h}}^T(\mathbf{x}_1),\dots,\widetilde{\mathbf{h}}^T(\mathbf{x}_N)]^T$, $\mathbf{G}(\mathbf{X}) = \mathrm{Blkdiag}\{\mathbf{g}(\mathbf{x}_1, \boldsymbol{\rho}_1),\dots,\mathbf{g}(\mathbf{x}_N, \boldsymbol{\rho}_N)\}$, $\mathbf{X} = [\mathbf{x}_1,\dots,\mathbf{x}_N]$, and $\mathbf{P} = [\boldsymbol{\rho}_1,\dots,\boldsymbol{\rho}_N]$ denote the overall free-space channel vector, in-waveguide channel matrix, PA position matrix, and PA power radiation matrix, respectively. Similar to \eqref{BF_problem_1}, the communication performance maximization problem in the multi-waveguide system is given by 
\begin{figure}[!t]
\centering
\includegraphics[height=0.22\textwidth]{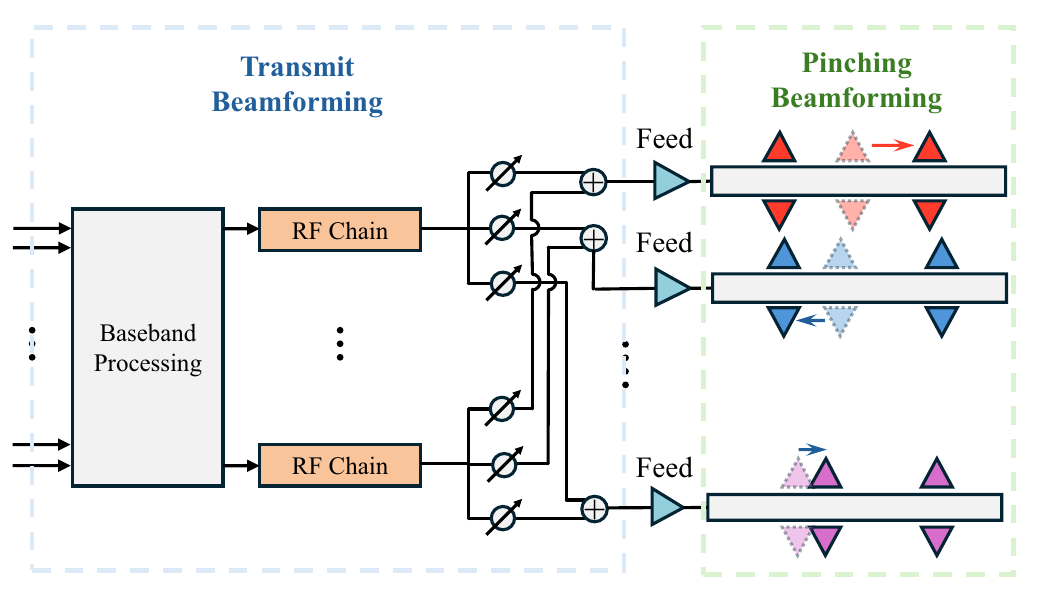}
\caption{Fully-connected architecture for joint transmit and pinching beamforming}
\label{Figure_fully-connected}
\end{figure}
\begin{center}
    \begin{tcolorbox}[title = Joint Transmit and Pinching Beamforming Optimization with Sub-Connected Architecture]
    {\setlength\abovedisplayskip{2pt}
    \setlength\belowdisplayskip{2pt}
    \begin{subequations} \label{BF_problem_2}
      \begin{align}
        \max_{\mathbf{w}, \mathbf{X}, \mathbf{P}} \quad & \left|\mathbf{\mathbf{h}}^H(\mathbf{X}) \mathbf{G}(\mathbf{X}, \mathbf{P}) \mathbf{w}\right|^2 \\
        \mathrm{s.t.} \quad & \|\mathbf{w}\|^2 \le P_t, \\ 
        & \mathbf{x}_n \in \mathcal{X}_n, \; \boldsymbol{\rho}_n \in \mathcal{P}_n, \forall n.
      \end{align}
    \end{subequations}
    }\end{tcolorbox}
\end{center}
Here, $P_t$ denotes the maximum transmit power. It is well known that for a single-user system, the optimal transmit beamforming solution is given by MRT, i.e., $ \mathbf{w}_{\mathrm{MRT}} = \sqrt{P_t} \frac{\left(\mathbf{\mathbf{h}}^H(\mathbf{X}) \mathbf{G}(\mathbf{X}, \mathbf{P})\right)^H}{\|\mathbf{\mathbf{h}}^H(\mathbf{X}) \mathbf{G}(\mathbf{X}, \mathbf{P})\|}$. By substituting the MRT solution, problem \eqref{BF_problem_2} reduces to 
\begin{subequations}
  \begin{align}
    \max_{\mathbf{X}, \mathbf{P}} \quad & \left\|\mathbf{\mathbf{h}}^H(\mathbf{X}) \mathbf{G}(\mathbf{X}, \mathbf{P}) \right\|^2 P_t \\
    \mathrm{s.t.} \quad & \mathbf{x}_n \in \mathcal{X}_n, \; \boldsymbol{\rho}_n \in \mathcal{P}_n, \forall n.
  \end{align}
\end{subequations}
Following \eqref{single_user_MISO_signal}, the objective function, which is essentially the overall receive power at the communication user, can be rewritten as 
\begin{align} \label{multi_user_receiver}
  P_r = & \left\|\mathbf{\mathbf{h}}^H(\mathbf{X}) \mathbf{G}(\mathbf{X}, \mathbf{P}) \right\|^2 P_t \nonumber \\
  = & \sum_{n=1}^N \left| \widetilde{\mathbf{h}}^H(\mathbf{x}_n) \mathbf{g}(\mathbf{x}_n, \boldsymbol{\rho}_n) \right|^2 P_t.
\end{align}
It can be observed that the joint beamforming problem for the single-user multi-waveguide system is equivalent to $N$ independent single-waveguide pinching beamforming problems. Therefore, based on \textbf{Theorem \ref{theorem_scaling_law}}, the maximum receive power under the assumption of equal power radiation and continuous activation can be tightly approximated as
\begin{equation}
  P_{r, \max} \approx \frac{2 \eta^2 P_t}{\Delta_{\min}} \sum_{n=1}^N \frac{1}{\zeta_n} f_{\mathrm{ub}} \left( \frac{M \Delta_{\min}}{2 \zeta_n} \right).
\end{equation} 

For more general cases, each of the multiple SISO problems can be solved individually. Thereby, high complexity search algorithms are no longer needed in this case, thus significantly reducing the optimization complexity.

\subsubsection{Fully-connected Architecture}

In the fully-connected architecture depicted in Fig. \ref{Figure_fully-connected}, each RF chain is connected to all waveguides via phase shifters (PSs), where $N_{\mathrm{RF}} \neq N$ becomes possible. Let $\mathbf{w}_{\mathrm{BB}} \in \mathbb{C}^{N_{\mathrm{RF}} \times 1}$ and $\mathbf{W}_{\mathrm{RF}} \in \mathbb{C}^{N \times N_{\mathrm{RF}}}$ denote the baseband beamforming vector and the PS-based analog beamforming matrix, respectively. For this setup, the received signal at the communication user is given by 
\begin{align}
  y(t)  = \mathbf{h}^H(\mathbf{X}) \mathbf{G}(\mathbf{X}, \mathbf{P}) \mathbf{W}_{\mathrm{RF}}\mathbf{w}_{\mathrm{BB}} s(t) + z(t).
\end{align}    
The corresponding optimization problem is given by 
\begin{center}
    \begin{tcolorbox}[title = Joint Transmit and Pinching Beamforming Optimization with Fully-Connected Architecture]
    {\setlength\abovedisplayskip{2pt}
    \setlength\belowdisplayskip{2pt}
    \begin{subequations} \label{BF_problem_3}
      \begin{align}
        \max_{\mathbf{W}_{\mathrm{RF}}, \mathbf{w}_{\mathrm{BB}}, \mathbf{X}, \mathbf{P}} \quad & \left|\mathbf{\mathbf{h}}^H(\mathbf{X}) \mathbf{G}(\mathbf{X}, \mathbf{P}) \mathbf{W}_{\mathrm{RF}} \mathbf{w}_{\mathrm{BB}}\right|^2 \\
        \mathrm{s.t.} \quad & \|\mathbf{w}_{\mathrm{BB}}\|^2 \le P_t, \\ 
        & \left|\left[ \mathbf{W}_{\mathrm{RF}} \right]_{ij}\right|^2 = \frac{1}{N}, \forall i,j, \\
        & \mathbf{x}_n \in \mathcal{X}_n, \; \boldsymbol{\rho}_n \in \mathcal{P}_n, \forall n.
      \end{align}
    \end{subequations}
    }\end{tcolorbox}
\end{center}
While the problem formulation resembles conventional hybrid beamforming, where energy efficiency is improved by using a few RF chains ($N_{\mathrm{RF}} < N$), the fully-connected architecture for PASS can, in contrast, operate with more RF chains than waveguides ($N_{\mathrm{RF}} > N$). The reasons are two-fold. On the one hand, physically increasing the number of waveguides is challenging because each waveguide occupies substantial space. A more practical approach is to deploy a small set of waveguides and pinch a larger number of PAs onto them, thereby enlarging the spatial DoFs. On the other hand, to exploit the spatial DoFs provided by the large number of PAs, employing fewer RF chains than waveguides may be insufficient to realize the full beamforming capability. 

Compared to the sub-connected architecture, the fully-connected architecture introduces additional challenges due to the coupling between the baseband and PS-based analog beamforming vectors. For the case of $N_{\mathrm{RF}} < N$, this problem can be solved in a similar way to the conventional hybrid beamforming problem. In particular, the baseband and PS-based analog beamforming vectors can be jointly optimized to approximate a optimal fully-dimensional transmit beamforming vector $\mathbf{w}^{\mathrm{opt}} \in \mathbb{C}^{N \times 1}$ by minimizing the least-square function $\|\mathbf{w}^{\mathrm{opt}} - \mathbf{W}_{\mathrm{RF}} \mathbf{w}_{\mathrm{BB}} \|$, which can be effectively addressed exploiting existing methods \cite{6717211, 7397861, 8332507}. On the contrary, for the case of $N_{\mathrm{RF}} > N$, simply extending the above approach is insufficient because it fails to leverage the extra DoFs provided by the RF chains. A straightforward workaround is to treat $\mathbf{G}(\mathbf{X}, \mathbf{P}) \mathbf{W}_{\mathrm{RF}} \mathbf{w}_{\mathrm{BB}}$ as a whole and design to approximate a high-dimensional equivalent PA beamformer $\widetilde{\mathbf{w}}^{\mathrm{opt}} \in \mathbb{C}^{MN \times 1}$ for the overall free-space channel $\mathbf{h}(\mathbf{X})$ by minimizing $\|\widetilde{\mathbf{w}}^{\mathrm{opt}} - \mathbf{G}(\mathbf{X}, \mathbf{P}) \mathbf{W}_{\mathrm{RF}} \mathbf{w}_{\mathrm{BB}} \|$. However, since both the free-space channel and the overall PA beamformer depend on the PA position matrix $\mathbf{X}$, decoupled optimization can incur substantial performance loss, underscoring the need for an efficient joint algorithm design to directly tackle the original problem \eqref{BF_problem_3} when $N_{\mathrm{RF}} > N$.

\subsection{Joint Beamforming Design for the Multi-user Case}
Next, we consider the general case with $K$ communication users and $N$ waveguides with $M$ PAs on each one of them and $N_{\mathrm{RF}}$ chains. The position of user $k$ is denoted by $\mathbf{r}_k = [x_{\mathrm{R},k}, y_{\mathrm{R},k}, z_{\mathrm{R},k}]^T$. The distance from the $m$-th PA on the $n$-th waveguide to the user $k$ is thus given by 
\begin{equation}
  r_{knm} = \|\mathbf{p}_{nm} - \mathbf{r}_k\| = \sqrt{\left(x_{nm} - x_{\mathrm{R},k}\right)^2 + \zeta^2_{nk}},
\end{equation}        
where $\zeta^2_{nk} = (y_{\mathrm{G},n} - y_{\mathrm{R},k})^2 + (z_{\mathrm{G},n} - z_{\mathrm{R},k})^2$. The overall free-space channel for user $k$ is given by 
\begin{align}
  &\mathbf{h}_k(\mathbf{X}) = \left[ \widetilde{\mathbf{h}}_k^T(\mathbf{x}_1),\dots, \widetilde{\mathbf{h}}_k^T(\mathbf{x}_N) \right]^T, \\
  & \widetilde{\mathbf{h}}_k(\mathbf{x}_n) = \left[ \frac{\eta e^{-\mathrm{j} \frac{2 \pi}{\lambda} r_{kn1}}}{r_{kn1}},\dots,\frac{\eta e^{-\mathrm{j} \frac{2 \pi}{\lambda} r_{knM}}}{r_{knM}} \right]^H.
\end{align}
In contrast to single-user systems, multi-user systems require both the enhancement of the desired signal and the mitigation of inter-user interference. To this end, we next introduce three beamforming protocols \cite{zhao2025pinchingantenna} that leverage the unique characteristics of the PASS.

\subsubsection{Waveguide Switching}
This strategy refers to switching waveguides to serve different users across distinct time slots, thereby eliminating inter-user interference. 
It is applicable to both sub-connected and fully-connected beamforming architectures. For clarity, we focus on the sub-connected architecture, where the received signal at user $k$ is expressed as 
\begin{equation}
  y_k(t) = \mathbf{h}_k^H (\mathbf{X}) \mathbf{G}(\mathbf{X}, \mathbf{P}) \mathbf{w}_k s_k(t) + z_k(t).
\end{equation}
Here, $s_k(t)$ is the signal indented for user $k$, $\mathbf{w}_k \in \mathbb{C}^{N \times 1}$ denotes the transmit beamforming vector for user $k$, and $z_k(t) \sim \mathcal{CN}(0, \sigma_k^2)$ is additive white Gaussian noise. 
While it is technically possible to configure PA positions individually for each user in each time slot, this strategy would require excessively rapid adjustments, especially with a large number of users. Therefore, we assume a single PA position configuration is shared among all users.
Assuming time is equally allocated to each user and defining $\mathbf{W} = [\mathbf{w}_1,\dots,\mathbf{w}_k]$, the corresponding weighted sum-rate (WSR) maximization problem can then be formulated as follows:
\begin{center}
    \begin{tcolorbox}[title = Multi-User Beamforming via Waveguide Switching]
    {\setlength\abovedisplayskip{2pt}
    \setlength\belowdisplayskip{2pt}
    \begin{subequations} \label{BF_problem_4}
      \begin{align}
        \max_{\mathbf{W}, \mathbf{X}, \mathbf{P}} \quad & \sum_{k=1}^K  \frac{\omega_k}{K} \log_2 \left( 1 + \gamma_k^{\mathrm{WS}} \right)  \\
        \label{BF_problem_4_constraint_1}
        \mathrm{s.t.} \quad & \|\mathbf{w}_k\|^2 \le P_t, \forall k, \\
        & \mathbf{x}_n \in \mathcal{X}_n, \; \boldsymbol{\rho}_n \in \mathcal{P}_n, \forall n \\
        & \gamma_k^{\mathrm{WS}} = \frac{\left| \mathbf{h}_k^H (\mathbf{X}) \mathbf{G}(\mathbf{X}, \mathbf{P}) \mathbf{w}_k \right|^2}{\sigma_k^2}, \forall k.
      \end{align}
    \end{subequations}
    }\end{tcolorbox}
\end{center}
In this problem, the weight factor $\omega_k > 0$ allows resource allocation to prioritize different users. Furthermore, since only a single user is served in each time slot, the transmit power for each user is individually constrained as specified in \eqref{BF_problem_4_constraint_1}. Since there is no inter-user interference for the waveguide switching strategy, the transmit beamforming $\mathbf{w}_k$ can be simply designed as the MRT beamformer for each user, as in the single-user systems, leading to the following equivalent optimization problem
\begin{subequations} \label{BF_problem_4_0}
      \begin{align}
        \max_{\mathbf{X}, \mathbf{P}} \quad & \sum_{k=1}^K \frac{\omega_k}{K} \log_2 \left( 1 + \frac{P_t}{\sigma_k^2} \| \mathbf{h}_k^H (\mathbf{X}) \mathbf{G}(\mathbf{X}, \mathbf{P}) \mathbf{w}_k \|^2 \right)  \\
        \mathrm{s.t.} \quad
        & \mathbf{x}_n \in \mathcal{X}_n, \; \boldsymbol{\rho}_n \in \mathcal{P}_n, \forall n.
      \end{align}
\end{subequations}
In this formulation, the pinching beamforming parameters are shared among all users and must therefore be jointly optimized to achieve balanced WSR performance. Compared to the single-user case, the optimization problem for the multi-user system is even more multi-modal, and thus, an element-wise one-dimensional search method can be employed to address this challenge.

\subsubsection{Waveguide Division}
In this strategy, each waveguide is fed with the signal of merely a single communication user, and thus $K = N$ users are scheduled at a time. Therefore, this strategy can only be applied to the sub-connected beamforming architecture. Without loss of generality, we assume that the signal of user $k$ is fed into the $k$-th waveguide. For this configuration, the signal received at user $k$ is given by
\begin{align}
  y_k(t) = & \underbrace{\widetilde{\mathbf{h}}_k^H (\mathbf{x}_k) \mathbf{g}(\mathbf{x}_k, \boldsymbol{\rho}_k) \sqrt{\nu_k} s_k(t)}_{\text{desired signal}} \nonumber \\
  & + \underbrace{\sum_{i=1, i\neq k}^K \widetilde{\mathbf{h}}_k^H (\mathbf{x}_i) \mathbf{g}(\mathbf{x}_i, \boldsymbol{\rho}_i) \sqrt{\nu_i} s_i(t)}_{\text{inter-user interference}} + z_k(t),
\end{align}
where $\nu_k$ is the power allocation factor of user $k$. The corresponding WSR maximization problem is given by 
\begin{center}
    \begin{tcolorbox}[title = Multi-User Beamforming via Waveguide Division]
    {\setlength\abovedisplayskip{2pt}
    \setlength\belowdisplayskip{2pt}
    \begin{subequations} \label{BF_problem_5}
      \begin{align}
        \max_{\boldsymbol{\nu}, \mathbf{X}, \mathbf{P}} \quad & \sum_{k=1}^K \omega_k \log_2 \left( 1 + \gamma_k^{\mathrm{WD}} \right)  \\
        \mathrm{s.t.} \quad & \sum_{k=1}^K \nu_k \le P_t, \\
        & \mathbf{x}_n \in \mathcal{X}_n, \; \boldsymbol{\rho}_n \in \mathcal{P}_n, \forall n \\
        & \gamma_k^{\mathrm{WD}} = \frac{\left| \widetilde{\mathbf{h}}_k^H (\mathbf{x}_k) \mathbf{g}(\mathbf{x}_k, \boldsymbol{\rho}_k) \right|^2 \nu_k }{\sum_{i \neq k} \left| \widetilde{\mathbf{h}}_k^H (\mathbf{x}_i) \mathbf{g}(\mathbf{x}_i, \boldsymbol{\rho}_i) \right|^2 \nu_i +\sigma_k^2 }, \forall k.
      \end{align}
    \end{subequations}
    }\end{tcolorbox}
\end{center}
In the above optimization problem for the waveguide division strategy, the transmit beamforming design reduces to a simplified power allocation problem. However, the presence of inter-user interference transforms the problem into a weighted sum-of-logarithms fractional optimization, which is generally challenging to solve directly. To tackle this, techniques such as the weighted minimum mean-squared error (WMMSE) method~\cite{4712693} or fractional programming~\cite{8314727} are commonly used to transform the problem into more tractable quadratic or linear subproblems. Then, the power allocation factor and the pinching beamforming can be jointly optimized based on the reformulated problem. 

If each waveguide is deployed in a geographically isolated region and exclusively serves a user located within that region, then the channels from this waveguide to other users are naturally blocked, i.e., $\widetilde{\mathbf{h}}_k(\mathbf{x}_i) = \mathbf{0}, \forall k \neq i$. In this case, inter-user interference is inherently eliminated, and the receive signal at user $k$ becomes
\begin{align}
  y_k(t) = & \widetilde{\mathbf{h}}_k^H (\mathbf{x}_k) \mathbf{g}(\mathbf{x}_k, \boldsymbol{\rho}_k) \sqrt{\nu_k} s_k(t) + z_k(t),
\end{align}  
which corresponds to a single-user SISO model. Therefore, the pinching beamforming can be designed following the approach discussed in Section~\ref{Optimization_SISO}, and user priorities can be easily controlled by adjusting the power allocation factors $\nu_k$.

\subsubsection{Waveguide Multiplexing}
In contrast to waveguide division, the waveguide multiplexing strategy multiplexes the signals of all users into each waveguide, which is applicable for both sub-connected and fully-connected architectures. For the purpose of exposition, we focus on the sub-connected architecture, where the received signal at user $k$ is expressed as 
\begin{align} \label{multi_user_signal_general}
  y_k(t) = & \underbrace{\mathbf{h}_k^H (\mathbf{X}) \mathbf{G}(\mathbf{X}, \mathbf{P}) \mathbf{w}_k s_k(t)}_{\text{desired signal}} \nonumber \\
  & + \underbrace{\sum_{i=1, i\neq k}^K \mathbf{h}_k^H (\mathbf{X}) \mathbf{G}(\mathbf{X}, \mathbf{P}) \mathbf{w}_i s_i(t)}_{\text{inter-user interference}} + z_k(t),
\end{align} 
The weighted summer rate (WSR) maximization problem for the waveguide multiplexing strategy is given by
\begin{center}
    \begin{tcolorbox}[title = Multi-User Beamforming via Waveguide Multiplexing]
    {\setlength\abovedisplayskip{2pt}
    \setlength\belowdisplayskip{2pt}
    \begin{subequations} \label{BF_problem_6}
      \begin{align}
        \max_{\mathbf{W}, \mathbf{X}, \mathbf{P}} \quad & \sum_{k=1}^K \omega_k \log_2 \left( 1 + \gamma_k^{\mathrm{WM}} \right)  \\
        \mathrm{s.t.} \quad & \sum_{k=1}^K \| \mathbf{w}_k \|^2 \le P_t, \\
        & \mathbf{x}_n \in \mathcal{X}_n, \; \boldsymbol{\rho}_n \in \mathcal{P}_n, \forall n, \\
        & \gamma_k^{\mathrm{WM}} = \frac{\left| \mathbf{h}_k^H (\mathbf{X}) \mathbf{G}(\mathbf{X}, \mathbf{P}) \mathbf{w}_k \right|^2 }{\sum_{i \neq k} \left| \mathbf{h}_k^H (\mathbf{X}) \mathbf{G}(\mathbf{X}, \mathbf{P}) \mathbf{w}_i \right|^2 +\sigma_k^2 }, \forall k.
      \end{align}
    \end{subequations}
    }\end{tcolorbox}
\end{center}
Compared to the waveguide switching and division strategies, the waveguide multiplexing problem is more general but also significantly more challenging, as both the transmit and pinching beamforming must be jointly optimized across all users. 
Several solution methods have been proposed to address this problem. For instance, the authors of \cite{wang2025modeling} developed a penalty-based approach that directly tackles the joint optimization by incorporating the highly coupled pinching beamforming terms into the objective function as penalties, which are then decoupled into a sequence of tractable subproblems. Additionally, the authors of \cite{sun2025multiuser} proposed an element-wise optimization framework based on heuristic beamforming strategies such as MRT, ZF, and MMSE, where the pinching beamforming variables are optimized via an element-wise one-dimensional search.

\begin{table*}[!t]
\centering
\caption{Comparison of Beamforming Protocols for PASS}
\resizebox{\linewidth}{!}{
\begin{tabular}{l|c|c|c|c}
\hline
\textbf{Protocol} 
% & \textbf{Control} 
& \textbf{Transmit Beamforming} 
& \textbf{Pinching Beamforming} 
& \textbf{Spatial Multiplexing} 
& \textbf{Design Complexity} \\
\hline

Waveguide Switching
% & Centralized
& \ding{52} 
& \ding{52}
& \ding{56}
& Low \\
\hline

Waveguide Division
% & Centralized 
& \ding{56} 
& \ding{52} 
& \ding{52} 
& Medium \\
\hline

Waveguide Multiplexing
% & Centralized
& \ding{52} 
& \ding{52}  
& \ding{52}  
& High \\
\hline
\end{tabular}
}
\label{tab:beamforming_protocol}
\end{table*}
In Table \ref{tab:beamforming_protocol}, we compare the different beamforming protocols.

\begin{figure}[!t]
  \centering
  \includegraphics[width=0.35\textwidth]{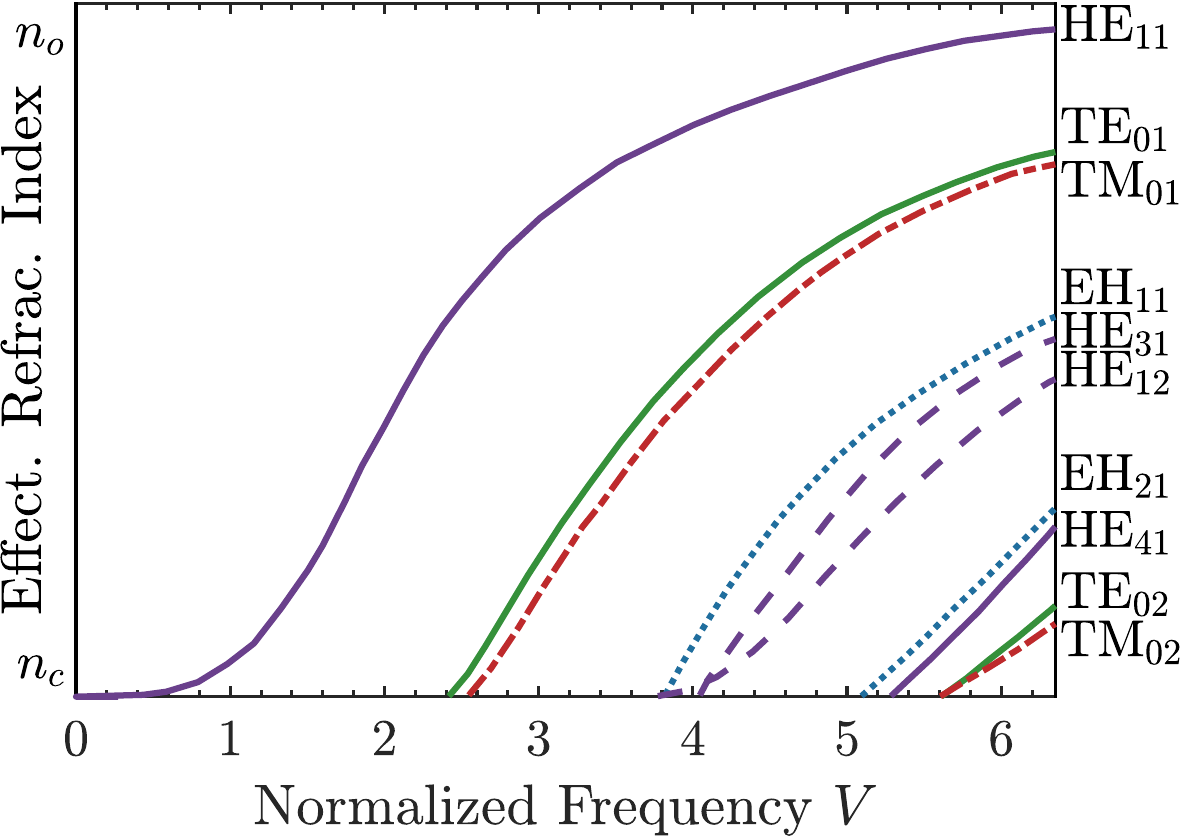}
  \caption{Effective refractive index $n_{\rm eff}$ versus the normalized frequency for selected low-order guided modes in a cylindrical dielectric step-index waveguide~\cite{cheo1989fiber}.}
  \label{Figure_effective_index}
\end{figure}

\subsection{Wideband OFDM Beamforming} \label{sect:wideband}
We now discuss the application of PASS in wideband systems, with a focus on conventional orthogonal frequency division multiplexing (OFDM), which is widely employed to address frequency-selective fading in modern wideband communication scenarios. Owing to the distinct in-waveguide signal propagation enabled by PASS, such systems exhibit several unique characteristics. In the following, we analyze these characteristics from three key perspectives: available bandwidth, waveguide dispersion, and frequency-dependent pinching beamforming.

\subsubsection{Available Bandwidth}
In conventional communication systems, the available bandwidth is theoretically unlimited, as free-space propagation imposes no inherent constraints on signal transmission. In contrast, PASS-based systems rely on in-waveguide propagation, where the usable bandwidth is inherently restricted by factors such as the waveguide’s material, geometry, and structural design.

The upper bound of the available bandwidth in PASS is governed by the modal characteristics of the waveguide, specifically, the transition between single-mode and multi-mode operations. Unlike free space, which supports fully transverse electromagnetic (TEM) waves, a waveguide can only support discrete electromagnetic modes, each existing at specific frequency ranges \cite{cheo1989fiber}. To elaborate, Fig. \ref{Figure_effective_index} illustrates the effective refractive index $n_{\mathrm{eff}}$ of different modes against the normalized frequency $V$ for a cylindrical dielectric waveguide with core radius $r_o$, where the $\text{HE}_{\text{11}}$ mode is the dominant mode. The normalized frequency is defined as $V = 2 \pi f r_o \sqrt{n_o^2 - n_c^2}/c$, where $n_c$ and $n_o$ are the refractive indices of the cladding and the core, respectively. When the cladding is air, we have $n_c = 1$. A mode is considered \emph{cut off} when its effective refractive index equals that of the cladding, i.e., $n_{\mathrm{eff}} = n_c$, and cannot propagate in the waveguide. As shown in Fig. \ref{Figure_effective_index}, each mode has a distinct cut-off frequency below which it cannot be supported. Although a waveguide may theoretically support multiple modes, practical designs often favor single-mode operation to simplify signal transmission and detection, and to ensure system stability. This requires all higher-order modes to be cut off, retaining only the dominant mode. For the waveguide in Fig. \ref{Figure_effective_index}, single-mode transmission requires the normalized frequency to satisfy 
\begin{equation}
  V \le 2.405 \; \Rightarrow \; f \le \frac{0.3828 c}{r_o\sqrt{n_o^2 - n_c^2}},
\end{equation}
which defines the upper bound of the available frequency band. For example, for $r_o = 2$ mm, $n_o = 1.4$, and $n_c = 1$, the maximum frequency for single-mode operation is approximately 58 GHz.  

In addition to the upper bound, the available bandwidth in PASS is also constrained from below by the cut-off frequency of the dominant mode. For the cylindrical dielectric waveguide discussed earlier, the dominant mode theoretically has no cut-off frequency, meaning it can propagate at arbitrarily low frequencies. However, this property does not hold for all types of waveguides. For instance, in a metallic rectangular waveguide with width $a$, the dominant mode $\text{TE}_{\text{10}}$ is cut off at $f = c/(2a)$ \cite{pozar2021microwave}. As a result, the signal frequency must exceed this threshold to ensure propagation. Note that even in waveguides where the dominant mode has no theoretical cut-off frequency, a practical lower frequency bound may still exist due to factors such as material losses, antenna coupling efficiency, and feed structure limitations. In general, PASS imposes a bounded frequency range to ensure the unique existence and stable transmission of the dominant mode. This constraint directly impacts the available bandwidth for wideband communication applications, and therefore, the carrier frequencies in OFDM systems must be carefully selected.

\subsubsection{Waveguide Dispersion}

Waveguide dispersion is another critical factor that must be considered in PASS-aided wideband systems. Dispersion refers to the frequency dependence of signal propagation velocity, which arises from the frequency-dependent effective refractive index $n_{\mathrm{eff}}(f)$ of the guided modes, as illustrated in Fig. \ref{Figure_effective_index}. Consequently, the signal velocity, given by $v_{\mathrm{G}}(f) = c/n_{\mathrm{eff}}(f)$, also varies with frequency, leading to inter-symbol interference (ISI) in wideband transmissions \cite{xiao2025ofdm, pozar2021microwave}.

In conventional wideband systems operating in free space, the signal velocity is constant, and ISI typically results from two main effects. The first is the frequency-wideband effect caused by multipath propagation, where reflections, diffusions, and diffractions generate multiple paths with different lengths and corresponding delays. The second is the spatial-wideband effect, which arises when antenna arrays are used. Even in LoS channels, signals from different antennas may experience varying propagation distances due to the large array aperture, introducing delay differences across antennas. In both cases, delay variations originate from differences in propagation distance. This is still relevant for the free-space segment of PASS. However, for in-waveguide propagation, the dominant source of delay variation is waveguide dispersion. Although the physical propagation distance inside the waveguide is the same for all frequencies, the frequency-dependent signal velocity introduces differential delays.

To mitigate ISI in OFDM systems, a cyclic prefix (CP) is typically added to each OFDM symbol, with a length no shorter than the maximum expected delay spread. In PASS-aided wideband systems, the CP length must be carefully chosen to account for all three sources of delay variation: the frequency-wideband effect from multipath propagation, the spatial-wideband effect due to the antenna array geometry, and the waveguide dispersion arising from in-waveguide propagation.

\subsubsection{Frequency-Dependent Pinching Beamforming}
In addition to the frequency-dependent effective refractive index discussed earlier, several other frequency-sensitive factors must be considered in PASS-aided wideband systems, necessitating a frequency-dependent beamforming design. To elaborate, consider a single-user OFDM system with $Q$ subcarriers, where the $q$-th subcarrier operates at frequency $f_q$. For clarity, we consider a single waveguide equipped with $M$ PAs. The LoS signal model for the $m$-th PA at frequency $f$ is given by
\begin{equation}
  y_m (f) = \frac{\eta(f) \sqrt{P_m(f)}}{r_m} e^{-\mathrm{j} \frac{2\pi f}{c} (r_m + n_{\mathrm{eff}}(f) x_m)} s(f),
\end{equation}    
where $s(f)$ denotes the signal modulated onto the subcarrier at frequency $f$. This model reveals several sources of frequency dependence. First, the free-space channel gain $\eta(f)$ varies with frequency due to path loss and radiation efficiency. Second, the radiated power $P_m(f)$ for the $m$-th PA depends on frequency through the frequency-varying coupling coefficients. Third, the effective refractive index $n_{\mathrm{eff}}(f)$ is also frequency-dependent, as previously discussed. 

Accordingly, the signal received on the $q$-th subcarrier at the user can be expressed as
\begin{align}
  y_q = & \sum_{m=1}^M y_m(f_q) + z_q \\
  = & \mathbf{h}^H(\mathbf{x}, f_q) \mathbf{g}(\mathbf{x}, f_q) s(f_q) + z_q, 
\end{align} 
where $z_q \sim \mathcal{CN}(0, \sigma^2)$ represents additive white Gaussian noise. 
Vectors $\mathbf{h}(\mathbf{x}, f)$ and $\mathbf{g}(\mathbf{x}, f)$ are defined as
\begin{align}
  & \mathbf{h}(\mathbf{x}, f) = \left[ \frac{\eta(f) e^{-\mathrm{j} \frac{2 \pi f}{c} r_1}}{r_1},\dots,\frac{\eta(f) e^{-\mathrm{j} \frac{2 \pi f}{c} r_M}}{r_M} \right]^H, \\
  & \mathbf{g}(\mathbf{x}, f) = \Big[ \sqrt{P_1(f)} e^{-\mathrm{j} \frac{2\pi f}{c} n_{\mathrm{eff}}(f) x_1}, \nonumber 
  \\&\hspace{3cm}\dots,\sqrt{P_M(f)} e^{-\mathrm{j} \frac{2f}{c} n_{\mathrm{eff}}(f) x_M} \Big]^T,
\end{align} 
The overall achievable rate of the OFDM system is given by
\begin{equation}
  R = \frac{1}{L_{\mathrm{CP}} + Q} \sum_{q=1}^Q \log_2 \left( 1 + \frac{\left| \mathbf{h}^H(\mathbf{x}, f_q) \mathbf{g}(\mathbf{x}, f_q) \right|^2}{\sigma^2} \right),
\end{equation}
where $L_{\mathrm{CP}}$ denotes the length of the CP as discussed in the previous section. The corresponding beamforming optimization problem for maximization of the overall achievable rate can be formulated as follows:
\begin{center}
    \begin{tcolorbox}[title = Wideband Pinching Beamforming Optimization]
    {\setlength\abovedisplayskip{2pt}
    \setlength\belowdisplayskip{2pt}
    \begin{subequations} \label{BF_problem_wideband}
      \begin{align}
        \max_{\mathbf{x}} \quad& \sum_{q=1}^Q \log_2 \left( 1 + \frac{\left| \mathbf{h}^H(\mathbf{x}, f_q) \mathbf{g}(\mathbf{x}, f_q) \right|^2}{\sigma^2} \right)\\
        \mathrm{s.t.} \quad & \mathbf{x} \in \mathcal{X}.
      \end{align}
    \end{subequations}
    }\end{tcolorbox}
\end{center}
In this formulation, we assume a fixed power radiation model and do not treat $P_m(f)$ as an optimization variable. This problem highlights that, in PASS-aided wideband systems, the PA positions $\mathbf{x}$ must be optimized to maximize beamforming gains across all subcarriers, rather than targeting a single frequency as in narrowband systems. Failure to account for this frequency dependence can lead to substantial performance degradation \cite{xiao2025ofdm}.

\subsection{Discussion and Outlook}
Above, we have discussed the optimization of PASS across a wide range of system configurations and explored potential solutions. What follows is a set of open research problems intended to inspire further investigation.

\begin{itemize}
    \item \emph{Wideband Pinching Beamforming:} To explore abundant bandwidth resources on high-frequency bands, wideband transmission is a key focus in PASS research. As discussed in Section \ref{sect:wideband}, the propagation characteristics of wideband signals are altered when transmitted through a waveguide. To fully harness the potential of PASS for wideband applications, it is crucial to develop practical and analytically tractable signal models. 
    These models will enable efficient protocol design and algorithm development for the practical deployment of wideband PASS systems.
    
     \item \emph{Latency-Aware Pinching Beamforming} 
    Early theoretical studies on PASS primarily explored its potential for flexible operation but gave limited attention to the delays associated with the flexibility.
    In fact, the flexibility of PASS relies on discrete port activation or mechanical movement of PAs. 
    These operations introduce latency that can degrade performance in fast‐varying or mobile communication networks. 
    Future work should therefore model activation delays and investigate control strategies that mitigate their impact on communication latency and reliability.
    
    \item \emph{Robust Pinching Beamforming:}
    Optimizing PASS depends heavily on accurate CSI from CE, as detailed in Section \ref{sect:channel_estimation}. This presents two key challenges. First, the pinching beamforming optimization is highly multi-modal, meaning small CE errors can cause large performance losses. This necessitates robust beamforming designs that account for imperfect CSI \cite{zeng2025robust}. Second, PASS involves a large number of candidate antenna positions, leading to prohibitive CE overhead. To address this, beamforming approaches using statistical CSI are worth investigating, as this data changes slowly and thus reduces the need for frequent estimation.
\end{itemize}

\section{PASS CSI Acquisition} \label{sect:channel_estimation}
With the enhanced spatial flexibility, the superiority of PASS can be realized by exploiting the optimization methods proposed in the previous section.
However, the most important prerequisite is the availability of CSI, which serves as the foundation for the implementation of the optimization algorithms.
For conventional fixed-position antenna systems, the acquisition of CSI has been extensively investigated and solved by a large number of effective approaches.
However, these methods cannot be directly applied to PASS scenarios, due to the position reconfiguration ability of PASS.
In particular, since the channel is parameterized by the locations of the activated PAs, CSI is closely related to the positions of the PAs.
Hence, once the positions of the PAs change, the previously obtained CSI becomes outdated, which can invalidate subsequent optimization.
Therefore, the objective of CSI acquisition in PASS is to effectively and efficiently obtain the CSI for each candidate PA position on the waveguides.
In this section, we will elaborate on two categories of CSI acquisition strategies for PASS, featuring the pilot-based CE and beam training.

\subsection{Pilot-Based Channel Estimation}
As the most straightforward method for CSI acquisition, pilot-based CE aims to recover CSI from dedicated known pilots using signal processing techniques.
To achieve this, the positions on the waveguides are discretized into grids, where each grid point represents a candidate position for a PA.
It is noted that the space between two adjacent grid points should be larger than $\lambda / 2$ to suppress the mutual coupling effect.
Here, we assume that time-division duplex (TDD) is valid and the channel is reciprocal.
Therefore, once the uplink channel is estimated through dedicated pilot transmission, the downlink channel can be correspondingly obtained through channel reciprocity.
Letting the uplink pilots be $\mathbf{s} = [s_1, s_2, ..., s_T]^{T} \in \mathbb{C}^{T\times 1}$ with $T$ being the length of the pilots, the received uplink pilots of the PASS can be expressed as follows:
\begin{align}
    \mathbf{Y} = \mathbf{G}^{H}\mathbf{h}\mathbf{s}^T + \mathbf{Z} \in \mathbb{C}^{N \times T},
\end{align}
where $\mathbf{Z} \in \mathbb{C}^{N\times T}$ denotes the complex-valued Gaussian noise matrix whose entries are identically and independently sampled from distribution $\mathcal{CN}(0, \sigma^2)$ with noise power $\sigma^2$.
It is noteworthy that, as in-waveguide channel matrix $\mathbf{G}$ is solely determined by the position of the PAs, this channel does not necessitate estimation.
Therefore, defining $\tilde{\mathbf{G}}^{H} \triangleq \mathbf{s} \otimes \mathbf{G}^{H}
$ as equivalent pilot matrix, the signal model for CE can be compactly expressed as 
\begin{center}
    \begin{tcolorbox}[title = Channel Estimation Signal Model]
    {\setlength\abovedisplayskip{2pt}
    \setlength\belowdisplayskip{2pt}
    \begin{equation} \label{far_field_vector}
        \mathbf{y}=\mathrm{Vect}\left\{ \mathbf{Y} \right\} =\left( \mathbf{s}\otimes \mathbf{G}^{H} \right) \mathbf{h}+\tilde{\mathbf{z}}=\tilde{\mathbf{G}}^{H}\mathbf{h}+\tilde{\mathbf{z}},
    \end{equation}
    }\end{tcolorbox}
\end{center}
where $\tilde{\mathbf{z}}=\mathrm{Vect}\left\{ \mathbf{Z} \right\} 
$ denotes the vectorized noise matrix.
The main challenge of estimation $\mathbf{h}$ lies in the rank deficiency of the equivalent pilot matrix, i.e., $\tilde{\mathbf{G}}^{H}$.
Specifically, according to the definition of the in-waveguide channel matrix, matrix $\mathbf{G}$ is a column full-rank matrix, i.e., $\mathrm{Rank}\left\{ \mathbf{G} \right\} =N$ \cite{petersen2012matrixcookbook}, thus resulting in $\mathrm{Rank} \{ \tilde{\mathbf{G}}^{H}\} =\mathrm{Rank}\{ \mathbf{s}^{}\otimes \mathbf{G}^{H} \} =N$.
Given $\mathbf{h}\in\mathbb{C}^{MN \times 1}$, unique recovery cannot be achieved due to the rank-deficient structure.
In this scenario, the conventional least-squares (LS) methods cannot yield a unique solution.
In addition, the length of the pilot is irrelevant to the rank of $\tilde{\mathbf{G}}^{H}$, due to the combination effects exerted by in-waveguide propagation.

To alleviate this issue, several methods can be employed, including sequential activation (SA), compressed sensing (CS), and parameter-sensing (PS) methods.
\begin{itemize}
    \item \textbf{Sequential Activation}: The SA method is the most straightforward approach, which sequentially activates individual PAs on each waveguide.
    In this case, the in-waveguide matrix is unfolded over time, making $\tilde{\mathbf{G}}$ a diagonal matrix, which facilitates CE.
    More specifically, for time instant $m \in\{1,2,...,M\} \triangleq \mathcal{M}$, the waveguides will activate their respective $m$-th candidate port, implying the resulting in-waveguide channel matrix $\mathbf{G}_{m} = \mathrm{blkdiag}\{[\mathbf{g}_1]_m, ..., [\mathbf{g}_N]_m\} \in \mathbb{C}^{N \times N}$ is of full rank.
    Under this condition, the optimization problem can be decomposed into $M$ unconstrained subproblems:
    \begin{tcolorbox}[title={Channel Estimation via Sequential Activation}]
        \begin{subequations} \label{obj:Ls}
        \begin{align}
            \min _{\mathbf{h}_m}\left\| \mathbf{y}_m-{\mathbf{G}}_{m}^{H}\mathbf{h}_m \right\| _{2}^{2}, \quad \mathrm{for}~m \in \mathcal{M},
        \end{align}
    \end{subequations}
    \end{tcolorbox}
    where $\mathbf{y}_m$ is the received pilot when the $m$-th candidate positions of all waveguides are activated, and $\mathbf{h}_m$ is the free-space channel corresponding to these activated positions.
    Finally, after the waveguide is traversed, the full-dimensional free-space channel can be recovered as
    \begin{align}
        \mathbf{h}=\sum\nolimits_{m=1}^M{\mathbf{e}_m\otimes}\mathbf{h}_m,
    \end{align}
    where $\mathbf{e}_m$ denotes the $m$-th column of identity matrix $\mathbf{I}_M$.
    As ${\mathbf{G}}_{m}^{H}$ for $\forall m\in\mathcal{M}$ are now full-rank, the closed-form solution to \eqref{obj:Ls} can be attained using the LS algorithm or the MMSE algorithm by further exploiting the prior knowledge of the channel.
    However, the CE overhead of this solution is proportional to the number of candidate ports on each waveguide, i.e., $M$.
    Thereby, when the number of candidate positions is large, the practicality of this solution will be reduced.
    In other words, there will be a tradeoff between the length of CE overhead and the resolution of CE.

    \item \textbf{Compressed Sensing}: To reduce this prohibitive overhead, the CS method can be applied.
    Due to the domination of the LoS PASS, $\mathbf{h}$ can be sparse in the wavenumber domain \cite{jiang2024sense}.
    In particular, the spherical waves in the near-field region can be approximated by a superposition of a finite number of planar waves \cite{sanguinetti2023wavenumber}. 
    Therefore, the channel needed for CE can be expressed as $\mathbf{h}=\boldsymbol{\Psi} \mathbf{x}$, with dictionary matrix $\boldsymbol{\Psi} \in \mathbb{C}^{MN \times L}$ with $L \ge MN$ and sparse vector $\mathbf{x}$ conforming to the condition $\|\mathbf{x}\|_0 \ll MN $.
    In this context, the CS-based CE problem can be formulated as
    \begin{tcolorbox}[title={Channel Estimation via Compressed Sensing}]
            \begin{subequations}
            \begin{align}
            \min _{\mathbf{x}} \quad& \left\| \mathbf{x} \right\| _0 \\
            \mathrm{s.t}. \quad &\left\| \mathbf{y}-\tilde{\mathbf{G}}_{}^{H}\mathbf{\Psi x} \right\| _{2}^{2}\le \epsilon,
        \end{align}
        \end{subequations}
     \end{tcolorbox}
    where $\epsilon >0$ denotes the error tolerance.
    The objective function of this problem aims at making $\mathbf{x}$ as sparse as possible, while the constraint enforces the estimation error to fall within the tolerable region.
    Under the restricted isometry property (RIP) condition, $\mathbf{h}$ can be recovered using CS methods, such as orthogonal matching pursuit (OMP), compressive sampling matching pursuit (CoSaMP), and maximum likelihood estimation techniques.
    The pitfall of this method lies in the design of the dictionary matrix, which should effectively exploit the channel sparsity of PASS channels.

    \item \textbf{Parameter Sensing}: The aforementioned methods can only obtain a CE with finite resolution, due to the discrete PA locations.
    In some cases, continuous CE over waveguides is needed for effective optimization.
    To achieve this, parameter sensing is a potential solution, which leverages the fact that the free-space channel vector is determined by a finite number of physical parameters related to the position of the users and scatterers.
    In particular, as demonstrated by the general narrowband PASS channel model in \eqref{eq:NLoS_PASS}, the PASS channel is determined by a parameter set $\mathcal{T}$, which contains amplitudes and phase shifts corresponding to the LoS and NLoS links.
    In this case, once $\mathcal{T}$ has been obtained via effective sensing algorithms, e.g., \cite{zhou2025channel}, the PASS channel can be reconstructed at any positions on the waveguide, indicating that a function of the CSI can be obtained.
    Based on this idea, the optimization problem for parameter sensing CE can be formulated as
    \begin{tcolorbox}[title={Channel Estimation via Parameter Sensing}]
        \begin{subequations}
            \begin{align}
            \mathcal{T} ^{\star}=arg\min_{\mathcal{T}} \quad \left\| \mathbf{y}-\tilde{\mathbf{G}}^H\mathbf{h}\left( \mathcal{T} \right) \right\| _{2}^{2}
                \end{align}
            \end{subequations}
            \end{tcolorbox}
    In the above problem, the main task is to extract the physical parameters from the received signal $\mathbf{y}$.
    This problem can be solved by maximum likelihood estimation (MLE) or subspace super-resolution algorithms, such as multiple signal classification (MUSIC).
    Moreover, the mobility of the PAs can help in parameter estimation \cite{zhou2025channel}, and the sparsity of the PAs can be potentially useful to enhance the sensing performance \cite{shen2015low}. 
    Compared to the former two categories, this parameter sensing method can achieve a continuous recovery of the CSI alongside the waveguide.
    However, the challenge of this method lies in the fact that the estimation accuracy will be degraded by random scattering effects.
    Moreover, the introduction of additional signal processing modules for sensing/detection purposes will also increase complexity.
\end{itemize}

\subsection{Beam Training}
\begin{figure}[h!]
    \centering
    \includegraphics[width=0.9\linewidth]{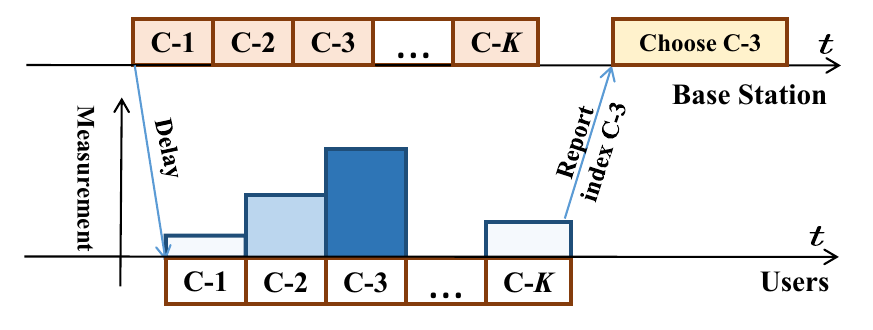}
    \caption{Illustration of beam training protocol, where $\text{C-k}$ denotes the $k$-th codeword.}
    \label{fig:illustration_of_beam_training_protocol}
\end{figure}
The primary advantage of PASS lies in creating LoS or sparse scatter conditions for transmission.
Therefore, beam training can be a promising candidate in such scenarios \cite{va2019online}.
The beam training framework consists of the following three steps:
1) \emph{Beam Sweeping}: The beam training is carried out by sweeping through predefined codebooks exhaustively or hierarchically.
The codewords in the codebooks correspond to beams pointing to different locations.
2) \emph{Measurement Report}: The users will locally measure the received beam gains and report the index of the codeword that produces the highest beam gain back to the transmitter.
This step can be conducted through established low-frequency control links between transceivers.
3) \emph{Beam Determination}: After receiving the reports from the users, the transmitter will assign the beams for downlink transmission. 
This process is illustrated by Fig. \ref{fig:illustration_of_beam_training_protocol}, where codeword ``C-3'' produces the highest beam gain measurement and thereby is selected for the upcoming data transmission stage.
It is essential to note that beam training cannot directly obtain the CSI.
Instead, it only produces the best transmission mode, i.e., the determined codeword. 

Due to its simplicity and computational efficiency, beam training has been widely utilized in mmWave MIMO scenarios and has been integrated into the 5G new radio (NR) beam management framework \cite{heng2022learning}.
The beam training problem for PASS can be formulated as

\begin{tcolorbox}[title={Beam Training for PASS}]
    \begin{subequations}
      \begin{align}
        \underset{\mathcal{F},\,\mathcal{W},\,f,\,g}{\mathrm{max}}&\quad
           \left|\mathbf{h}^{H} \mathbf{G}(\mathbf{X}_i)\,\mathbf{w}_j\right|^2 
        \label{eq:beam_training_opt-objective}\\
        \mathrm{s.t.} \quad& i = f(\mathbf{m}), \\
        &j = g(\mathbf{m}), \\
        &\|\mathbf{w}_p\|^2 \le P_{\max}, \quad \forall\,\mathbf{w}_p\in\mathcal{W}, \\
        &\mathbf{x}_n \in \mathcal{X}_n, \; \boldsymbol{\rho}_n \in \mathcal{P}_n, \forall n. 
      \end{align}
    \end{subequations}
\end{tcolorbox}
The optimization objective in \eqref{eq:beam_training_opt-objective} is to maximize beam gains for any given free-space channel vector $\mathbf{h}$.
To achieve this, two types of codewords in the antenna position codebook $\mathcal{F}=\{\mathbf{X}_1, ..., \mathbf{X}_{|\mathcal{F}|}\}$ and the beam codebook $\mathcal{W}=\{\mathbf{w}_1, ..., \mathbf{w}_{|\mathcal{W}|}\}$ need to be identified, i.e., $\mathbf{X}_i \in \mathcal{F}$ and $\mathbf{w}_j \in \mathcal{W} $.
As mentioned above, the transmission mode is chosen according to the users' local measurements of swept codewords.
Thus, the indices $i$ and $j$ are output by two functions $f(\cdot)$ and $g(\cdot)$, whose inputs are the beam measurement results $\mathbf{m}$ reported by the users.
Generally speaking, $\mathbf{m} \in \mathbb{R}^{| \mathcal{F}||\mathcal{W}| \times 1}$ is obtained after the base station sweeps through codebooks $\mathcal{F}$ and $\mathcal{W}$, which will subsequently be reported from the users to the base station via established, reliable feedback channels.
To this end, we need to optimize functions $f(\cdot)$ and $g(\cdot)$ to ensure that the correct beam index can be extracted efficiently.
Besides, codebooks $\mathcal{F} $ and $\mathcal{W}$ also need to be designed to satisfy:
1) \textbf{Compactness:} This requirement indicates that the codebook size should be reduced as much as possible to reduce the beam training overheads. 
2) \textbf{Scalability:} Different from fixed antenna configurations, PASS has enhanced flexibility, as the number of PAs and the position of PAs can be altered, indicating that the size of the codebooks constantly changes.

In what follows, several possible solutions to the formulated beam training problem are listed:
\begin{itemize}
    \item \textbf{Three Stage Hierarchical Beam Training \cite{lv2025beam}}: This beam training follows the idea of the classic beam training procedure, and is composed of three stages.
    In the first stage, the positions or divided grids in the serving areas of PASS are coarsely searched along the $x$-direction in a hierarchical fashion.
    Subsequently, given the best $x$-coordinate, the $y$-direction is searched hierarchically in a similar manner.
    Finally, the best grid obtained through the preceding steps will be divided into finer grids, where an exhaustive search is applied.
    This method is straightforward for deployment.
    However, high beam training overheads are introduced when the served area is large, which is typically true when multiple waveguides are deployed.
    
    \item \textbf{Site-Specific Codebook Design \cite{heng2022learning}}: To truncate the size of the codebook, the environment information of the PASS's service area needs to be exploited.
    In particular, the codewords in codebooks $\mathcal{F}$ and $\mathcal{W}$ can be considered as trainable parameters.
    The beam selection functions $f(\cdot)$ and $g(\cdot)$ can be modelled as neural networks.
    Then, given environment information as a training dataset, these NNs can be trained with the objective function in \eqref{eq:beam_training_opt-objective}.
    This method employs a stationary channel map for a given served area to largely reduce the size of the codebook.
    However, this stationary assumption might not hold when a dynamic communication network is considered.
    Furthermore, when the layout of the served area is altered for some reason, the codebook will lose its effectiveness and thus needs to be redesigned.
    
    \item \textbf{Sensing Aided Beam Training \cite{prelcic2024the}}: This approach mainly exploits the prior knowledge of the positional information of the users in the service area.
    With such prior knowledge, only a subset of codebooks related to positions needs to be traversed, thus greatly reducing the training overhead.
    This approach is more suitable for mobile networks, as it can eliminate the need for repeated beam training.
    However, this method requires additional hardware for sensing functionality, thus complicating the system design and signal processing procedure.
    Additionally, when scattering paths are included in the system model, sensing becomes more challenging and may even become inapplicable.
    \end{itemize}

\subsection{Discussion and Outlook}
While the foregoing discussion highlights several potential pilot-based CE and beam training methods for PASS, additional avenues remain to be explored. 
In the following, we outline some promising research directions for CSI acquisition in PASS.

\begin{itemize}
    \item \emph{Continuous CSI Acquisition}: Most existing work on PASS optimization assumes gridless or continuous CSI availability along the waveguide. 
    In practice, however, current CE methods for PASS borrow directly from conventional MIMO methods and can only estimate CSI at a discrete set of antenna positions. 
    This introduces an unavoidable performance–complexity tradeoff.
    In particular, a long waveguide may contain hundreds or even thousands of potential PA positions, so sampling them all leads to prohibitive CE overheads.  
    To overcome this challenge, dedicated continuous CE methods that directly estimate a continuous CSI function of the PA positions, rather than relying on dense discrete grids, are urgently needed.

    \item \emph{CSI Acquisition in Multi‑Path Scenarios}: Although the LoS component often dominates in PASS links, NLoS paths can be significant in environments such as indoor networks. 
    In these setups, greedily aligning beams to the strongest LoS direction may not yield the best overall channel quality. 
    Consequently, CE must capture both LoS and NLoS components. 
    Parameter‑sensing approaches, which can reduce overhead and provide continuous CSI, may be problematic when multiple NLoS paths introduce separate directions of arrivals that are hard to resolve. 
    Likewise, beam training faces a difficult codebook‑design problem, since the codebook design must balance the coverage of the direct link against that of the scattered NLoS links, without incurring excessive sweep lengths or sacrificing scalability. 
    Dedicated CE and beam‑training designs for NLoS scenarios are, therefore, essential for unlocking the full potential of PASS.
\end{itemize}

\section{Machine Learning For PASS}\label{sect:ml_for_pass}
PASS enables new reconfigurability at the cost of increased system complexity, operation overheads, and design intricacies during both system optimization and CSI acquisition. 
In this section, we will discuss the motivations for exploiting ML in PASS, and then explore promising ML solutions for cost-efficient PASS optimization and CSI acquisition. 

\subsection{Motivation for Exploiting ML in PASS}
There are several critical challenges to unlock the full potential of PASS. 
1) \textit{High complexity:} PASS relies on high-dimensional pinching beamforming optimization and rank-deficient CSI estimation for realistic multi-waveguide multi-PA designs. Although low-complexity optimization algorithms or compressed sensing algorithms have been explored, they typically converge with a polynomial-time complexity or require relatively high pilot overheads. As practical sub-6 GHz and mmWave systems usually exhibit millisecond-level channel coherence time, it is difficult to meet the responsiveness requirements of real-world deployments. 2) \textit{Suboptimality:} Owing to the deep coupling between the pinching beamforming and the other parameters of the wireless system (e.g., transmit beamforming and power allocation), the solution spaces for PASS optimization may comprise numerous suboptimal solutions. Hence, current gradient-based optimization algorithms are prone to being stuck in local optima with undesirable qualities \cite{xu2025joint,gan2025joint}. To avoid this drawback, the branch-and-bound algorithm (BnB) \cite{xu2025pinchingopt} has been proposed to find the globally optimal solution, and particle swarm optimization (PSO) algorithms \cite{gan2025joint,jiang2025pinching,zhu2025pinching} have been proposed to search for high-quality local optima. However, both BnB and PSO algorithms cannot guarantee even polynomial-time convergence as they rely on a large number of bound/fitness evaluations. 
3) \textit{Environment dynamics:} The PA configuration must follow real-time changes in both user mobility and radio environment. However, estimating the dynamically evolving channels for all potential locations of PAs incurs high overheads and spatial nonstationary issues.  
4) \textit{Simulation cost:} The traditional antenna design workflow relies heavily on full-wave EM simulation software (e.g., ANSYS HFSS) and parameter search, which are inherently time-consuming. Conventional optimization approaches rely on specific system parameters (e.g., coupling coefficient) estimated by full-wave EM simulation, making the overall decision-making process computationally expensive. 

To address the aforementioned challenges, ML techniques can be leveraged for PASS in the following ways:
\begin{itemize}
    \item \textbf{ML-enabled accelerated PASS optimization and CSI acquisition}: ML can accelerate both nonconvex system optimization and CSI acquisition for PASS, while enhancing the achieved performance. 
    
    For \textit{PASS} optimization, the principles of conventional optimization theory can be integrated into neural networks (NNs) consisting of a fixed number of layers, thereby reducing computational burdens compared to conventional iterative optimizers (e.g., alternating optimization and majorization–minorization methods). 
    Moreover, unlike conventional optimizers that only explore solutions around a given initialization, ML can learn the global mapping between system objectives and primal/dual variables from large-scale training samples. Built on the optimality conditions from Karush–Kuhn–Tucker (KKT) theory, ML can avoid poor local optima in a hybrid model-driven and data-driven manner, while performing fast inference with millisecond-level response speed \cite{xu2025joint,guo2025gpass,karagiannidis2025deep}.
    For \textit{CSI acquisition}, NNs can be trained to recover highly accurate spherical wave channels \cite{xiao2025channel}. Moreover, ML can predicatively configure the locations of the PAs during channel measurement to reduce CSI acquisition overheads and to refine prediction results. 
    \item \textbf{ML-enabled autonomous PASS configuration}: Empowered by ML models deployed at the edge or the cloud, PASS can intelligently interact with dynamic environments for autonomous configuration. Leveraging the power of deep models to represent complex nonlinear system behaviors, PASS can detect outdated/inaccurate channels, identify hardware imperfections, and predict user movements and traffic variations. 
    Moreover, emerging large generative AI models can construct surrogate models, such as high-fidelity wireless digital twins, to approximate computationally expensive full-wave EM simulations. These models can synthesize real-world radio environments using multimodal sensing inputs (e.g., cameras, LiDAR, GPS), enabling PASS to learn to mitigate blockages and track mobile users cooperatively via distributed waveguides. Such intelligence allows for on-demand, proactive configuration and deployment of PASS infrastructure.
\end{itemize}

\subsection{ML-empowered Optimization}

Deep learning is a major branch of ML methods that can be exploited for PASS for optimization. In what follows, we first introduce the policies resulting from the optimization problems in Section \ref{sec:pass-opt}. A policy is defined as the mapping from known parameters (say the positions of users) to the decisions (say the pinching beamforming) of a problem. Then, we introduce deep neural network (DNN) architectures that can be used to learn these policies. We consider general multi-user systems in the sequel, because the single-user system is a special case of multi-user systems.

\subsubsection{Policies from Optimization Problems}\label{sec:prob-policy}
We next introduce two policies from the optimization problems in PASS, i.e., beamforming and power allocation.

\begin{itemize}[
    leftmargin=0pt,          
    labelsep=0em,            
    itemindent=1.5em,    
    labelwidth=1em,         
    align=left,    
    listparindent=0pt 
]
\item \textbf{Beamforming}:
We take jointly optimizing multi-user pinching and fully digital beamforming shown in problem \eqref{BF_problem_6} as an example. For simplicity, the PA radiating power is assumed to be fixed as $\omega_1=\cdots=\omega_K=1$ and $\sigma_1=\cdots\sigma_K=\sigma_0^2$.

Given a set of user positions $\bm\Phi=[\bm\psi_1,\cdots,\bm\psi_K]$, problem \eqref{BF_problem_6} can be solved by optimizing $\mathbf{X}$ and $\mathbf{W}$. The mapping from the user positions to the optimized variables is called the \emph{beamforming policy}, which is formulated by
\begin{align}\label{eq:bf-policy}
    \{\mathbf{X}^\star,\mathbf{W}^\star\}=F_B(\bm\Phi),
\end{align}
where the inputs of the policy are called the \emph{environmental parameters}.

\item\textbf{Power allocation in Wideband Systems}: In wideband multi-user systems, the power allocated to the users and subcarriers can also be optimized to maximize the system performance, measured by the spectral efficiency or energy efficiency. The optimization problem is not formulated here because it is similar to the beamforming optimization problem in \eqref{BF_problem_6}. The mapping from the user positions to the optimized pinching antenna positions and the power allocation is called the \emph{power allocation policy}, which is formulated as,
\begin{align}\label{eq:pa-policy}
    \{\mathbf{X}^\star,\mathbf{p}^\star\}=F_P(\bm\Phi),
\end{align}
where $\mathbf{p}^\star$ is the optimized vector containing the power allocated to the users in each subcarrier.
\end{itemize}

\subsubsection{DNN Architectures for Learning the Policies}
We next introduce the DNN architectures for learning the policies.
\begin{itemize}[
    leftmargin=0pt,          % 整个 item 块的左边距
    labelsep=0em,            % 项目符号和文本之间的距离
    itemindent=1.5em,        % 首行缩进距离（比 labelsep 稍大）
    labelwidth=1em,          % 项目符号所占宽度
    align=left,              % 左对齐段落
    listparindent=\parindent        % 段内段落不缩进
]
\item \textbf{Fully-Connected Neural Networks and Convolutional Neural Networks}:
 DNN architectures commonly used in the literature include fully-connected neural networks (FNNs) and convolutional neural networks (CNNs). It was empirically observed that these architectures suffer from high training complexity (in terms of the number of samples and time). Specifically, DNNs require high training complexity (in terms of samples and time) when the problem scale is large. 

\item \textbf{Graph Neural Networks}:
Graph neural networks (GNNs) have been designed for wireless communication applications recently. They have been demonstrated to be able to learn policies more efficiently with respect to the following two aspects,
\begin{itemize}
    \item \emph{Better size generalizability}: This indicates that a trained GNN can be applied to different problem scales (say the number of users) without the need of re-training.
    \item \emph{Better scalability}: This indicates that the GNN can be trained with low complexity even for large problem scales.
\end{itemize}
\indent It has been noticed that the advantages of GNNs stem from harnessing the permutation properties that widely exist in wireless policies \cite{Eisen2020,SYF, GJ_TWC_GNN,LSJ_MultiDim_GNN_2022}. For example, both the beamforming policy in \eqref{eq:bf-policy} and the power allocation policy in \eqref{eq:pa-policy} are not affected by permuting the users, waveguides, or PAs on each waveguide. 

A GNN learns over a graph with vertices and edges between them. A representative class of GNNs are graph convolutional networks (GCNs), where the convolution can be computed in the spatial or spectral domain \cite{GNN_survey}. Taking a spatial GCN as an example, the representation of every vertex or edge is updated in each layer of the GCN, by first extracting information from every neighbored vertex with a \emph{processor}, and then aggregating the extracted information with a \emph{pooling function}, and finally combining with the representation of the vertex itself with a \emph{combiner}.

When the processor is a parameterized linear function, the pooling function is a summation, and the combiner is a parameterized linear function cascaded by an activation function, the update equation of the GCN for learning over a homogeneous complete graph with $K$ vertices in the $\ell$-th layer can be expressed as
\begin{equation}\label{eq:upd-gcn}
    \mathbf{d}_k^{(\ell+1)} = \textstyle\sigma\Big(\mathbf{U}\mathbf{d}_k^{(\ell)}+\sum_{j=1,j\neq k}^K\mathbf{V}\mathbf{d}_j^{(\ell)}\Big),
\end{equation}
where $\mathbf{d}_k^{(\ell)}$ is the updated representation of the $k$-th vertex in the $\ell$-th layer, $\mathbf{U}$ and $\mathbf{V}$ are respectively the weight matrix in the combiner and processor, and $\sigma(\cdot)$ is the activation function. 

We can see that parameter sharing is introduced into the GCN, i.e., $\mathbf{U}$ and $\mathbf{V}$ are identical for all $K$ vertices. This guarantees that the input-output (I-O) relation of the update equation is equivariant to the permutations of the vertices.

% Since the processor and combiner are identical among all the vertices, the input-output relation of the GCN is equivariant to the permutations of vertices. 

% By judiciously modeling the PASS as a graph, where users and pinching antennas are vertices and the wireless links are edges, a GNN that learns over the graph is with the permutation property of the beamforming policy. Hence, the GNN has the potential of scalable and generalizable to large problem scales.

The beamforming and power allocation policies can be learned over graphs with GNNs. The permutation property of the I-O relation of the GNN should match that of the policy to be learned. Otherwise, performance degradation and poor generalizability may result \cite{GJ_TWC_GNN}. To this end, the graph and the GNN architecture need to be judiciously designed \cite{LSJ_MultiDim_GNN_2022}. For example, the PASS can be modeled as a heterogeneous graph, where the users and PAs can be defined as two types of vertices, and the edges are the links between the users and the PAs. A heterogeneous GNN that learns over the graph can be designed to learn the beamforming or power allocation policy \cite{guo2025gpass}.

In Fig. \ref{fig:perf}, we provide simulation results of learning beamforming in PASS with a GNN, which we refer to as GPASS. The learning performance is compared with two baseline methods: i) FR-BCD: This is the algorithm proposed in \cite{bereyhi2025mimo}, where fractional programming
and block coordinate descent were employed to solve the problem \eqref{BF_problem_6}, ii) CMIMO: This is the performance of fully digital beamforming in a conventional MIMO system. It can be seen that GPASS can achieve a performance close to FP-BCD, while further results demonstrate that GPASS has much lower complexity during execution than FP-BCD \cite{guo2025gpass}. GPASS performs much better than CMIMO, and the performance gain comes from adjusting the PA positions to mitigate the impact of path loss. GNN has also been designed for learning beamforming in ISAC systems \cite{gj_pass_isac_gnn}, where the GNN is trained in an unsupervised manner.
\begin{figure}[t!]
    \centering
    \includegraphics[width=.72\linewidth]{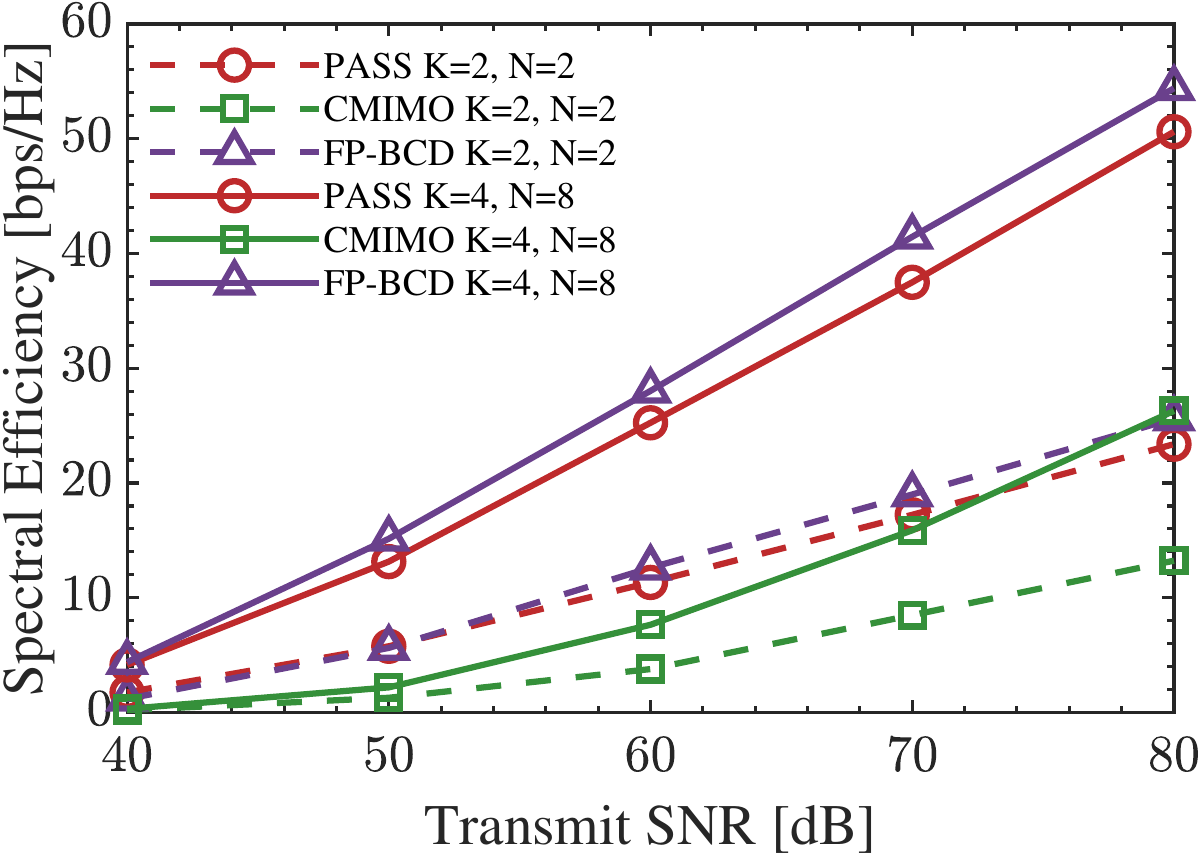}
    \caption{Spectral efficiency (SE) versus SNR. There are three PAs on each waveguide, and $K$ and $N$ are respectively the numbers of users and PAs. The users are deployed in a region of $10\times 10$ m$^2$, the height of the waveguides is $3$ m.}
    \label{fig:perf}
\end{figure}

\item \textbf{Transformers}:
Transformers 
%has been widely used in the fields of natural language processing (NLP), computer vision, and large language models, which 
have been introduced to wireless communications for tasks such as channel prediction/estimation and channel compression. 
They leverage the attention mechanism to capture global dependencies and correlations of elements (also known as ``tokens'') in sequential data, and can achieve superior performance compared to conventional recurrent neural networks (RNNs).
% Recent works have tailored the Transformer architectures for learning beamforming \cite{li2024hpe}. Such a architecture is only with an encoder, and the positional encoding that is originally used to distinguish the order of tokens (say words in a sentence) was omitted. Then, the equation of updating the representation of each of the $K$ tokens (say the $k$-th token) in the $\ell$-th layer can be expressed as \cite{DYX},
% \begin{align}\label{eq:upd-transformer}
% 			&\mathbf{d}_{k}^{(\ell)} = \sigma\Bigg(\mathbf{W}^{\mathsf{F}} \mathbf{d}_{k}^{(\ell-1)}+\notag\\
% 			&\mathbf{W}^{\mathsf{F}}\mathbf{W}^{\mathsf{V}}\sum_{i=1}^{K}  \xi\Big(\big(\mathbf{W}^{\mathsf{Q}} \mathbf{d}_k^{(\ell-1)}\big)^{\mathsf{T}} \big(\mathbf{W}^{\mathsf{K}} \mathbf{d}_i^{(\ell-1)}\big)\Big) \mathbf{d}_i^{(\ell-1)}\Bigg),
% 		\end{align}
% where $\mathbf{W}^{\mathsf{F}}, \mathbf{W}^{\mathsf{V}},\mathbf{W}^{\mathsf{Q}},\mathbf{W}^{\mathsf{K}}$ are weight matrices, and $\mathbf{d}_{k}^{(\ell)}$ is the updated representation of the $k$-th token in the $\ell$-th layer.
A transformer-based learning framework, termed \emph{KKT-guided dual learning Transformer} (KDL-Transformer), has been recently proposed to address the joint transmit and pinching beamforming optimization in PASS, which is a highly non-convex and tightly coupled problem \cite{xu2025joint}. 
KDL-Transformer integrates conventional optimization principles, i.e., KKT theory, to guide the data-driven deep learning, which significantly improves learning efficiency, reduces path loss, and mitigates multi-user interference for PASS beamforming design. 
The CSI-to-beamforming mapping can be modeled as a sequence-to-sequence prediction task, which can be solved by transformers using an encoder-decoder structure. Specifically, the encoder employs self-attention to capture inter-user CSI dependencies, while the decoder leverages cross-attention to characterize the interactions among PAs and the dependencies of beamforming solutions on the CSI. 
In each decoder layer \(\ell\), the PA positions \(\mathbf{X}^{(\ell)}\) and the dual variables \(\bm{\lambda}^{(\ell)}\) are updated based on the decoder hidden state \(\mathbf{Z}_{\mathrm{D}}^{(\ell-1)}\) through a cross-attention mechanism followed by a feed-forward NN $\mathcal{F}(\cdot)$: 
\begin{equation}
\mathbf{Z}_{\mathrm{D}}^{(\ell)}\!\!=\!\left\{ \!\mathbf{X}^{(\ell)},\bm{\lambda}^{(\ell)}\right\} \!=\!\mathcal{F}\!\left(\mathrm{softmax}\!\left(\!\frac{\mathbf{Q}^{(\ell)}\!\left(\mathbf{K}^{(\ell)}\right)^{T}}{\sqrt{N^{(\ell)}}}\!\right)\!\mathbf{V}^{(\ell)}\!\!\right)\!\!,
\end{equation}
where the query matrix $\mathbf{Q}^{(\ell)} = \mathbf{Z}_{\mathrm{D}}^{(\ell-1)} \mathbf{W}_{Q}^{(\ell)}$ models the decoder’s attention to PAs' states captured by $\mathbf{Z}_{\mathrm{D}}^{(\ell-1)}$, and the key and value matrices $\mathbf{K}^{(\ell)} = \mathbf{Z}_{\mathrm{E}} \mathbf{W}_{K}^{(\ell)}$ and $\mathbf{V}^{(\ell)} = \mathbf{Z}_{\mathrm{E}} \mathbf{W}_{V}^{(\ell)}$ are derived from the encoder output $\mathbf{Z}_{\mathrm{E}}$ that contains the embedded CSI features. Matrices $\mathbf{W}_{Q}^{(\ell)}$, $\mathbf{W}_{K}^{(\ell)}$, and $\mathbf{W}_{V}^{(\ell)}$ are learnable  NN weights that transform the encoder/decoder states into the attention subspaces. 
Moreover, $N^{(\ell)}$ denotes the dimensionality of the attention subspace at layer $\ell$.
From the predicted $\mathbf{X}^{(\ell)}$ and $\bm{\lambda}^{(\ell)}$, the transmit beamforming is further reconstructed via interpretable and differentiable closed-form solutions derived from KKT conditions.
\begin{table*}[!tb]
\centering
\footnotesize
\caption{Summary of ML for Optimization}
\label{tab:summary-ml}
\resizebox{\textwidth}{!}{%
\begin{tabular}{l|c|c|c|c|c}
\hline
\textbf{Task} & \textbf{Input} & \textbf{Output} & \textbf{Architecture} & \textbf{Training Manner} & \textbf{Characteristics} \\
\hline
% \multirow{3}{*}{\begin{tabular}[c]{@{}c@{}} Machine learning \\ for optimization \end{tabular}} & 
Beamforming & User position & \begin{tabular}[c]{@{}c@{}} PA positions and transmit \\ beamforming matrix \end{tabular} & {\begin{tabular}[c]{@{}c@{}} GNN, Transformer,\\ Graph Transformer
\end{tabular}
} & \multirow{3}{*}{Unsupervised learning} & \multirow{3}{*}{\begin{tabular}[l]{@{}l@{}}\textbullet~Real-time implementation\\\textbullet~Robust to channel \\~~estimation errors\end{tabular}}  \\
\cline{1-4}
Power allocation &  User position & \begin{tabular}[c]{@{}c@{}} PA positions and \\ power allocation vector \end{tabular} & Spatial GCN & ~ & ~ \\
\hline
\end{tabular}
}
\end{table*}
\begin{figure}[t!]
    \centering
    \includegraphics[width=.38\textwidth]{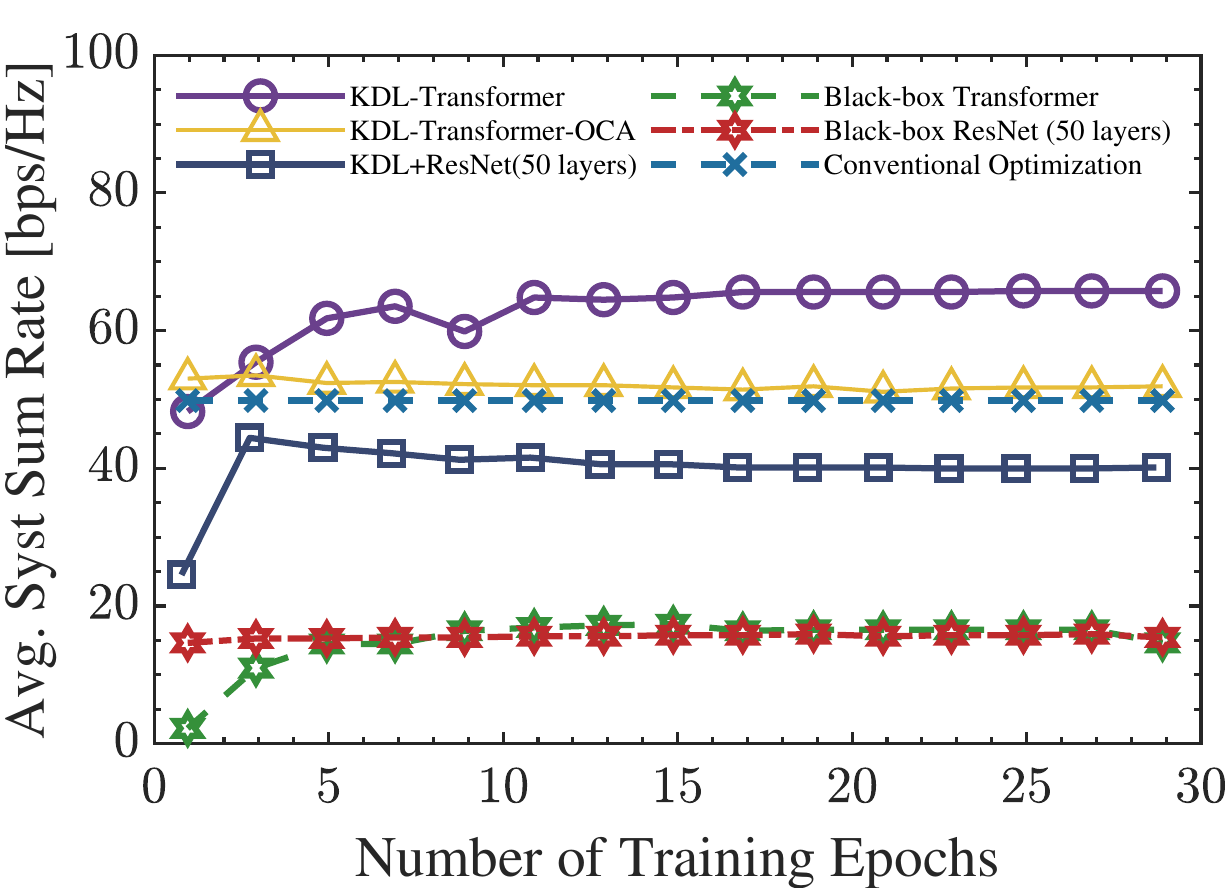}
    \caption{Comparison of the proposed KDL-Transformer and conventional gradient-based/learning-based optimization algorithms \cite{xu2025joint}.}
    \label{fig:PASS_transformer}
\end{figure}

Fig. \ref{fig:PASS_transformer} compares the convergence performance of conventional gradient-based optimization and the KDL-Transformer algorithm. 
Specifically, the black-box learning methods based on ResNet and transformer perform poorly compared to both the conventional optimization algorithm and the KDL approach, due to the non-convex coupled nature of the joint beamforming optimization. In contrast, KDL-Transformer effectively approximates the KKT points and thus significantly improves the learning performance. 
The KDL-Transformer achieves more than $20\%$ improvement over the conventional gradient-based optimization method, and outperforms KDL-ResNet and KDL-Transformer without cross-attention (KDL-Transformer-OCA) by over $40\%$ and $18\%$, respectively.
This highlights its efficiency in characterizing both inter-PA/user and CSI-beamforming dependencies.

The application of GNN and transformer is not mutually exclusive, but complementary. 
Specifically, by regarding each token as
a vertex, the encoder-only transformer without positional encoding can be regarded as a GNN that learns over a homogeneous complete graph with only one type of vertices and edges, where the attention mechanism is employed in the processor to model the correlation among representations of vertices. Hence, the transformer is also featured with the permutation equivariance (PE) property. 
By doing so, graph transformers can be designed to exploit the advantages of both the PE properties and the attention mechanism. The policies for PASS can be learned by graph transformers. The attention mechanism can model the multi-user interference, which is important for the generalizability of the number of users when learning beamforming in interference networks \cite{guo2025attention}.

\end{itemize}
 
\subsubsection{Training the DNN Architectures}
The DNNs can be trained in supervised, unsupervised, and reinforced manners, which are introduced in the following.

\begin{itemize}[
    leftmargin=0pt,       
    labelsep=0em,            
    itemindent=1.5em,      
    labelwidth=1em,     
    align=left,     
    listparindent=\parindent  
]
\item \textbf{Supervised learning}: When an optimal or suboptimal solution of the resource allocation problem is available, it can serve as a label for training the DNN. Each training sample consists of the input (i.e., environmental parameters) and the expected output (i.e., labels). Taking the learning beamforming policy in \eqref{eq:bf-policy} as an example, the $i$-th training sample can be expressed as $\{\mathbf{\Phi}^{[i]}, \mathbf{X}^{\star[i]},\mathbf{W}^{\star[i]}\}$. The DNN can be trained to minimize the difference between the learned beamforming and the labels (say MSE) averaged over all the training samples. The loss function can be expressed as
\begin{align}
    \mathcal{L}(\bm{\theta}) = \frac{1}{N_{\sf tr}}\sum_{i=1}^{N_{\sf tr}} \Big(\big(\mathbf{X}^{[i]}-\mathbf{X}^{\star[i]}\big)^2 + \big(\mathbf{W}^{[i]}-\mathbf{W}^{\star[i]}\big)^2 \Big), \notag
\end{align}
where $\mathbf{X}^{[i]}$ and $\mathbf{W}^{[i]}$ are respectively the learned pinching and transmit beamforming matrices for the $i$-th sample, $N_{\sf tr}$ is the number of training samples, and $\bm{\theta}$ denotes all the trainable parameters.

The optimization problems for PASS are usually non-convex problems, and their optimal solutions are not easy to obtain. If the DNNs are trained in a supervised manner to approximate the sub-optimal solution, the learning performance may be degraded due to the sub-optimality of the labels.

\item \textbf{Unsupervised learning}:  As an alternative to supervised learning, unsupervised learning can be used for training the DNNs \cite{guo2025gpass} without relying on labels. Taking learning the beamforming policy as an example, the loss function can be designed as the negative objective function averaged over all the training samples, which can be expressed as
\begin{align}
    \mathcal{L}(\bm{\theta})=\frac{1}{N_{\sf tr}}\!\sum_{i=1}^{N_{\sf tr}}\!\sum_{k=1}^K \!\textstyle\log_2 \!\left(\!1 \!+\! \frac{\left| \mathbf{h}_k^H (\mathbf{X}^{[i]}) \mathbf{G}(\mathbf{X}^{[i]}) \mathbf{w}_k^{[i]} \right|^2 }{\sum_{i \neq k} \left| \mathbf{h}_k^H (\mathbf{X}^{[i]}) \mathbf{G}(\mathbf{X}^{[i]}) \mathbf{w}_i^{[i]} \right|^2 +\sigma_0^2 }\!\right). \notag
\end{align}
The constraints in problem \eqref{BF_problem_6} can be satisfied by designing activation functions in the output layers of the DNNs \cite{guo2025gpass}. 

\item \textbf{Reinforcement learning}: If the optimization problem can be modeled as a Markov decision process with multiple steps, where the action of one step affects the state of the next step, then reinforcement learning can be adopted to learn the policy from the problem. Since both of the problems introduced in Section \ref{sec:prob-policy} are instantaneous decision problems where the actions (i.e., beamforming or power allocation) are only affected by the state (i.e., user positions) in the current time slot, supervised or unsupervised learning is more suitable to learn the policies. 

{
To be more specific, in reinforcement learning (RL), an agent learns a policy by interacting with the environment to maximize the expected cumulative reward, where each action affects the subsequent states. In contrast, for the beamforming and power allocation problems that only aim to maximize the reward within one time slot, the actions in every time slot do not affect future states. Therefore, supervised or unsupervised learning methods are more suitable than RL for learning such single-step policies. More discussions about the differences between RL and supervised/unsupervised learning was discussed in \cite{gj_dm_mag}.}

Due to the above reason, we do not introduce reinforcement learning here in detail.
\end{itemize}

\emph{Discussions}: For learning power allocation policies in PASS, a GCN can be employed, which can leverage the PE properties of the policies that enable scalability and size generalizability. However, for learning beamforming policies (specifically, transmit beamforming) in PASS, a GCN may yield poor performance, because interference is not reflected in the processor of the GCN, which is important for size generalizability according to the analytical results in \cite{guo2025attention}. Instead, the beamforming policy can be learned by transformers or graph transformers, where the attention mechanism can address the interference in the PASS. The GPASS proposed in \cite{guo2025gpass} adopts a processor that can learn the ``attention'' among users, which can also be regarded as a type of graph transformer.

For learning the beamforming and power allocation policies in PASS, unsupervised learning for training the DNNs seems preferable, because the optimal solutions to PASS problems are usually not available to be used as labels, due to the non-convexity and highly-coupled nature of the problems.

\begin{figure}[t!]
    \centering
    \includegraphics[width=0.95\linewidth]{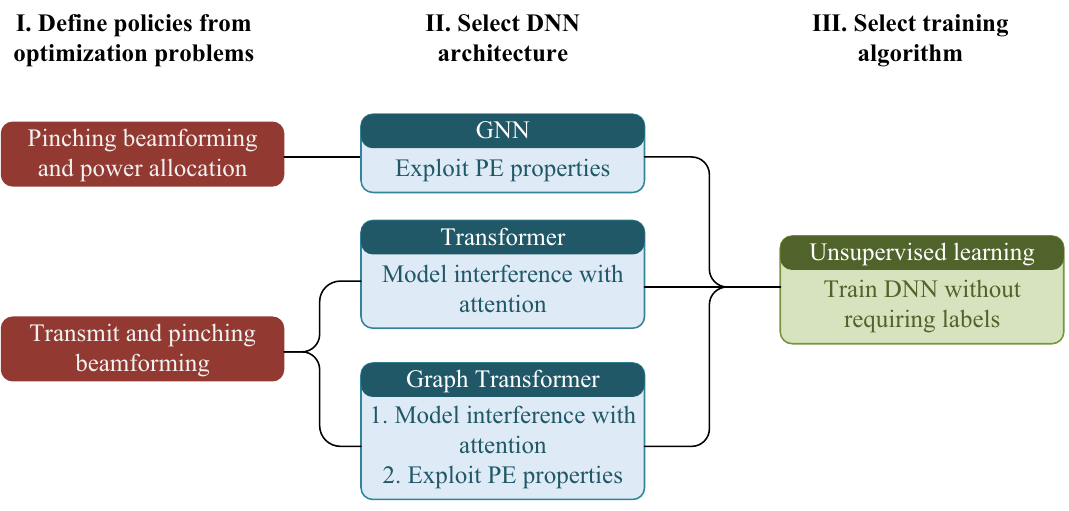}
    \caption{Illustration of deep learning for optimization}
    \label{fig:dl}
\end{figure}

The selection of DNN architectures and training algorithms is summarized in a flow chart in Fig. \ref{fig:dl}.

\subsection{ML-empowered CSI Acquisition}
\subsubsection{Machine Learning for CE}
As previously discussed, PASS needs to estimate high-dimensional channels with limited pilot measurements and RF chains. 
To address the resulting under-determined system and to accommodate various PASS deployments, multiple NNs, named ``experts'', can be trained to model the channel under different PA layouts \cite{xiao2025channel}. To inform the model of the antenna geometry, the measurement positions of the PAs, i.e., $\mathbf{x}$, can be encoded using Fourier features. 
The resulting spatial embedding captures geometric patterns and enables the model to infer high-dimensional channels from low-dimensional pilot signals $\mathbf{s}$.
The mixture of experts (MoE) is realized by a gating network, which assigns weight $\alpha_i$ to each expert $\mathcal{E}_{i}\left(\cdot\right)$ and combines the predicted outputs of $E$ experts as  
$\hat{\mathbf{h}} = \sum_{i=1}^E \alpha_i \mathcal{E}_i\left(\mathbf{x},\mathbf{s}\right)$. 
The expert NN can be realized by a transformer, where PAs are represented by multiple tokens, and their spatial relationships are learned by a self-attention mechanism. Moreover, a transformer can process a variable number of tokens, thus supporting the activation of any arbitrary number of PAs. 

The accurate PASS CSI can be probed by activating PAs at different positions to obtain pilot measurements. 
% Considering a single-waveguide PASS with $M$ PAs, multiple pilot signals are received by the user at time slot $t=1,2,\dots,T$:
% \begin{equation}
% y_{t} = \mathbf{h}_{t}^T\left(\mathbf{x}_{t}\right) \mathbf{g}_{t}\left(\mathbf{x}_{t}\right)s_{t} + z_{t}.
% \end{equation}
% where $\mathbf{g} \in \mathbb{C}^{N\times 1}$ is the in-waveguide channel, $\mathbf{h}_k \in \mathbb{C}^{N\times 1}$ is the wireless channel from PAs to user $k$, and $s_{k}[t]$ is the pilot symbol. 
% If the number of pilot training slots $T < N$, the channel estimation problem becomes:
% \begin{equation}
% \mathbf{y}_k = \mathbf{A} \mathbf{h}_k, \quad \mathbf{A} = \left[s_{k,1} \mathbf{g}, s_{k,2} \mathbf{g}, \dots, s_{k,T} \mathbf{g}\right]^T 
% \in \mathbb{C}^{T \times N}.
% \end{equation}
% It can be observed that $\text{rank}(\mathbf{A}) \leq 1$ since $\mathbf{A}$ consists of scaled vectors of $\mathbf{g}^T$. This rank deficiency renders the under-determined system, where the solution for $\mathbf{h}_k$ is non-unique. As classical estimators such as Least Squares (LS), i.e., 
% \[
% \hat{\mathbf{h}}_{\text{LS}} = (\mathbf{A}^H \mathbf{A})^{-1} \mathbf{A}^H \mathbf{y}_k,
% \]
% require $\mathbf{A}^H \mathbf{A}$ to be full rank, they result in an highly ill-posed problem and become ineffective. 
Compressed sensing methods (e.g., OMP) can be employed to estimate both angle and distance parameters, thereby recovering LoS and NLoS multi-path components. However, the inherent rank deficiency of the system typically necessitates a higher CSI acquisition overhead to achieve acceptable accuracy. Furthermore, as the number and locations of active PAs may vary dynamically along the waveguide, the channel estimator should effectively deal with diverse PA configurations. 
Particularly, the large-scale span of the dielectric waveguide results in spatially non-stationary behavior of PASS channels. Hence, it is necessary to reconstruct a high-fidelity channel from low-precision pilot measurements in response to the changes of PAs' positions.
The diffusion model \cite{xu2024generative} offers a promising solution to deal with these difficulties and to achieve channel refinement.
A diffusion model comprises a forward process that gradually injects noise into the accurate channel, and a reverse process that iteratively refines the channel estimate using the learned denoising neural network (NN). 
Let $\mathbf{h}_{\mathrm{real}}\left(\mathbf{x}\right)$ denote the perfect channel vector at the queried PA positions $\mathbf{x}$. 
Specifically, the forward process adds Gaussian noise to $\mathbf{h}_{\mathrm{real}}\left(\mathbf{x}\right)$ over $I$ steps, i.e., 
$q\left(\mathbf{h}_{i}\left(\mathbf{x}\right) \mid \mathbf{h}_{\mathrm{real}}\left(\mathbf{x}\right)\right) = \mathcal{N}\left(\mathbf{h}_{i}\left(\mathbf{x}\right); \sqrt{\bar{\alpha}_{i}} \mathbf{h}_{\mathrm{real}}\left(\mathbf{x}\right), (1 - \bar{\alpha}_{i}) \mathbf{I}\right)$,
where $\bar{\alpha}_{i} = \prod_{j=1}^i \alpha_j$ controls the noise level. 
Conditioned on the queried PA positions $\mathbf{x}$, the measurement PA positions $\mathbf{x}_{\mathrm{prob}}$, and the corresponding pilot measurement $\mathbf{y}\left(\mathbf{x}_{\mathrm{prob}}\right)$,    
the channel estimate is recovered using a denoising NN $\epsilon_\theta(\mathbf{h}_{i}\left(\mathbf{x}\right), \mathbf{x}, \mathbf{x}_{\mathrm{prob}}, \mathbf{y}\left(\mathbf{x}_{\mathrm{prob}}\right), i)$, which iteratively refines channel $\mathbf{h}_{i-1}$ in denoising step $i=I, I-1, \dots, 1$ as
\begin{equation}
\mathbf{h}_{i-1} = \mu_\theta(\mathbf{h}_{i}, \mathbf{x}, \mathbf{x}_{\mathrm{prob}}, \mathbf{y}(\mathbf{x}_{\mathrm{prob}}), i) + \boldsymbol{\Sigma}_i^{1/2} \mathbf{z}, 
\quad \mathbf{z} \sim \mathcal{N}(\mathbf{0}, \mathbf{I}),
\end{equation}
where $\mu_\theta (\cdot)$ is derived from noise prediction $\epsilon_\theta (\cdot)$, $\boldsymbol{\Sigma}_i$ denotes the covariance matrix in denoising step $i$, and $\boldsymbol{\Sigma}_i^{1/2}$ represents its matrix square root. 
By doing so, the PASS channel can be modeled as a function of the PA positions, which captures the scattering environment of users with different LoS blockage states and visible regions along the PA activation areas. 

\begin{table*}[!htb]
\centering
\footnotesize
\caption{Summary of ML for CSI Acquisition}
\label{tab:ml-csi-acquisition}
\resizebox{\textwidth}{!}{%
\begin{tabular}{l|c|c|c|c|c}
\hline
\textbf{Task} & \textbf{Input} & \textbf{Output} & \textbf{Method} & \textbf{Training Manner} & \textbf{Characteristics} \\
\hline
\multirow{2}{*}{Channel estimation} & 
\multirow{2}{*}{\begin{tabular}[c]{@{}c@{}} Pilot measurement,\\PA positions\end{tabular}} & LoS/NLoS channels & Transformer, MoE & \multirow{2}{*}{Supervised learning} & Scalable \\
\cline{3-4}\cline{6}
& & Nonstationary channels & Diffusion model & & High fidelity \\
\hline
\multirow{2}{*}{\begin{tabular}[c]{@{}c@{}}Pinching alignment \end{tabular}} & 
\multirow{2}{*}{\begin{tabular}[c]{@{}c@{}} Symbol measurement,\\initial codebook \end{tabular}} & Discrete PA activation & MAB, MLP & \multirow{2}{*}{Semi-supervised learning} & Fast convergence, low cost \\
\cline{3-4}\cline{6}
 & & Adaptive PA activation & Two-stage NNs & & Accurate, flexible \\
\hline
\multirow{3}{*}{\begin{tabular}[c]{@{}c@{}}Pinching tracking \end{tabular}} & \multirow{3}{*}{\begin{tabular}[c]{@{}c@{}} Previous observation,\\symbol measurement \end{tabular}} &  \multirow{3}{*}{Future PA positions} & EKF & None (Bayesian inference) & Low-dynamic scene, low cost\\
\cline{4-6}
~ & ~ & ~ &LSTM/GRU, MLP& DRL & 
\begin{tabular}[c]{@{}c@{}} High-dynamic scene, \\nonlinear mobility 
\end{tabular}\\
\hline
\multirow{2}{*}{\begin{tabular}[c]{@{}c@{}}Pinching prediction \end{tabular}} & Previous observation & \multirow{2}{*}{\begin{tabular}[c]{@{}c@{}} Future PA positions\\(multi-step) \end{tabular}} & Transformer & 
\multirow{2}{*}{Semi-supervised learning} & Model global dependency \\
\cline{2}\cline{4}\cline{6}
~ & Multimodal sensing & ~ & LLM (e.g., GPT) & & Zero-shot generalization \\
\hline
\end{tabular}
}
\end{table*}
\subsubsection{Machine Learning for Beam Training}
Beam training configures the pinching beamforming without requiring full and explicit CSI estimation in PASS. 
As shown in Fig. \ref{fig:ML_beamtraining}, ML can assist in three key beam training tasks for PASS, namely pinching  \textit{alignment}, pinching \textit{tracking}, and pinching \textit{prediction}, which is elaborated in the following.

\noindent $\bullet$ \textbf{Pinching Alignment:} 
Different from conventional MIMO that necessitates beam alignment at a specific angle, PASS requires both large-scale pathloss minimization (i.e., distance-domain alignment) and the small-scale phase alignment among PAs. The joint distance-phase alignment leads to the \textit{pinching alignment} problem. 
Specifically, the phase alignment aims to align the phase differences between PAs to achieve coherent combining among PAs:
\begin{equation}
\phi_{m}(x_{m,n}) - \phi_{m}(x_{m',n}) = 2k\pi,
\quad \forall m \ne m', ~ 
k \in \mathbb{Z},
\end{equation} 
where $k\in\mathbb{Z}$ is an integer, and $\phi_{m,n}(x_{m,n})$ denotes the phase of the signal radiated from PA $m$:
\begin{equation}
\phi_{m,n}(x_{m,n}) = \frac{2\pi}{\lambda}\left(\left\Vert\mathbf{r} - \mathbf{p}_{m,n}\right\Vert + n_{\mathrm{eff}}x_{m,n} \right).
\end{equation}
Two ML-based strategies can be explored. 
\textit{(i) Beam selection:} The beam selection strategy achieves pinch alignment by selecting PA activation patterns from a predefined candidate set. This process can be modeled as a stochastic multi-armed bandit (MAB) problem, where each arm corresponds to a unique PA configuration. Multi-arm bandit algorithms \cite{wei2022fast}, such as Upper Confidence Bound (UCB) and Thompson Sampling, can be employed to efficiently explore the configuration space and quickly converge to high-performing PA patterns, thereby minimizing the required number of probing attempts.  Alternatively, supervised classifiers (e.g., MLPs) can also be trained to score candidate beams (PA positions) based on limited channel measurement feedback. 
\textit{(ii) Beam Adaptation:} In contrast to selecting from a discrete codebook, the beam adaptation strategy enables NNs to generate arbitrary PA activation positions for both measurement and data transmission. Specifically, in each measurement time slot $i = 1, 2, \dots, P$,  the measurement PA position   $\mathbf{x}_{\mathrm{prob},i}$
is successively adapted by a codebook adaptation neural network parameterized by $\bm{\theta}_{1}$, thus forming a set of measurement beams 
$\mathcal{X}_{\bm{\theta}_{1}} = \left[\mathbf{x}_{\mathrm{prob},1}, \mathbf{x}_{\mathrm{prob},2}, \dots, \mathbf{x}_{\mathrm{prob},P}\right]$. This approach allows for dynamic and efficient exploration of the pinching beamforming space, eliminating the need for a predefined codebook. 
The resulting measurements $\mathbf{y}(\mathcal{X}_{\bm{\theta}_{1}})$ are used to infer the PA activations for data transmission. 
A beam alignment neural network $\mathcal{F}_{\bm{\theta}_{2}}(\cdot)$ is trained to generate activation positions $x_{m,n}$ that pursue both the distance and phase alignments. Defining $\lambda_{1}$ and $\lambda_{2}$ the weight factors, the NNs' parameters $\bm{\theta} = \left[\bm{\theta}_{1},\bm{\theta}_{2}\right]$ are trained to minimize the following loss function:
% \begin{tcolorbox}[title={Loss Function of ML-based Pinching Alignment}]
\begin{equation}
\begin{split}
\mathcal{L}_{\bm{\theta}}\!\left(\mathbf{x}\right)\!&=\!\underset{\text{distance alignment}}{\lambda_{1}\underbrace{\sum_{m=1}^{M}\left\Vert \mathbf{r}\!-\!\mathbf{p}_{m,n}\left(x_{m,n}\right)\right\Vert}}\!-\!\lambda_{2}\underset{\text{phase alignment}}{\underbrace{\left|\sum_{m=1}^{M}e^{j\phi(x_{m,n})}\right|^{2}}}, \\
x_{m,n} & \triangleq \mathcal{F}_{\bm{\theta}_{2}}\left(\mathcal{X}_{\bm{\theta}_{1}},\mathbf{y}\left(\mathcal{X}_{\bm{\theta}_{1}}\right)\right), 
\quad \forall m, n.
\end{split}
\end{equation}
% \end{tcolorbox}

\begin{figure}[t!]
    \centering
    \includegraphics[width=1\linewidth]{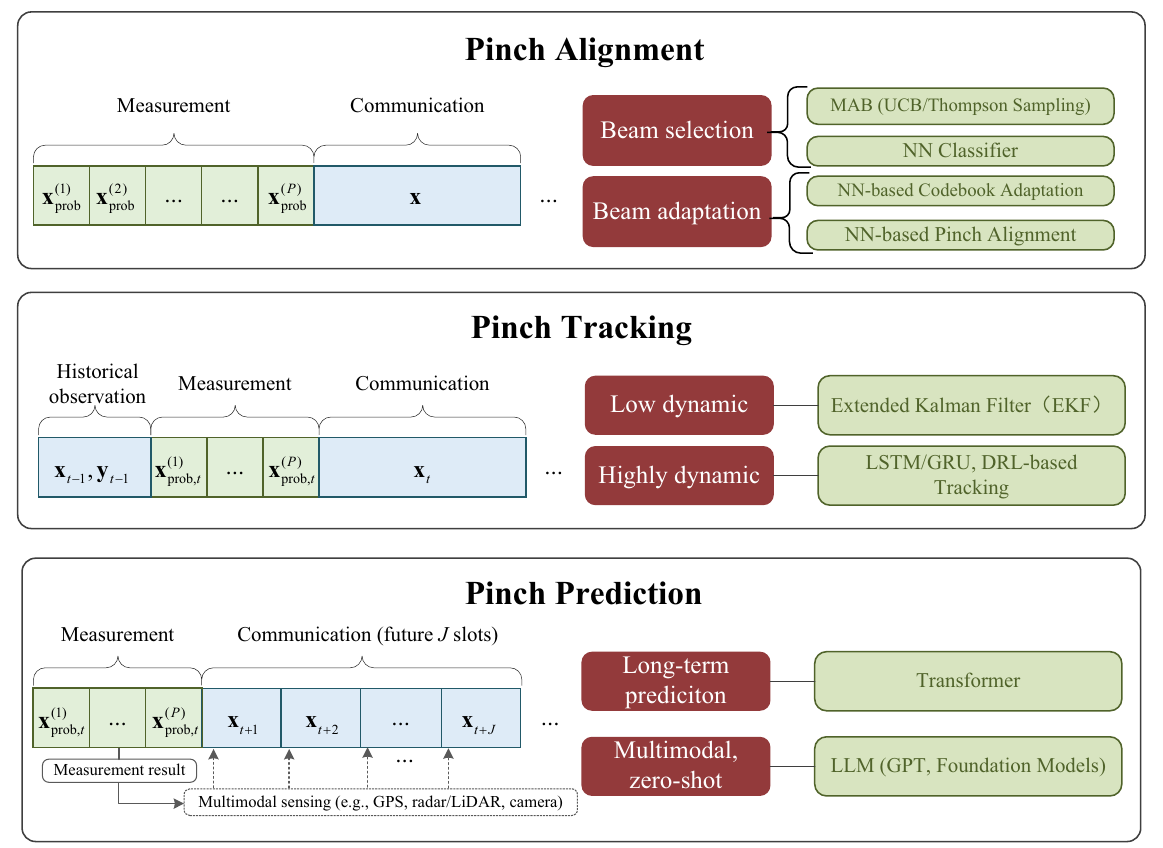}
    \caption{ML-based beam training for PASS.}
    \label{fig:ML_beamtraining}
\end{figure}

\noindent $\bullet$ \textbf{Pinching Tracking:} 
Pinching tracking addresses real-time adaptation of the PAs' positions to react to user mobility and environmental changes. It requires predictive models that exploit spatio-temporal continuity in the channel statistics.
% \textit{(i) Kinematic Modeling with Kalman Filtering:} 
For low-speed or slowly varying users, the existing kinematic models depict user motion using a linear model:
\begin{equation}\label{eq:Kinematic}
\boldsymbol{\psi}_{t+1} = \mathbf{F} \boldsymbol{\psi}_t + \mathbf{z}_t, \quad \mathbf{y}_t\left(\boldsymbol{\psi}_t\right) = \mathbf{h}^{T}\!\left(\boldsymbol{\psi}_t\right)\mathbf{g}\left(\boldsymbol{\psi}_t\right) + \mathbf{n}_t,
\end{equation}
where $\boldsymbol{\psi}_t$ represents the position of the user to be predicted, $\mathbf{F}$ denotes the state transition matrix, 
$\mathbf{y}_{t}$ denotes the received reference signal, and $\mathbf{z}_t$ and $\mathbf{n}_t$ are Gaussian noises. 
The extended Kalman Filter (EKF) \cite{chen2023mmwave} provides a lightweight and robust solution for real-time state estimation in PASS for low-dynamic scenarios. Compared to the vanilla Kalman filter, EKF can handle the nonlinear observation model $\mathbf{y}_t\left(\boldsymbol{\psi}_t\right)$, as defined in \eqref{eq:Kinematic}. EKF recursively estimates the state vector $\boldsymbol{\psi}_{t+1}$ by combining a known state transition model with the latest observation $\mathbf{y}_{t}$. The prediction step forecasts the state and its associated uncertainty, while the update step refines this estimate using the Kalman gain to effectively fuse prediction and measurement. 
However, in rapidly varying or blockage-prone scenarios, users or scatters may exhibit complicated motion dynamics that are difficult to capture using purely linear kinematic models. To address this issue, deep sequence models, such as long short-term memory (LSTM) and gated recurrent unit (GRU) networks, can be employed to characterize the temporal evolution of states for both users and scatters. These models can be trained on historical beam-RSS sequences to predict the future activation positions of PAs, effectively capturing time-sequential dependencies and nonlinear variations in user trajectories. Furthermore, by integrating LSTM/GRU models into the actor network of a deep reinforcement learning (DRL) framework, the agent can optimize PA configurations dynamically, thus balancing between the system performance and stability.
In highly dynamic environments, distributed waveguides need to be deployed and coordinated to track users for reliable communications. 
Hence, cooperative multi-agent learning algorithms need to be developed for efficient waveguide selection and pinching beamforming tracking.

\noindent $\bullet$ \textbf{Pinching Prediction:}
Unlike pinching tracking, which reacts to real-time channel measurement feedback in each time slot, \textit{pinching prediction} proactively determines pinching beamforming patterns several time slots in advance without requiring immediate feedback. By leveraging structured historical sequences, this approach improves robustness and significantly reduces beam training overhead, which is particularly beneficial for high-mobility users or latency-sensitive applications. 
Compared to RNNs, transformer models with self-attention layers can capture long-range temporal dependencies in PAs' position sequences and learn user mobility patterns effectively. 
Input features may include prior PA positions, received signal quality, and blockage indicators, while the outputs consist of single-step or multi-step forecasts of future PA activation patterns.
While both transformers and LLMs share attention-based architectures, LLMs  (e.g., GPT)  can be pre-trained on large-scale sequences and then fine-tuned for PASS-specific tasks \cite{zhu2025wireless}. LLMs offer strong generalization and zero-shot capabilities, allowing them to adapt to previously unseen scenarios or mobility patterns. 
Furthermore, LLMs are especially useful when rich environmental data is available. They can integrate multimodal sensing information from radar, GPS, and camera images by mapping these modalities into high-dimensional tokens to empower beam prediction. 
In contrast, transformers trained from scratch are typically more compact and computationally efficient than LLMs, though they require more task-specific training data to achieve comparable performance.

\subsection{Discussion and Outlook}
In Table \ref{tab:summary-ml} and Table \ref{tab:ml-csi-acquisition}, we summarize the ML methods for PASS optimization and CSI acquisition, including potential ML architectures/methods, training manners, and characteristics of different ML methods.
Despite significant advances in applying ML to PASS design, several critical open challenges remain to be resolved:
\begin{itemize}
    \item \emph{Generalization Across Heterogeneous Scenarios:}  
    Current ML models are typically trained on fixed system configurations and often fail to generalize to a case with an arbitrary numbers of users, PAs, or waveguides. Achieving generalization across network scales, layouts, and mobility patterns remains a major open issue. Architectures that leverage PE and structural priors, such as GNNs and graph-based transformers, require further theoretical development and empirical validation.
    \item \emph{Integration of Physical Constraints into Learning:}  
    Although ML offers powerful function approximation capabilities, incorporating physical laws and system-level constraints (e.g., electromagnetic compatibility, antenna geometry, hardware limitations) into learning remains challenging. Hybrid frameworks that embed optimization principles (e.g., KKT conditions) into NNs represent a promising direction, yet a generalizable and principled approach for physics-constrained learning in large-scale and real-time settings is still lacking.
    \item \emph{Reliability and Scalability in Dynamic Environments:}  
    PASS must maintain performance under highly dynamic, uncertain, and resource-constrained conditions. This includes user mobility, time-varying channels, partial CSI, and hardware impairments. Ensuring reliability under such conditions necessitates adaptive, uncertainty-aware models enabling online learning and continual adaptation. Moreover, scalability becomes critical as the numbers of users and system elements grow. Lightweight architectures, model compression, and distributed inference schemes will be essential for enabling real-time, large-scale PASS deployments.
\end{itemize}
\begin{figure*}[t!]
    \centering
    \includegraphics[width=0.95\linewidth]{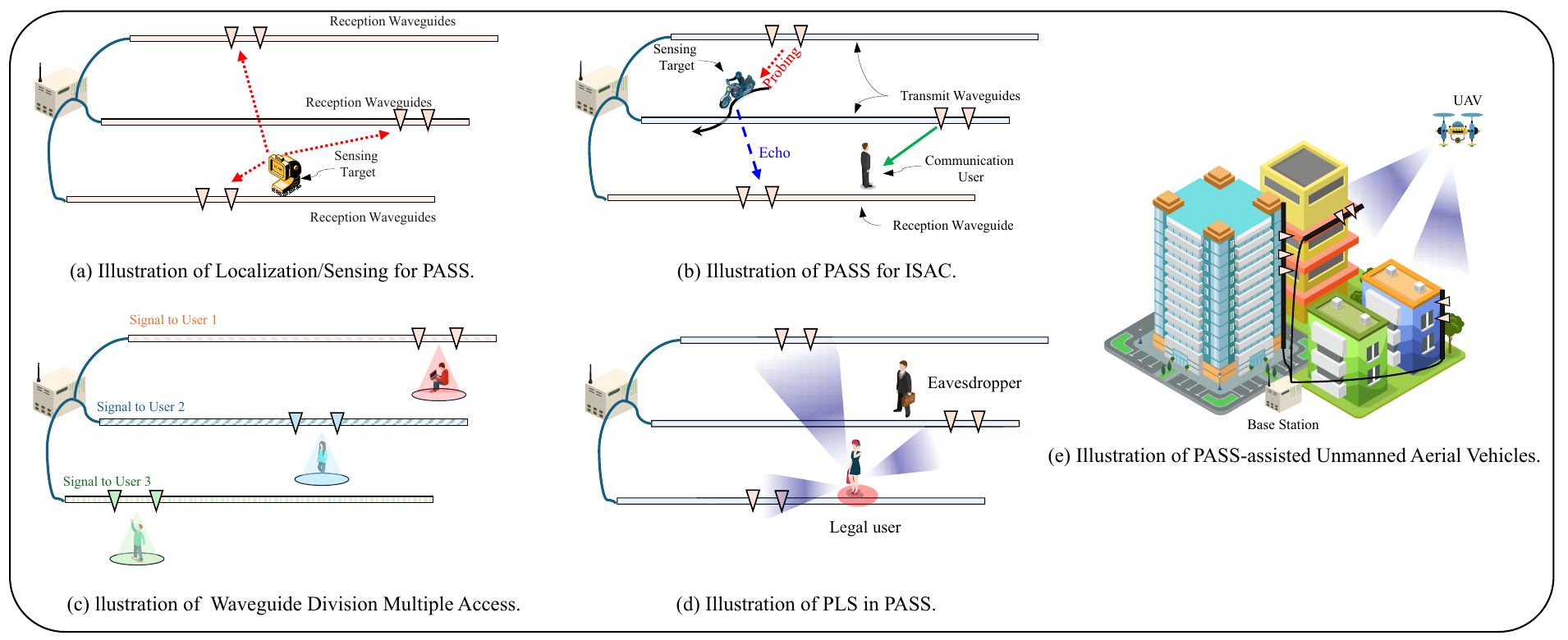}
    \caption{Illustration of the potential application for PASS.}
    \label{fig:pass-application}
\end{figure*}
\section{Application and Deployment}\label{sect:applications}
In this section, we will elaborate on the possible applications and deployment scenarios for PASS in next-generation wireless networks.

\subsection{Localization/Sensing for PASS }
The higher flexibility provided by PASS is advantageous for communications shown in Fig. \ref{fig:pass-application}(a), motivating the exploration of PASS’s sensing capabilities. 
In particular, PASS can create LoS propagation conditions by bypassing blockages, thus establishing strong direct links for sensing.
Moreover, as PAs can be sparsely distributed along long waveguides, PASS can achieve a large aperture size with a moderate number of antennas. This large aperture enhances near-field effects, enabling full-dimensional capture of the mobility status of sensing targets \cite{wang2025near}.
Recent studies have examined the sensing functionality of PASS in both uplink \cite{ding2025pinching} and downlink \cite{bozanis2025cramer} scenarios. Their results indicated that PASS can achieve higher sensing accuracy and robustness compared to conventional fixed-position MIMO systems.
On the parallel, to mitigate inter-antenna coupling, LCX has also been explored for signal reception, which enhances the sensing performance via fixed-position slots \cite{wang2001theory, wang2024antenna}. 
Although many research endeavors have been devoted to exploring the sensing capability of PASS, there are many challenging yet practical problems worth investigating.
First, when multi-PAs are leveraged for the reception of sensing signals, the coupling between received signals from different PAs should be properly modeled in light of the coupling theory.
Furthermore, it is worthwhile to study the effects of coupling on the sensing performance of PASS.
Second, despite the fact that the Cramér-Rao bound (CRB) is a widely adopted sensing performance metric, it is parameterized by unknown parameters, e.g., a user's location, that require estimation \cite{hue2006posterior}.
Thus, Bayesian CRB (BCRB) can provide a more fundamental understanding of PASS sensing performance \cite{jiang2025pinching_bcrb}. 
Unlike the conventional CRB, BCRB remains valid regardless of estimator bias and avoids dependence on unknown sensing parameters by exploiting prior position distributions, making it a suitable metric for sensing optimization.
Therefore, BCRB analysis and optimization can be a practical solution for sensing in PASS.
The BCRB derivations in \cite{jiang2025pinching_bcrb} further reveal a unique mismatch between the sensing centroid (optimal PA position) and the prior distribution centroid of the target, underscoring the need for dynamic PA repositioning. 
This originates from the near-field wavefront curvature effect: moving PAs slightly away from the target can enhance sensitivity, improving sensing accuracy, while creating a trade-off with communication throughput, which prefers alignment above the user.

\subsection{PASS for ISAC}
% \begin{figure}[t!]
%     \centering
%     \includegraphics[width=0.85\linewidth]{eps/ISAC-eps-converted-to.pdf}
%     \caption{Illustration of PASS for ISAC.}
%     \label{fig:pass-isac}
% \end{figure}
By allowing the shared use of resource blocks between sensing and communication functionalities, ISAC is a promising technical trend for the next-generation communication networks \cite{lu2024integrated, 9737357}.
As both the sensing and communication superiority of PASS have been proven in the existing works, PASS for ISAC can be a promising future direction, which is illustrated in Fig. \ref{fig:pass-application}(b). 
A few existing works have recently shed light on this issue.
For instance, the authors of \cite{zhang2025integrated} aimed to maximize illumination power while ensuring communication quality-of-service (QoS). 
Separately, the authors of \cite{ouyang2025rate} analyzed the performance limits of PASS-based ISAC systems and derived closed-form expressions for the achievable communication rate (CR) and sensing rate (SR).
The authors of \cite{jiang2025pinching_pls} developed a PASS-aided tracking scheme designed to determine the position of a malicious user and thereby maintain transmission covertness.
Due to the flexibility introduced by PASS, integrating communications and sensing can be done in a number of different ways.
First, following the idea of waveguide switching, we can assign some time slots for sensing and other slots for communications.
This design will introduce a sensing-communication tradeoff hidden in the time slot allocation procedure.
Second, according to the idea behind waveguide multiplexing, we can introduce waveguides as an additional resource block for the shared usage between communication and sensing.
Last but not least, PASS for wideband ISAC can also be a promising direction, as adequate bandwidth resources can not only enable high-fidelity sensing but also high-throughput communication. 
In a nutshell, future research may focus on how to orchestrate communication and sensing through the new sharing dimension introduced by the flexibility of PASS.

% \begin{figure}[t!]
%     \centering
%     \includegraphics[width=0.8\linewidth]{eps/wdma-eps-converted-to.pdf}
%     \caption{Illustration of waveguide division multiple access.}
%     \label{fig:pass-wdma}
% \end{figure}
\subsection{Waveguide Division Multiple Access}
Given the rapidly increasing number of connected devices, supporting efficient multi-user communications is a fundamental task in wireless networks. 
Nevertheless, since PAs activated on a given waveguide can be fed only with the same signal source, the design of PASS-based multi-user communications becomes more challenging.
Therefore, existing research contributions \cite{ding2025flexible,wang2024antenna} have explored the employment of NOMA to facilitate single-waveguide PASS-based multi-user communications. Moreover, for multiple-waveguide PASS-based multi-user communications, baseband transmit beamforming was introduced in \cite{wang2025modeling} to be jointly designed with the pinching beamforming for enabling multi-user communications through spatial division multiple access (SDMA). This, however, requires complex baseband signal processing capabilities and potentially high computational complexity to address the highly coupled joint transmit and pinching beamforming problem. To facilitate a simple yet efficient multiple-waveguide PASS-based multi-user communication framework, a novel waveguide division multiple access (WDMA) design was proposed in \cite{zhao2025waveguide}, as illustrated in Fig. \ref{fig:pass-application}(c). The key principle of WDMA is to treat the waveguide as a new type of radio resource similar to time/frequency resource blocks, and thus each user is served by a dedicated waveguide. To mitigate the inter-waveguide interference, pinching beamforming is used to enhance the desired signal reception at the served users and mitigate the interference imposed on other users, i.e., realizing the nearly orthogonal transmission in the spatial domain. Compared to the joint transmit and pinching beamforming design, the advantage of WDMA is that the baseband signal processing unit only needs to carry out simple power allocation among users, which reduces the design complexity. However, as WDMA relies on analog pinching beamforming for inter-user interference mitigation, the DoFs for communication design are limited. A notable performance degradation may occur when employing WDMA in dense user distributions, where the co-channel interference is significant. Therefore, further research efforts are required to develop an adaptive multi-user transmission protocol for PASS to strike the trade-off between the achieved performance and the required design complexity.    

\subsection{Physical Layer Security}
% \begin{figure}[t!]
%     \centering
%     \includegraphics[width=0.8\linewidth]{eps/PLS-eps-converted-to.pdf}
%     \caption{Illustration of PLS in PASS.}
%     \label{fig:pass-pls}
% \end{figure}
The LoS-dominated and short-range transmission link provided by PASS may facilitate the potential information eavesdropping. To address this issue, physical layer security (PLS) is a promising technique to safeguard PASS-based transmissions, as shown in Fig. \ref{fig:pass-application}(d). As demonstrated by many existing works, pinching beamforming can be exploited for simultaneous desired signal enhancement and interference mitigation. Motivated by this, several works have investigated the performance gain of PASS for PLS \cite{zhu2025pinching,Papanikolaou2025pls,sun2025pls}. Focusing on a fundamental secure communication system with one legitimate user and one eavesdropper, the authors of \cite{zhu2025pinching} investigated both the single-waveguide and dual-waveguide cases for PASS-based PLS. For the single-waveguide case, a PA-wise successive tuning algorithm was proposed to achieve constructive signal enhancement at the legitimate user and destructive signal mitigation at the eavesdropper. For the dual-waveguide case, artificial noise (AN) is further employed for both waveguide multiplexing and waveguide division transmission structures. It was demonstrated that the secrecy rate can be significantly improved by PASS compared to conventional fixed antenna technologies.
Moreover, the authors of \cite{Papanikolaou2025pls} studied a multiple-waveguide PASS-based PLS system with the aid of AN, where the baseband legitimate beamforming, AN beamforming, and the positions of the PAs were jointly optimized to maximize the secrecy rate. As a further advance, the authors of \cite{sun2025pls} studied a multiple-waveguide PASS-based AN-aided PLS system with multiple legitimate users and eavesdroppers, where a fractional programming-based block coordinate descent (BCD) algorithm was developed to maximize the system weighted secrecy sum-rate. Note that existing works on PASS-based PLS design all assumed that perfect CSI can be obtained, which is quite challenging in practice, especially for eavesdroppers. How to optimize robust PASS-based PLS designs constitutes an interesting but challenging research direction, which needs more research efforts in the future.

% \begin{figure}[t!]
%     \centering
%     \includegraphics[width=0.6\linewidth]{eps/UAV-eps-converted-to.pdf}
%     \caption{Illustration of PASS-assisted Unmanned Aerial Vehicles.}
%     \label{fig:pass-uav}
% \end{figure}
\subsection{PASS-assisted UAVs} 
{While most existing studies have focused on indoor deployments, PASS also offer significant potential for outdoor mobile scenarios, such as aerial, space, and vehicular communications \cite{liu2025pinching}. In such cases, dielectric waveguides can be installed along the sides of buildings, integrated into roadside infrastructure, or mounted on other supporting structures to enable flexible and efficient signal transmission. A representative example is the application of PASS in UAV-assisted outdoor wireless systems, which play a pivotal role in supporting the emerging low-altitude economy, as illustrated in Fig. \ref{fig:pass-application}(e).} By enabling dynamic and real-time control over signal propagation characteristics, such as LoS availability and path loss, PASS transcends the limitations of conventional static or mechanically steered antenna systems. These unique capabilities are particularly valuable in UAV scenarios, where mobility, energy constraints, and highly dynamic environments pose significant challenges for reliable and efficient system performance \cite{7470937}. Intelligent beamforming and blockage-aware trajectory design can be exploited to maintain robust LoS links in complex urban or obstacle-rich environments \cite{8038869}. To achieve this goal, PASS can be deployed in different areas, such as urban roadways,  building facades, or other infrastructures, to effectively extend both air-to-air and air-to-ground communication coverages.
To serve UAVs operating at high speeds, distributed waveguides can be coordinated to ensure consistent and low-latency connectivity. Furthermore, ML techniques can be leveraged to learn environment-specific beamforming, predict channel variations, and adaptively configure radiation properties to meet diverse communication and sensing requirements.
Since PASS enables on-the-fly reconfiguration of path loss, phase front, and radiation pattern, it also empowers UAVs to sense their environment in a directionally selective and multi-perspective fashion \cite{9737357}. Joint optimization of waveform design and PASS configuration can further enhance the sensing resolution and reliability. 
As UAVs can serve as airborne mobile edge computing servers \cite{9305267,10462480}, PASS-assisted UAV networks further facilitate link-quality-aware computation offloading, task partitioning, and latency-constrained smart computing.

\section{Conclusions }\label{sect:conclusions}
This paper presented a comprehensive tutorial on PASS, which is a newly emerging flexible-antenna technology.
In contrast to conventional antenna technologies, PASS enables large-scale antenna reconfigurations, LoS channels, scalable implementation, and near-field benefits.
First, the fundamentals of PASS were presented to explain the channel and signal models (narrow‑band LoS and NLoS), analyze hardware and radiation via coupling theory, and review activation methods for different scenarios.
Then, information-theoretical capacity limits of PASS were characterized together with other common performance metrics.
Building upon these foundations, pinching beamforming design was investigated for both the single- and multi-waveguide cases, followed by a section proposing two approaches to CSI acquisition.
Furthermore, ML-based solutions to PASS design were elaborated to overcome the high complexity and sub-optimality encountered with conventional optimization approaches.
Finally, several promising applications of PASS were discussed.

The introduction of PASS represents a significant paradigm shift in flexible-antenna technologies, extending antenna flexibility from the immediate vicinity of antenna arrays to distances of several meters.
With this unprecedented level of flexibility, PASS enables a wide range of novel use cases for 6G and beyond, providing fertile ground for interdisciplinary research spanning physics, antenna design, communications, machine learning, and more.
As PASS remains in its early stages, this tutorial is dedicated to researchers eager to explore the untapped potential of this promising new frontier.

\begin{appendices}
\section{Multiport Network-based Model}\label{Appendix_1}
Let $\mathbf{v}^{+} = [v_1^+, v_2^+, v_3^+]^T$ and $\mathbf{v}^{-} = [v_1^-, v_2^-, v_3^-]^T$ denote the incident and reflected voltage waves at the three ports of the PA, respectively. These are related by the scattering matrix as follows:
\begin{equation}
  \label{appendix_A_0}
  \mathbf{v}^{-} = \mathbf{S} \mathbf{v}^{+}.
\end{equation}
According to the waveguide propagation characteristics, the input and output voltages at the source network are $e^{-\gamma_{\mathrm{G}} L_1} v_1^-$ and $e^{\gamma_{\mathrm{G}} L_1} v_1^+$, respectively, with $L_1$ denoting the length of the waveguide connected to the port 1. Let $v_0$ be the source voltage at the transmitter.
Then, the voltage relationship at the source is given by 
\begin{equation}
  \label{appendix_A_1}
  e^{\gamma_{\mathrm{G}} L_1} v_1^+ = v_0 + \Gamma_{\mathrm{S}}e^{-\gamma_{\mathrm{G}} L_1} v_1^-,
\end{equation}
where $\Gamma_{\mathrm{S}}$ is the reflection coefficient of the source impedance $Z_{\mathrm{S}}$ and is given by $\Gamma_{\mathrm{S}} = \frac{Z_{\mathrm{S}} - Z_0}{Z_{\mathrm{S}} + Z_0}$. 
Similarly, the load conditions at ports 2 and 3 yield:
\begin{align}
  \label{appendix_A_2}
  e^{\gamma_{\mathrm{G}} L_2} v_2^+ &= \Gamma_{\mathrm{L}}e^{-\gamma_{\mathrm{G}} L_2} v_2^-, \\
  \label{appendix_A_3}
  v_3^+ &= \Gamma_{\mathrm{R}} v_3^-, 
\end{align}
where $L_2$ denotes the length of the waveguide connected to the port 2, and $\Gamma_{\mathrm{L}}=\frac{Z_{\mathrm{L}} - Z_0}{Z_{\mathrm{L}} + Z_0}$ and $\Gamma_{\mathrm{R}}=\frac{Z_{\mathrm{R}} - Z_0}{Z_{\mathrm{R}} + Z_0}$ are the reflection coefficients of the termination and radiation loads, respectively. Combining \eqref{appendix_A_1}–\eqref{appendix_A_3}, we obtain:
\begin{align}
  \label{appendix_A_4}
  \mathbf{v}^+ =  \mathbf{v}_0 + \mathbf{\Gamma} \mathbf{v}^-,
\end{align}
where 
\begin{align}
  \mathbf{\Gamma} = \mathrm{diag} \left\{ \Gamma_{\mathrm{S}} e^{-2\gamma_{\mathrm{G}} L_1}, \Gamma_{\mathrm{L}} e^{-2\gamma_{\mathrm{G}} L_2}, \Gamma_{\mathrm{R}} \right\}, \\
  \mathbf{v}_0 = \left[ e^{-\gamma_{\mathrm{G}} L_1} v_0, 0, 0 \right]^T.
\end{align}
Substituting \eqref{appendix_A_0} into \eqref{appendix_A_4} yields:
\begin{align}
  \mathbf{v}^+ &= e^{-\gamma_{\mathrm{G}} L_1} (\mathbf{I}_3 - \mathbf{\Gamma} \mathbf{S})^{-1} \mathbf{e}_1 v_0, \\
  \mathbf{v}^- &= e^{-\gamma_{\mathrm{G}} L_1} \mathbf{S} (\mathbf{I}_3 - \mathbf{\Gamma} \mathbf{S})^{-1} \mathbf{e}_1 v_0,
\end{align}
where $\mathbf{e}_1 = [1,0,0]^T$.

The overall source voltage $v_{\mathrm{S}}$  is given by the sum of the input and reflected components at port 1 as follows:
\begin{align} \label{appendix_A_5}
  v_{\mathrm{S}} = & e^{\gamma_{\mathrm{G}} L_1} v_1^+ + e^{-\gamma_{\mathrm{G}} L_1} v_1^- \nonumber \\
  = & \mathbf{e}_1^H \left(\mathbf{I}_3 + e^{-2\gamma_{\mathrm{G}}L_1} \mathbf{S}\right) \left( \mathbf{I} - \mathbf{\Gamma} \mathbf{S} \right)^{-1} \mathbf{e}_1 v_0.
\end{align}      
The radiation voltage $v_{\mathrm{R}}$, defined as the total voltage at port 3, is given by
\begin{align}
  \label{appendix_A_6}
  v_{\mathrm{R}} = & v_3^+ + v_3^- \nonumber \\
  = & e^{-\gamma_{\mathrm{G}} L_1} \mathbf{e}_3^H \left( \mathbf{I}_3 + \mathbf{S} \right)\left( \mathbf{I}_3 - \mathbf{\Gamma} \mathbf{S} \right)^{-1} \mathbf{e}_1 v_0,
\end{align}
where $\mathbf{e}_3 = [0,0,1]^T$. Combining \eqref{appendix_A_5} and \eqref{appendix_A_6} leads to \eqref{multiport_voltage_relation}.

\section{A Full Derivation of \eqref{OP_Calculation_Expression}}\label{Appendix_OP_Calculation_Expression}
The outage probability in \eqref{OP_Calculation_Expression_Basic} can be expanded as follows:
\begin{equation}
\begin{split}
{\mathcal{P}}_{\rm{PASS}}&=\Pr(\log_2(1+\gamma)<R_{\rm{target}}|\varepsilon=0)\Pr(\varepsilon=0)\\
&+\Pr(\log_2(1+\gamma)<R_{\rm{target}}|\varepsilon=1)\Pr(\varepsilon=1).
\end{split}
\end{equation}
Applying \emph{Bayes' theorem}, the outage probability becomes
\begin{equation}\label{OP_Basic_After}
\begin{split}
{\mathcal{P}}_{\rm{PASS}}&=\Pr(\varepsilon=0)+\Pr(\sqrt{y_{\rm{R}}^2+z_{\rm{G}}^2}>\tau_1)\\
&\times\Pr(\varepsilon=1|\sqrt{y_{\rm{R}}^2+z_{\rm{G}}^2}>\tau_1).
\end{split}
\end{equation}
Given that ${\mathcal{D}}=\{(x,y)|y^2+z_{\rm{G}}^2>\tau_1^2,y\in[-\frac{D_y}{2},\frac{D_y}{2}],x\in[-\frac{D_x}{2},\frac{D_x}{2}]\}$, it follows that $\Pr(\sqrt{y_{\rm{R}}^2+H^2}>\tau_1)=\iint_{{\mathcal{D}}}\frac{1}{D_xD_y}{\rm{d}}x{\rm{d}}y$ and $\Pr(\alpha=1|\sqrt{y_{\rm{R}}^2+z_{\rm{G}}^2}>\tau_1)=\iint_{{\mathcal{D}}}\frac{{\rm{e}}^{-\beta \sqrt{y^2+z_{\rm{G}}^2}}}{D_xD_y\Pr(\sqrt{y_{\rm{R}}^2+z_{\rm{G}}^2}>\tau_1)}{\rm{d}}x{\rm{d}}y$. Substituting this into \eqref{OP_Basic_After} yields \eqref{OP_Calculation_Expression}.

\section{Derivation of the Maximum Received Power and Power Scaling Law} \label{theorem_scaling_law_proof}
Under the assumptions of equal power radiation and continuous activation, the receive power is upper bounded as follows:
\begin{align}\label{Array_Gain_Upper_Bound_Maximization_Pre}
    P_r \leq \frac{\eta^2 P_t}{M} \left\lvert\sum_{m=1}^{M}\frac{1}{\sqrt{(x_m - x_{\mathrm{R}})^2 + \zeta^2}}\right\rvert^2.
\end{align}
As shown in \cite{xu2025rate,ouyang2025array}, the right-hand side of \eqref{Array_Gain_Upper_Bound_Maximization_Pre} is maximized when the PAs are uniformly placed with minimum spacing $\Delta_{\min}$, and the array aperture is centered at the projection of the user along the waveguide, which yields $x_m=(m-1)\Delta_{\min}+x_1, \forall m>1,$ and $x_M+x_1=2x_{\mathrm{R}}$. Assuming $M$ is even for simplicity, the upper bound simplifies to 
\begin{align}\label{Array_Gain_Upper_Bound_Maximization_Again}
P_r\leq\frac{\eta^2 P_t}{M}
\left\lvert\sum_{m=1}^{{M}/{2}}\frac{2}{\sqrt{(\Delta_{\min}/2+(m-1)\Delta_{\min})^2 + \zeta^2}}\right\rvert^2\triangleq P_r^{\star}.
\end{align}
As demonstrated in \cite{ouyang2025array}, this upper bound can be tightly approached using the antenna position refinement method proposed in \cite{xu2025rate}. The method aligns the total phase shifts---accounting for both free-space and waveguide propagation---by initially placing antennas uniformly with spacing $\Delta_{\min}$, then applying small perturbations to optimize constructive signal combination at the receiver. Given that a propagation delay of one wavelength induces a $2\pi$-phase shift, these slight perturbations serve to align phase contributions, which effectively approximates the array gain of a uniformly spaced PASS without phase compensation. Consequently, $P_r^{\star}$ provides a tight approximation of the maximum achievable receive power, i.e., $P_{r, \max}\triangleq\max_{ \mathbf{x} \in{\mathcal{X}}}P_r\approx P_r^{\star}$. An approximate expression for $P_r^{\star}$ was derived in \cite{ouyang2025array} and given in \eqref{maximum_receive_power}. 

For completeness, we also derive a lower bound on the maximum receive power. When the receive power is maximized, all antennas are phase-aligned, which yields
\begin{align}\label{Array_Gain_Maximum}
P_{r, \max}=\frac{1}{M}\left(\sum_{m=1}^M \frac{\eta  }{\sqrt{(x_m^{\star} - x_{\mathrm{R}})^2 + \zeta^2}}\right)^2   P_t,
\end{align}
where $\{x_m^{\star}\}_{m=1}^{M}$ denotes the set of optimized antenna positions. Let $\Delta_{\max}$ denote the largest inter-antenna spacing among these positions. By enforcing uniform spacing $\Delta_{\max}$ in \eqref{Array_Gain_Maximum}, we obtain the following lower bound:
\begin{align}\label{Array_Gain_Maximum_Lower_Bound}
P_{r, \max}\geq\left(\sum_{m=1}^{M/2}\frac{2\sqrt{\eta}}
{\sqrt{M}\sqrt{\left(m-1/2\right)^2\Delta_{\max}^2+\zeta^2}}\right)^2P_t.
\end{align}
This result follows from using a uniformly spaced array centered at the user location. According to the design of the antenna position refinement algorithm, $\Delta_{\max}$ is on the order of the wavelength. Following the same steps as used to obtain the upper bound in \eqref{maximum_receive_power}, the lower bound can be approximated as $\frac{2 \eta^2 P_t}{\zeta \Delta_{\max}} f_{\mathrm{ub}} \left( \frac{M \Delta_{\max}}{2 \zeta} \right)$. Taken together, both the upper and lower bounds of the maximum received power scale like ${\mathcal{O}}\left(\frac{\ln^2{M}}{M}P_t \right)$ as $M\rightarrow\infty$. Hence, the power scaling law is rigorously characterized via the squeeze theorem \cite{sohrab2003basic}.
\end{appendices}

\bibliographystyle{IEEEtran}
\bibliography{mybib}

\end{document}